\documentclass[usenatbib]{mn2e}
\usepackage[pdftex]{graphicx}
\newcommand{\be}{\begin{equation}}
\newcommand{\ee}{\end{equation}}
\newcommand{\bea}{\begin{eqnarray}}
\newcommand{\eea}{\end{eqnarray}}
\newcommand{\bc}{\begin{center}}
\newcommand{\ec}{\end{center}}

\renewcommand{\vec}[1]{ {\bmath #1} }

\def\gsim{ \lower .75ex \hbox{$\sim$} \llap{\raise .27ex \hbox{$>$}} }
\def\lsim{ \lower .75ex \hbox{$\sim$} \llap{\raise .27ex \hbox{$<$}} }

\renewcommand{\thefootnote}{\fnsymbol{footnote}}

\title[The hydrodynamics of galaxy formation]{Moving mesh cosmology:
  the hydrodynamics of galaxy formation}

\author[Sijacki et al.]
       {\parbox{18cm}{Debora~Sijacki$^{1}$\footnotemark[1], Mark
           Vogelsberger$^{1}$, Du\v{s}an Kere\v{s}$^{2,3}$\footnotemark[2], Volker
           Springel$^{4,5}$, and Lars Hernquist$^{1}$}\vspace{0.3cm}\\ 
         $^1$ Harvard-Smithsonian Center for Astrophysics, 60 Garden Street,
         Cambridge, MA, 02138, USA\\ 
         $^2$ Department of Astronomy and Theoretical Astrophysics Center, University
         of California, Berkeley, CA 94720-3411, USA\\
         $^3$ Department of Physics, Center for Astrophysics and Space Sciences,
         University of California at San Diego, \\\hspace{0.2cm} 9500 Gilman Drive, La Jolla, CA
         92093, USA\\
         $^4$ Heidelberg Institute for Theoretical Studies, Schloss-Wolfsbrunnenweg 35, 69118 Heidelberg, Germany\\
         $^5$ Zentrum f\"{u}r Astronomie der Universit\"{a}t Heidelberg, ARI,
         M\"onchhofstr. 12-14, 69120 Heidelberg, Germany\\}

\begin{document}

\maketitle
\begin{abstract} 
  We present a detailed comparison between the well-known smoothed particle
  hydrodynamics (SPH) code {\small GADGET} and the new moving-mesh code
  {\small AREPO} on a number of hydrodynamical test problems. Through a
  variety of numerical experiments with increasing complexity we establish a
  clear link between simple test problems with known analytic solutions and
  systematic numerical effects seen in cosmological simulations of galaxy
  formation. Our tests demonstrate deficiencies of the SPH method in several
  sectors. These accuracy problems not only manifest themselves in idealized
  hydrodynamical tests, but also propagate to more realistic simulation setups
  of galaxy formation, ultimately affecting local and global gas properties in
  the full cosmological framework, as highlighted in companion papers by
  \citet{Vogelsberger2011} and \citet{Keres2011}. We find that an inadequate
  treatment of fluid instabilities in {\small GADGET} suppresses entropy
  generation by mixing, underestimates vorticity generation in curved shocks
  and prevents efficient gas stripping from infalling substructures. Moreover,
  in idealized tests of inside-out disk formation, the convergence rate of gas
  disk sizes is much slower in {\small GADGET} due to spurious angular
  momentum transport. In simulations where we follow the interaction between a
  forming central disk and orbiting substructures in a massive halo, the final
  disk morphology is strikingly different in the two codes. In {\small AREPO},
  gas from infalling substructures is readily depleted and incorporated into
  the host halo atmosphere, facilitating the formation of an extended central
  disk. Conversely, gaseous sub-clumps are more coherent in {\small GADGET}
  simulations, morphologically transforming the central disk as they impact
  it. The numerical artifacts of the SPH solver are particularly severe for
  poorly resolved flows, and thus inevitably affect cosmological simulations
  due to their inherently hierarchical nature. Taken together, our numerical
  experiments clearly demonstrate that {\small AREPO} delivers a physically
  more reliable solution.
\end{abstract}


\section{Introduction}
\renewcommand{\thefootnote}{\fnsymbol{footnote}}
\footnotetext[1]{Hubble Fellow. E-mail: dsijacki@cfa.harvard.edu}
\footnotetext[2]{Hubble Fellow.}

Numerical simulations have become an indispensable tool for studying astrophysical
phenomena.  Over the last decade, the numerical accuracy and
fidelity of simulation methods have undergone drastic improvements, both in terms
of resolvable dynamic range and of the complexity of the physical processes
that are now routinely incorporated into the codes. This remarkable progress in
numerical techniques coupled with the ever increasing power of high
performance computing platforms has led to major advances in a number of
topics, such as star formation \citep{Klessen2009}, accretion disk dynamics
and associated jet phenomena \citep{Gammie2003, DeVilliers2003}, supernova
explosions \citep{Janka2007}, black hole coalescence \citep{Pretorius2007},
and cosmic structure formation \citep{Millennium}.    

Clearly, an in-depth understanding of many astrophysical
processes, particularly those that are inherently non-linear, relies
crucially on the accuracy and realism of the numerical approach. While
the latter is largely determined by the degree of adequateness,
comprehensiveness and consistency of the physical model assumed, the
former depends sensitively on the discretization scheme adopted for
the equations, which with increasing resolution should converge to the
correct solution. Nonetheless, even for some elementary physical
problems, with known analytic solutions, simulation methods can
sometimes produce poorly converged results, or even converge to a
solution which is, however, different from the expected result
\citep{SpringelRev2010}.

In the context of hydrodynamical cosmological simulations, an
important example is given by the seminal work of \citet{Frenk1999}
(the Santa Barbara Comparison Project). This study performed a detailed
comparison between twelve different simulation codes that tracked the
non-radiative evolution of a forming galaxy cluster in a cold dark
matter (CDM) cosmology. The initial conditions were generated
independently by each group from a provided linear theory density or
displacement field. Very different numerical methods, which can be
broadly classified into smoothed particle hydrodynamics (SPH) and
mesh-based techniques, employing also different gravity solvers and
different effective resolutions, were then used by the groups to
follow the formation and evolution of the target object. Reassuringly,
the Santa Barbara Comparison Project has shown that the global
properties of the simulated object, both in terms of dark matter and
gas distribution, were in reasonable agreement among the
codes. However, detailed gas properties of the Santa Barbara cluster,
notably in the central region, exhibited a much poorer level of
consistency. In particular, \citet{Frenk1999} pointed out that there
is a systematic discrepancy in the central entropy profiles predicted
by SPH and mesh-based codes, with the former producing power-law
entropy profiles all the way to the centre and the latter yielding
some form of entropy core.

Even though progress in numerical techniques has led to substantial
improvements in current simulation codes compared to those considered
in \citet{Frenk1999}, the systematic difference in the predicted
central cluster entropy still persists today, and its origin has not
been fully understood thus far. \citet{Agertz2007} have performed a
set of numerical experiments with SPH and mesh codes, where they have
simulated the evolution of a cold, dense blob in a hot windtunnel. By
comparing the outcomes from different methods (and at various
resolutions) against the characteristic blob disruption timescale
expected analytically, they conclude that in SPH codes the development
of fluid instabilities can be numerically hampered. The suppression of
dynamical fluid instabilities, such as Kelvin-Helmholtz,
Rayleigh-Taylor, and Richtmyer-Meshkov instabilities, leads to less
efficient mixing of fluid elements with different specific entropy,
and hence also inhibits entropy generation through mixing in the
simulated system. In simulations of colliding isolated galaxy
clusters, \citet{Mitchell2009} have shown that the central entropy
profiles obtained with the SPH code {\small GADGET} and the mesh code
{\small FLASH} show a similar level of discrepancy as found by
\citet{Frenk1999}, which they attribute to the different levels of
mixing.

There have been a number of attempts to improve the description of
mixing in SPH by modifying the standard set of discretized equations
\citep[see e.g.][]{Price2008, Wadsley2008, Hess2010} or by
incorporating a Riemann solver in place of the artificial viscosity
\citep[e.g.][]{Inutsuka2002, Cha2003, Murante2011}. For example,
\citet{Price2008} has shown that the introduction of an artificial
thermal conductivity term can improve the behavior of fluid elements
at contact discontinuities, which in turn leads to better developed
Kelvin-Helmholtz instabilities. More generally, \citet{Wadsley2008}
suggested that a physical modeling of heat diffusion is necessary when
simulating high Reynolds number flows (both for SPH and mesh-based
methods), and that by applying such an approach a flat entropy core is
likely a more correct solution for non-radiative galaxy cluster
simulations.

Differences in the hydrodynamical solver between SPH and mesh-based codes are
possibly not the only reason for the systematically different central entropy
profiles in the Santa Barbara Comparison Project. As mentioned above, the
groups involved in this study did not use identical initial conditions, and
did not perform their simulations at equal numerical resolutions and with
identical gravity solvers, which opens up the possibility of additional
sources for discrepancies.  A more recent code comparison study by
\citet{Heitmann2008} evolved uniform, dark matter only cosmological boxes in a
$\Lambda$CDM universe with $10$ different codes, starting from the same
initial conditions. They showed that for large systems there is a reassuring
agreement in the halo mass functions between the codes, as well as in the
internal structures of haloes in the outer regions. However, the study by
\citet{Heitmann2008} also revealed significant discrepancies for small halos,
demonstrating that the typical root grid resolution commonly adopted in
adaptive mesh refinement (AMR) codes for simulations of cosmic structure
formation is overly coarse and leads to a suppression of low mass halo
formation \citep[see also][]{OShea2005}. This emphasizes the need to
simultaneously strive for high accuracy both in the gravity solver and in
  the hydrodynamics. Additionally to the high accuracy of the gravity and
  hydro solver, it is also of prime importance to adopt sufficiently high mass
  and spatial resolution for the studied problem at hand. For example, if
  besides pure hydrodynamics other physical processes, such as star formation
  and associated feedback, are modelled in cosmological simulations, it is
  necessary to resolve all star-forming haloes with a sufficiently large
  number of resolution elements to reach convergent results
  \citep{Springel2003}.

In the present work we perform a detailed comparison study between the widely
used SPH code {\small GADGET} and the novel moving-mesh code {\small AREPO} on
a number of hydrodynamical test problems with increasing levels of
complexity. We have adopted a combination of existing test problems and newly
devised numerical experiments\footnote{High-resolution images and movies of
  various numerical experiments are available for download at the website
  http://www.cfa.harvard.edu/itc/research/movingmeshcosmology} in order to
provide a clear link between the results of our test problems and those of
full cosmological simulations, which are discussed in detail in our companion
papers \citep[][hereafter Paper I]{Vogelsberger2011} and \citep[][hereafter
  Paper II]{Keres2011}. Our work thus contributes to the understanding
of the significant differences in baryon properties found in cosmological
simulations between mesh-based and SPH techniques, both at a global level (see
Paper I) and at the scale of individual galaxies (see Paper II).
 
A unique advantage of our comparison study lies in the fact that simulations
with {\small GADGET} and {\small AREPO} can be started from identical initial
conditions and that both codes employ the same gravity solver. Note that
  in case of {\small AREPO} gravitational softenings for the gas can be kept
  fixed as is the case of {\small GADGET} or can be computed adaptively
  determined by the cell size. In the numerical experiments with gas
  self-gravity we explore both fixed and adaptive gas softening, where in the
  case of adaptive softenings a floor equal to the {\small GADGET} gas
  softenings is set. We find that these different choices of gravitational softenings in
  {\small AREPO} do not affect our findings. This allow us to isolate cleanly how the
different hydro solvers used by {\small GADGET} and {\small AREPO} impact gas
properties. We first consider elementary hydrodynamical numerical tests such
as a strong $1$D Sod shock tube test, a $2$D implosion test (which is widely
used for benchmarking mesh codes, but has rarely been considered in SPH), and
the ``blob'' test \citep{Agertz2007}. We then perform a number of isolated or
merging halo simulations in the non-radiative regime, aimed at understanding
how shocks and fluid instabilities affect their gaseous atmospheres. Finally,
we study the differences between {\small GADGET} and {\small AREPO} in
radiative simulations, where we follow inside-out disk formation and
interactions between the central disks and orbiting substructures.

This paper is organized as follows. In Section~\ref{Methodology}, we
provide a brief overview of the numerical codes used and the types of
simulations performed. Section~\ref{Results} represents the core of
the paper where all of our numerical experiments are
discussed. Finally, we summarize our findings in
Section~\ref{Conclusions}.

\section{Methodology} \label{Methodology}

\subsection{Numerical codes}

\subsubsection{{\small GADGET}}

{\small GADGET} is a massively parallel TreePM-SPH code widely used in
numerical astrophysics. In this study we adopt the latest {\small
  GADGET-3} version. A detailed description of an earlier version can
be found in \citet{Gadget2}. {\small GADGET} is fully adaptive in time
and space, and in its entropy formulation for SPH \citep{SH2002}
manifestly conserves both energy and entropy in the absence of
artificial viscosity. Gravitational forces are computed with an
octtree method \citep{Barnes1986, Hernquist1987}. To speed up the
computation, long-range forces can be optionally evaluated with a PM
method, with the tree being restricted to short-range forces only.

In all our tests, we adopt as standard value for the artificial
viscosity strength $\alpha = 1.0$, and we use $64$ neighbours for
kernel interpolation in three dimensional simulation, unless we
specifically vary these parameters to assess their effect.

\subsubsection{Other SPH implementations}\label{OtherSPH}

As mentioned in the introduction, there have been a number of recent
proposals to improve the standard SPH implementation in various ways,
for example by invoking a time-dependent artificial viscosity
\citep{Morris1997, Dolag2005}, a modified density estimate
\citep[e.g.][]{Ritchie2001, Hess2010, Saitoh2012}, a decoupling of the
hot and cold neighbours in multiphase flows \citep{Marri2003,
  Okamoto2003}, an artificial thermal conductivity \citep{Price2008},
a modified equation of motion \citep{Hess2010, Abel2011}, an explicit
modelling of mixing \citep{Wadsley2008}, an enlarged neighbour number
combined with a different kernel shape \citep{Read2010}, or a
replacement of the artificial viscosity by a Riemann solver
\citep[e.g.][]{Inutsuka2002, Cha2003, Murante2011}. While some of
these modifications of the standard SPH formalism deliver more
accurate results in targeted numerical experiments, no consensus has
yet emerged whether any of these approaches (or a combination thereof)
is sufficiently robust for cosmological applications and leads to
universally more accurate results in galaxy formation
simulations. Therefore, in this work we focus on the traditional SPH
implementation rather than on the various possible modifications
proposed recently. We note that this standard formulation of SPH has
also been typically employed in state-of-the art cosmological SPH
calculations \citep[e.g.][]{Crain2009, DiMatteo2012}.  Furthermore, as
discussed in Paper I, there are other, generic issues with SPH that
have not been resolved by any of the above modifications.  These
issues ultimately mean that SPH does not currently have a formal
convergence condition, which also complicates rigorous evaluations of
variants of the standard SPH algorithm.

\subsubsection{{\small AREPO}}

{\small AREPO} \citep{Springel2010} is a new massively parallel
simulation code, which uses the same gravity solver as {\small GADGET}
 (augmented with the possibility of adaptive gravitational
  softenings for the gas), but employs a completely different method
for the evolution of the fluid. It adopts a second-order accurate
finite volume technique, where the solution of the Euler equations is
computed by an unsplit Godunov method equipped with an exact Riemann
solver. Throughout we use the default choice of the slope limiter
  in {\small AREPO} which prevents the linear reconstruction to over-
  or undershoot the maximum/minimum values of neighbouring cells, as
  described in detail in \citet{Springel2010}. Unlike standard
finite-volume codes used in numerical astrophysics, {\small AREPO}
solves the equations on an unstructured Voronoi mesh, which is allowed
to freely move with the fluid. The resulting quasi-Lagrangian nature
of {\small AREPO} automatically guarantees spatial adaptivity and
greatly reduces numerical diffusivity even in the presence of large
bulk flows. Compared to standard Eulerian mesh codes, {\small AREPO}
has the advantage of being fully Galilean invariant (as is {\small
    GADGET} as well), it is less prone to advection errors and
over-mixing, and preserves contact discontinuities better.

In this study, we employ a mesh regularization method\footnote{Note that
    in the presence of mesh regularization the Galilean invariant nature of the
{\small AREPO} code is not violated.} in {\small
  AREPO} by default, based on a Lloyd algorithm \citep[for details
see][]{Springel2010}, which ensures that the geometric centre of each
Voronoi cell is sufficiently close to the cell's mesh-generating point to
ensure good accuracy of the spatial reconstruction. In cosmological
simulations (see Paper I), an alternative regularization criterion has
proven to be advantageous, based on the maximum opening angle under
which a cell face is seen from the mesh-generating point. We have
checked for a number of test runs presented in this study that this
alternative regularization method does not affect any of the results
described here. 

For most of the simulations presented here we do not use the possibility of
mesh refinement and de-refinement operations \citep[see][]{Springel2010},
except in our numerical experiments with star formation presented in
Section~\ref{GeneralBlobCooling}, where we employ it for verification of our
findings. Mesh de-/refinements are used to constrain the mass of cells
to a small range around a target value (equal to the gas particle mass in the
matching {\small GADGET} run). This restricts the spectrum of star particle
masses which are generated from gaseous cells, and thus assures that N-body
heating effects are minimized. Also, as our default choice we use the energy
formulation of {\small AREPO}. We verified for each numerical experiment that
there is no significant spurious transfer of kinetic into thermal energy.

\subsection{Types of simulations}

\subsubsection{Physical processes}
We perform both radiative and non-radiative hydrodynamical simulations, where
in the former case gas is represented by a primordial mixture of hydrogen and
helium in an optically thin limit \citep{Katz1996}. In simulations with
radiative cooling, we employ a subresolution multi-phase model for star
formation and associated supernova feedback \citep{SpringelH2003}. We slightly
modify the behaviour of this model for gas elements which are hot but already
above the density threshold for star formation, by allowing them to settle
quickly onto the effective equation of state: if their newly estimated
  temperature from radiative cooling would fall below the temperature of the
  multiphase medium, we set it equal to the multiphase temperature. This change
has been introduced to make the sub-grid star formation module consistent with
its current implementation in {\small AREPO}. We also perform some simulations
with the subresolution star formation model where spawning of star particles
is intentionally prevented, but the cold, dense gas above the density
threshold for star formation is still governed by the effective equation of
state. These simulations are particularly useful for understanding the
development of dynamical instabilities between cold and hot media, and have
the advantage of not being prone to fragmentation which might
affect pure cooling runs.

  Note that we deliberately consider only very simplified baryonic
  physics implementations in all presented numerical experiments, as
  detailed above. This is for two reasons. First, we need to make sure
  that gas cooling and star formation processes are treated on an as
  equal footing as possible in order to allow a meaningful comparison
  of the two codes. Thus, we adopt a simple multi-phase model for star
  formation which is largely insensitive to the detailed structure of
  the gas on very small scales where gas joins the interstellar medium
  and is described by a relatively stiff effective equation of
  state. Second, our aim is to isolate in an as clean manner as
  possible the differences between simulated systems with {\small
    GADGET} and {\small AREPO} stemming from the discretization of
  the fluid equations alone. This goal is given precedence over trying to
  reproduce observational findings through a more sophisticated
  modelling of additional physics.  While it is likely that such
  additional physical processes will change the properties of some of
  our simulated systems by possibly different degrees in {\small
    GADGET} and {\small AREPO}, it is of significant interest in its
  own right to disentangle numerical uncertainties from uncertainties
  in the physical modelling of star formation and associated feedback
  processes.

\subsubsection{Gravitational softenings}
For simplicity, many of the numerical experiments presented in this study have
been evolved without gas self-gravity. For simulations where self-gravity of
the gas is nevertheless included, we note that {\small GADGET} employs
constant gravitational softening for gas particles in the manner of
\citet{Hernquist1989}, while {\small AREPO} uses either the same constant
  gravitational softening or an adaptive softening determined by the cell
size with a minimum softening value set to the fixed softening of the
  matching {\small GADGET} run. We have verified that this does not
lead to substantial differences for any of the tests presented in this study.

\subsubsection{Effective hydro resolution}

  Even though {\small GADGET} and {\small AREPO} calculations use the
  same gravity solver and can be initiated from identical initial
  conditions, due to the very different nature of the hydro solver it
  is not straightforward to define unambiguously a comparison strategy
  at the same or equivalent hydro resolution. In this study we
  choose to keep the number of resolution elements the same in the
  both codes, i.e. to have the same number of SPH particles as Voronoi
  cells, corresponding to a comparable mass resolution in the gas. This
  allows us to adopt indeed the same initial conditions and to also
  have similar mass resolution in the stellar component in those
  simulations where star formation is included.  Furthermore, the CPU
  costs of the two codes are then roughly comparable, as discussed in
  Paper I.

  We note, however, that this choice implies that the ``effective''
  spatial resolution of {\small GADGET} is lower than that of the
  matching moving-mesh calculation, given that fluid properties in SPH
  are evaluated by kernel averaging over a somewhat larger number of
  neighbours than needed in {\small AREPO} for its stencil of
  neighbouring cells, for example for gradient estimates. For this
  reason, we perform all of our numerical experiments at a number of
  different resolutions, which also help us to gain some insight into
  the convergence properties of the two codes. Nonetheless, it is
  important to stress that the convergence properties of the SPH
  method are still not well understood. For example, one would
  ultimately require that the number of neighbours be increased with
  increasing total particle number \citep[see e.g.][and discussion in
  Paper I]{Rasio2000, Read2010, Robinson2011}, but the appropriate
  scaling of the neighbour number with increasing resolution is
  currently unknown. It is common practice, which we adopt as well, to
  simply always keep the number of neighbours fixed when the total
  particle number is varied, even though the discretized representation of
  the density field is not guaranteed to converge to its underlying
  smooth distribution with this choice.

\subsection{Initial conditions generation}

For all the tests presented in this study \citep[except for the
``blob'' experiment of][for which we take publicly available initial
conditions]{Agertz2007} we generate the initial conditions in terms of
simple particle/point distributions.  Depending on the simulated
problem at hand, we either populate the simulated domain with
particles uniformly spaced on a regular lattice or we adopt Poisson
sampling of a density field. For simulations with {\small AREPO}, the positions of
particles in the initial conditions define mesh generating points. All
purely hydrodynamical numerical experiments have been performed with
periodic initial conditions. In the case of simulated problems where
we follow the evolution of an isolated gaseous halo or merging haloes,
we select vacuum boundary conditions for {\small GADGET}. In
simulations with {\small AREPO}, it is however necessary to enclose
the simulation in a finite volume in order to make the tessellation
with a mesh well posed. If needed, we thus add a low resolution
background grid to our initial conditions, which is chosen
sufficiently large as to encompass the spatial extent of our simulated
objects at all times. The procedure adopted for this background grid
generation, which also reduces the Poisson noise in the initial mesh
geometry, is described in detail in Section 9.4 of
\citet{Springel2010}.
      
\section{Results}\label{Results}

\subsection{Strong shocks and interacting curved shocks in multi-dimensions}

\subsubsection{Strong shock in 1D}\label{1DShock}

As an introductory problem, we consider a strong shock with a Mach number
$M=6.3$ in one dimension. The initial conditions have been constructed as
follows: in a computational domain of length $L_{\rm x} = 10.0$, $N = 200$
particles (cells) have been placed on a regular grid such that for $x < 5.0$
the pressure and density are $P=30.0$, $\rho = 1.0$, while for $x \ge 5.0$
they are $P=0.14$ and $\rho = 0.125$, respectively. We adopt an adiabatic
index $\gamma = 1.4$, and assume that the fluid is initially at rest. In the
test run with {\small GADGET}, we adopt a standard value of the artificial
viscosity equal to $\alpha = 1.0$, and we vary the neighbor number $N_{\rm
  ngb}$ by setting it to $5$, $7$, $11$, or $15$, appropriate for the 1D
  nature of the test.

\begin{figure*}\centerline{
\includegraphics[width=6.6truecm,height=12truecm]{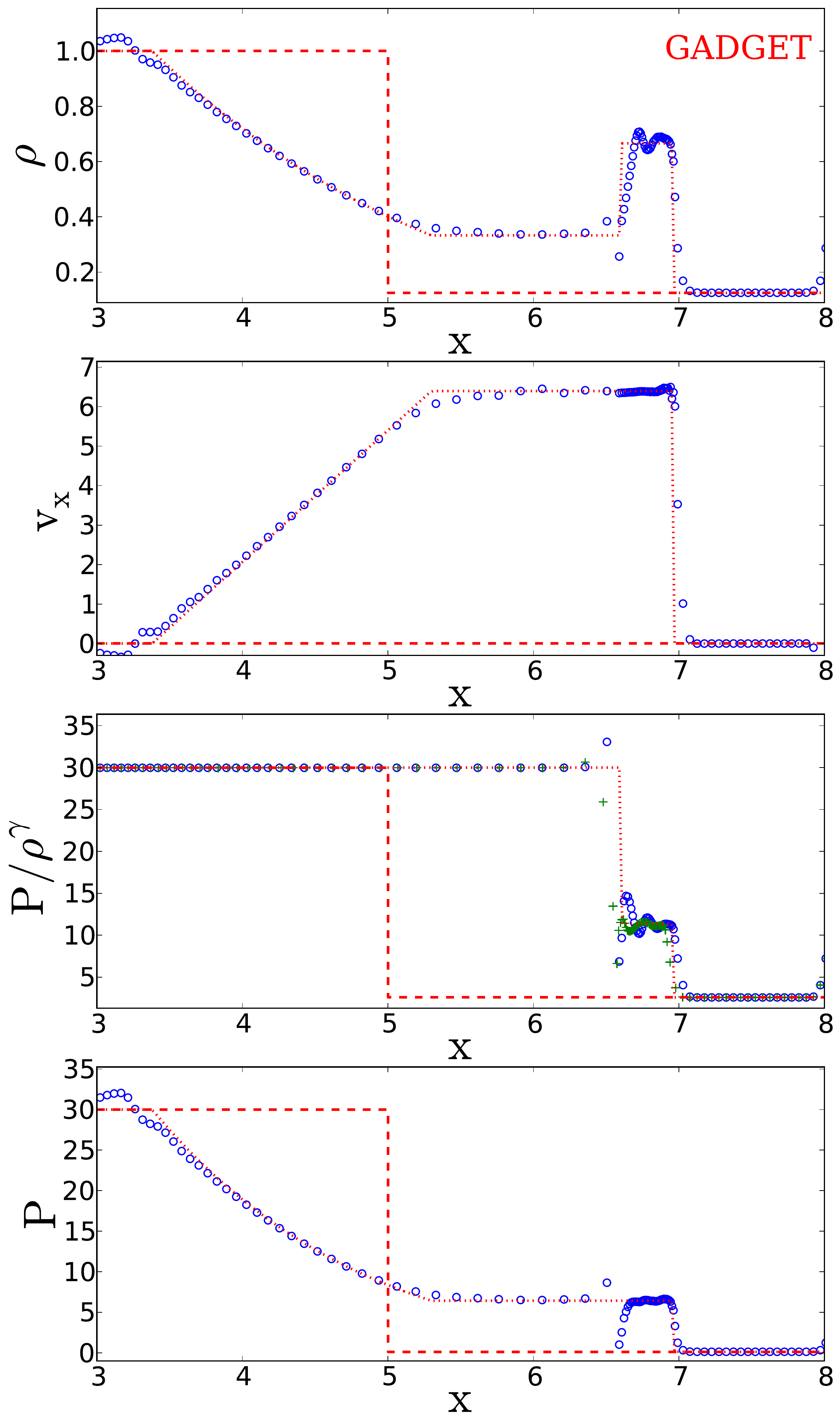}
\includegraphics[width=6.6truecm,height=12truecm]{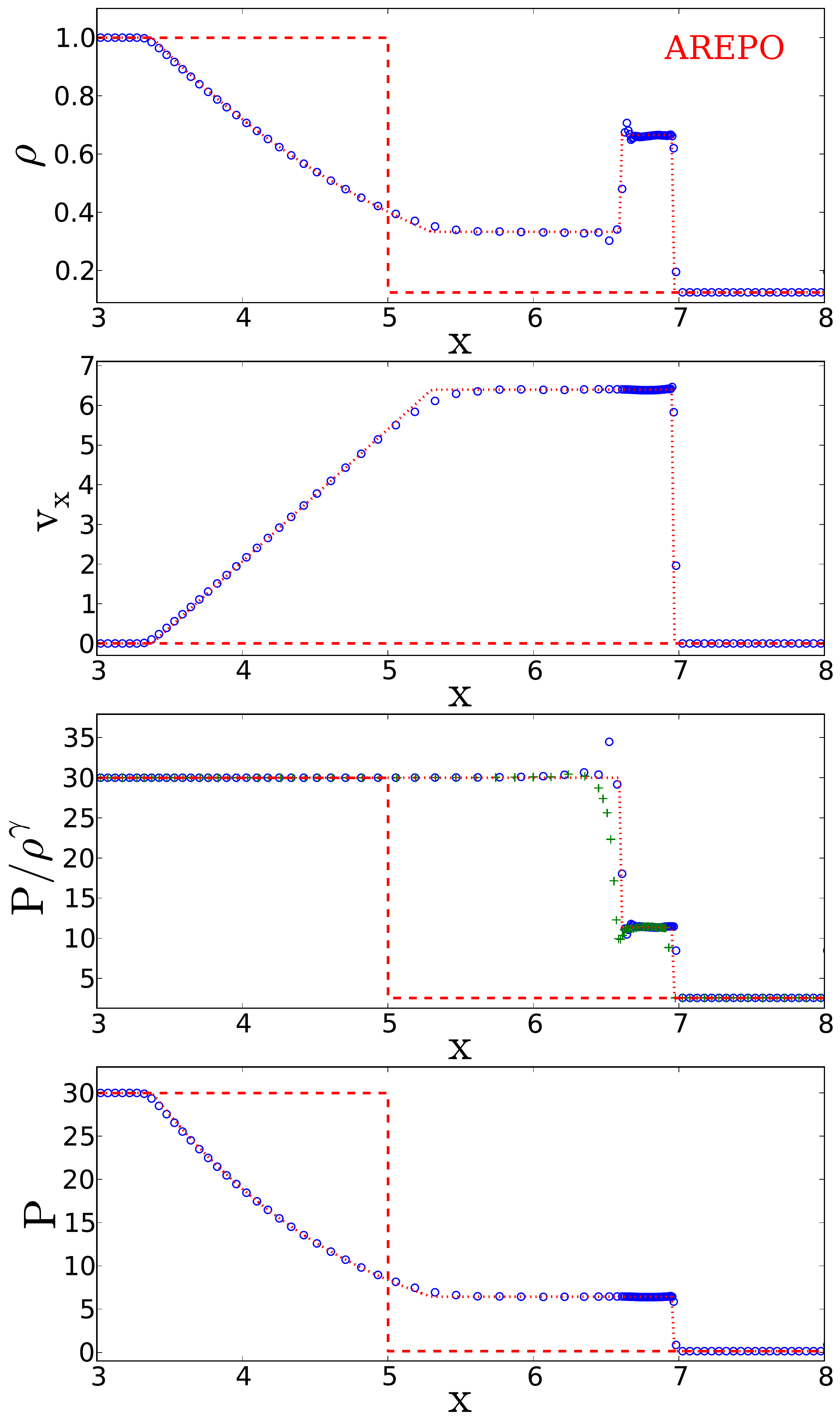}
\includegraphics[width=3.9truecm,height=12truecm]{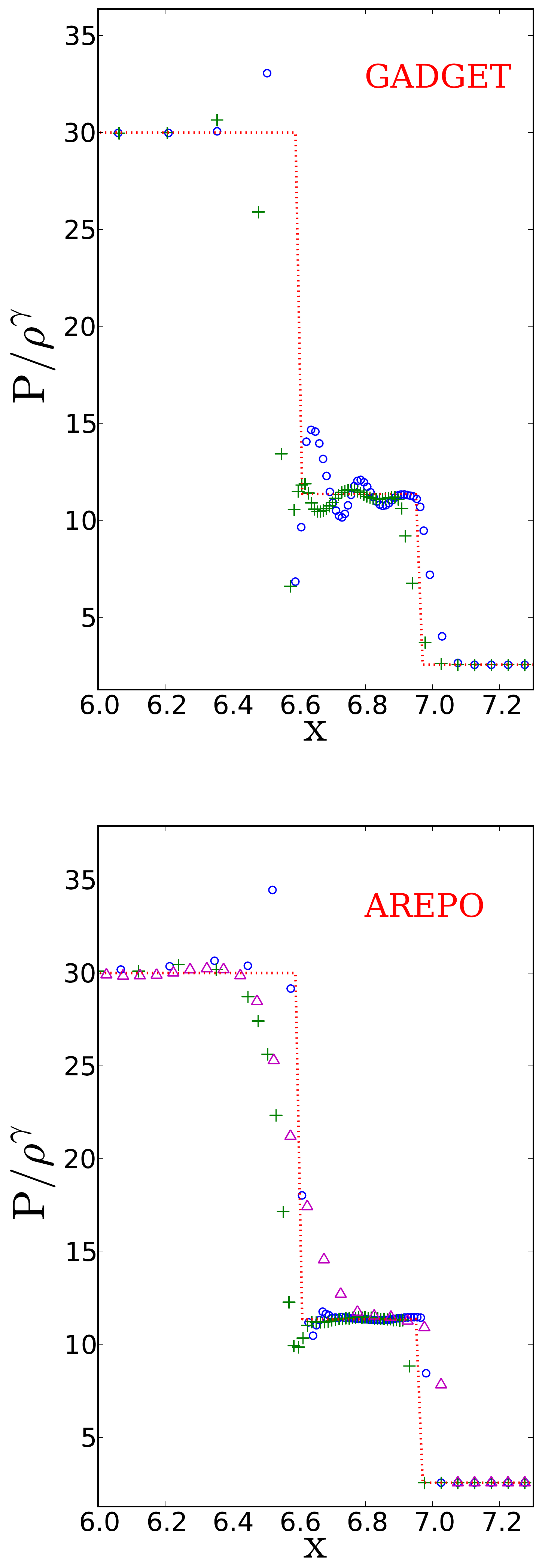}}
\caption{One dimensional shock tube problem with Mach number $M = 6.3$ at time
  $t = 0.25$. Left-hand panels: {\small GADGET} with standard artificial
  viscosity $\alpha = 1.0$ and $N_{\rm ngb} = 5$. Middle panels: {\small
    AREPO} with moving mesh. Dashed red lines denote initial conditions, while
  dotted red lines represent the analytic solution. In the third row, green
  cross symbols illustrate gas entropy in a shock tube of the same strength
  but with initially smoothed contact discontinuity adopting a Hann
  window. Right-hand panels: zoom into the entropy profile around the shock -
  here magenta triangle symbols in the lower panel are for the
    {\small AREPO} run with a static mesh.}
\label{Fig1DShock}
\end{figure*}

In Figure~\ref{Fig1DShock}, we show the gas density, velocity, entropy
(i.e. $P/\rho^\gamma$) and pressure at time $0.25$ for {\small GADGET} with
$N_{\rm ngb} = 5$ (left-hand panels), and for {\small AREPO} (middle
panels). Blue symbols give the values of individual particles/cells, dashed
red lines represent the initial conditions, while dotted red lines are the
analytic solution to this Riemann problem. It can be seen that in both {\small
  GADGET} and {\small AREPO} the post-shock properties of the fluid are
captured well, but the shock and the contact discontinuity are significantly
broader in {\small GADGET} which also shows a characteristic
  'pressure-blip' at the contact discontinuity \citep[see
  also][]{SpringelRev2010}. Contrary to what one may perhaps suppose, the
post-shock oscillations present in {\small GADGET} are not caused by
inadequate artificial viscosity, but are instead induced by an inaccurate
treatment of the sharp contact discontinuity of the initial conditions. To
demonstrate this point, we have performed exactly the same shock tube test but
this time smoothing the initial contact discontinuity over $5$ particles in
density and internal energy with a Hann window function, so as to reduce its
sharpness. Green cross symbols in the third row of Figure~\ref{Fig1DShock}
(see also the right-hand panels where we zoom into the region around the
shock) illustrate how the gas entropy is affected by this choice of smoothed
initial conditions. Post-shock oscillations and the ``spike'' in entropy at
the contact discontinuity are greatly reduced.

It is interesting to note that also in the case of {\small AREPO} the
``spike'' in the entropy at the contact discontinuity is cured by our smoothed
initial conditions. This feature is absent in the results of standard grid
codes, as we have verified by running this test with a static mesh option
  in {\small AREPO} (and by running it with {\small ENZO}), which tends to
  broaden contact discontinuities (depending on the advection speed) and thus
  largely wash-out this feature. This can be seen in the lower right-hand
    panel of Figure~\ref{Fig1DShock} where we show the results from the
    {\small AREPO} run with a static mesh with magenta triangle symbols. The
  much lower numerical diffusivity of {\small AREPO} for contact
  discontinuities preserves the initial start-up feature to much higher
  accuracy, but at the same time this can lead to a larger `wall heating'
    effect \citep{Rider2000} than in static grid codes, which tend to
    smooth out at some level the initial start-up errors at the contact
    discontinuity \citep[see also the description of the Noh problem
      in][]{Springel2010}.

When we increase the number of neighbors from $N_{\rm ngb} = 5$ (as
shown in Figure~\ref{Fig1DShock}) to $15$ in the shock tube tests
performed with {\small GADGET} (but keeping the total number of
particles constant), the spatial region over which the contact
discontinuity and the shock are broadened increases progressively, but
the pressure jump between the post-shock and pre-shock gas outside of
the broadened region remain the same, yielding consistent Mach
numbers. Based on these tests we later discuss in more detail the
consequences of shock broadening in {\small GADGET} when simulating
the radial infall of gas into static dark matter haloes in
Section~\ref{ColdInflow}.
 
\subsubsection{Interacting shocks in 2D}\label{2DShock}

\begin{figure*}\centerline{\vbox{\hbox{
\includegraphics[width=7.5truecm,height=6.6truecm]{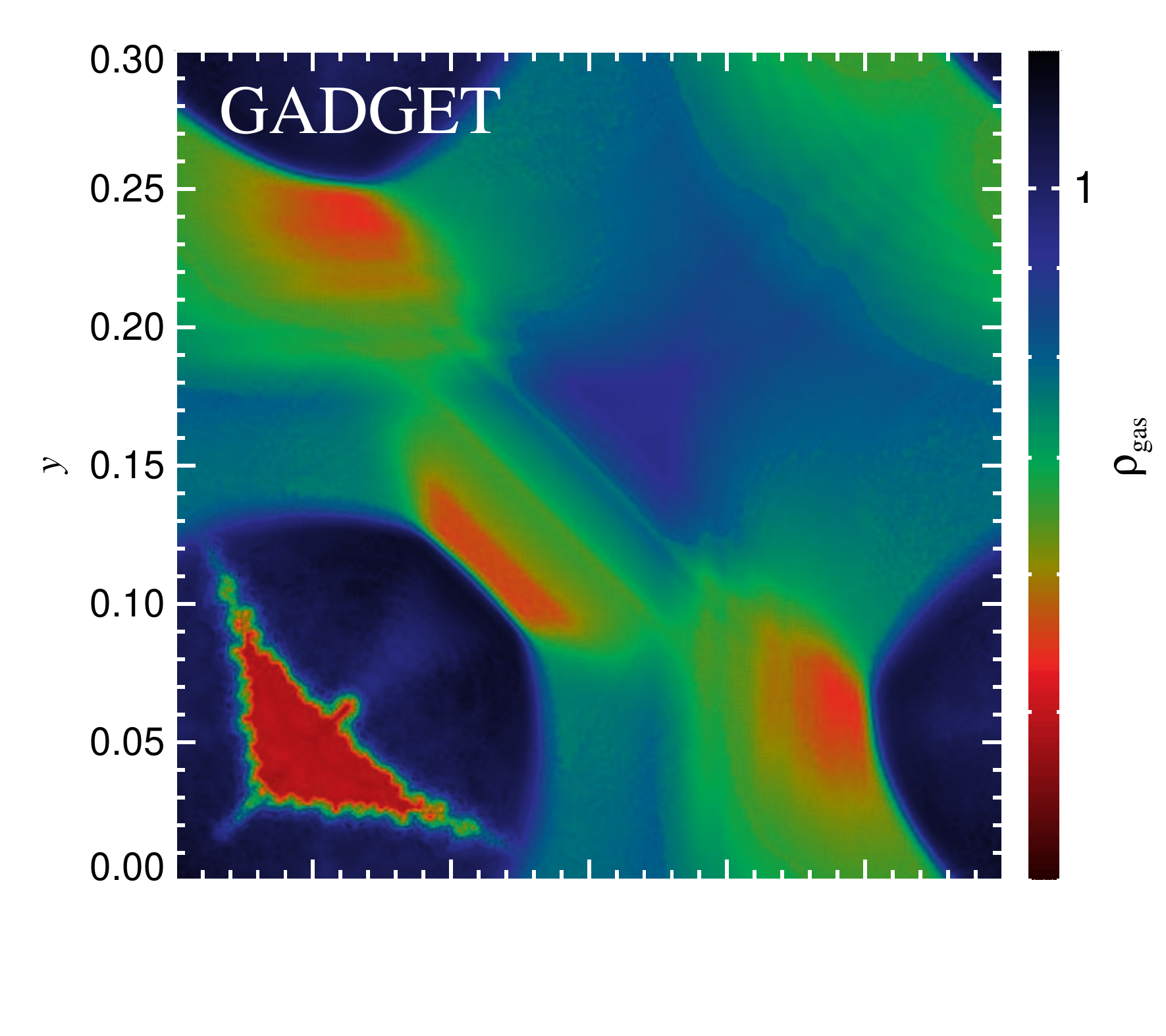}
\hspace{0.3cm}
\includegraphics[width=7.5truecm,height=6.6truecm]{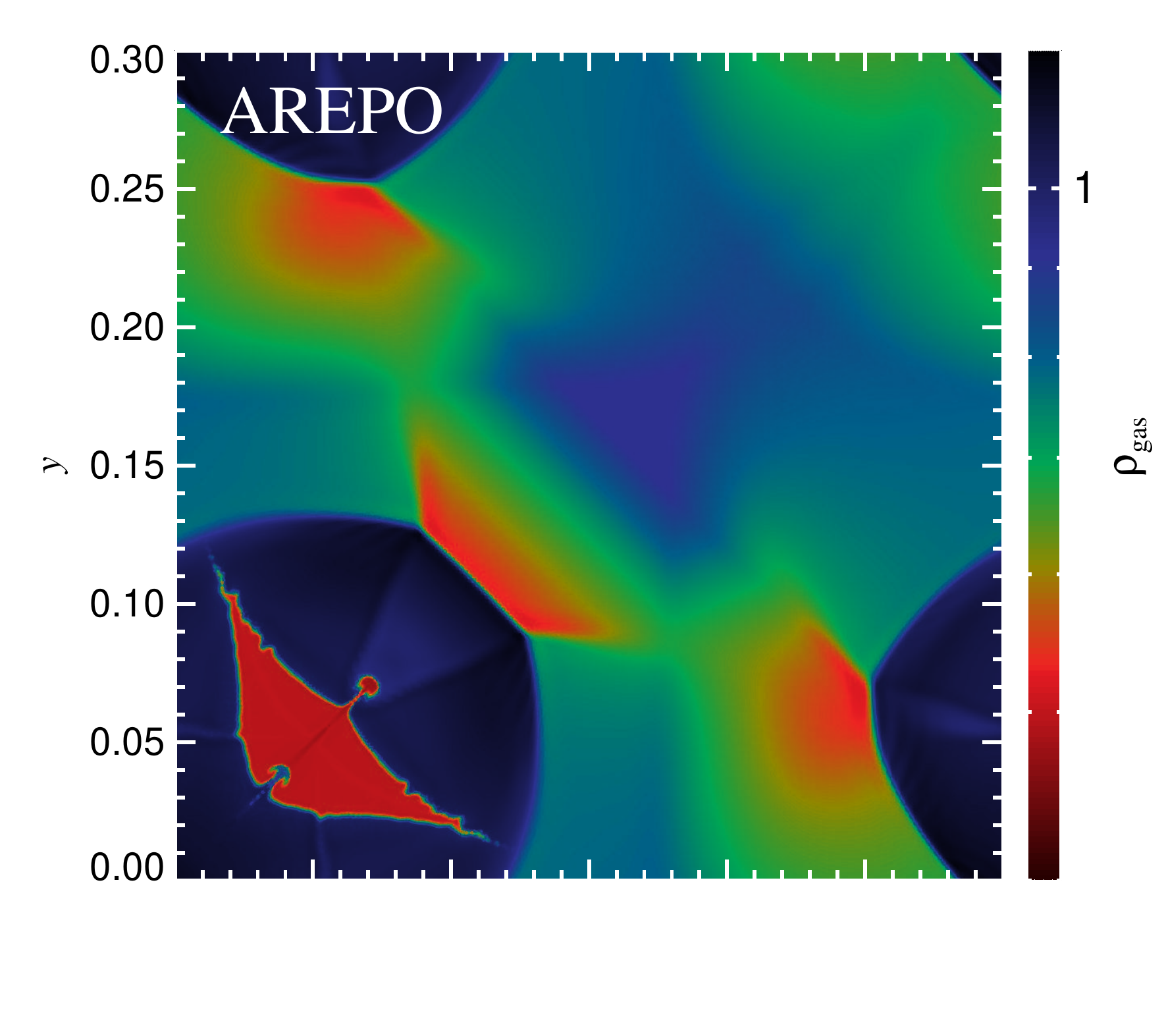}}
\vspace{-1cm}
\hbox{
\includegraphics[width=7.5truecm,height=6.6truecm]{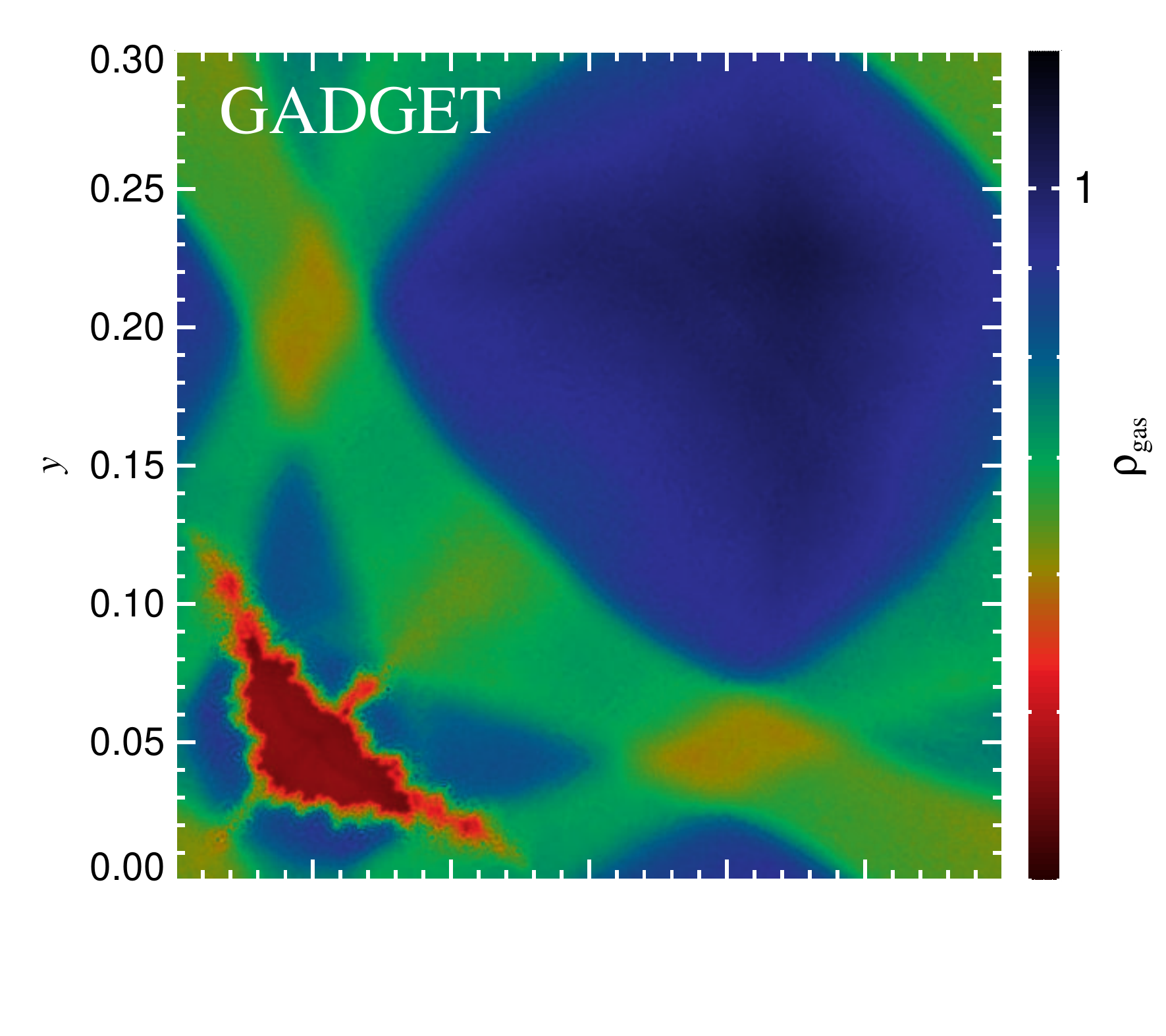}
\hspace{0.3cm}
\includegraphics[width=7.5truecm,height=6.6truecm]{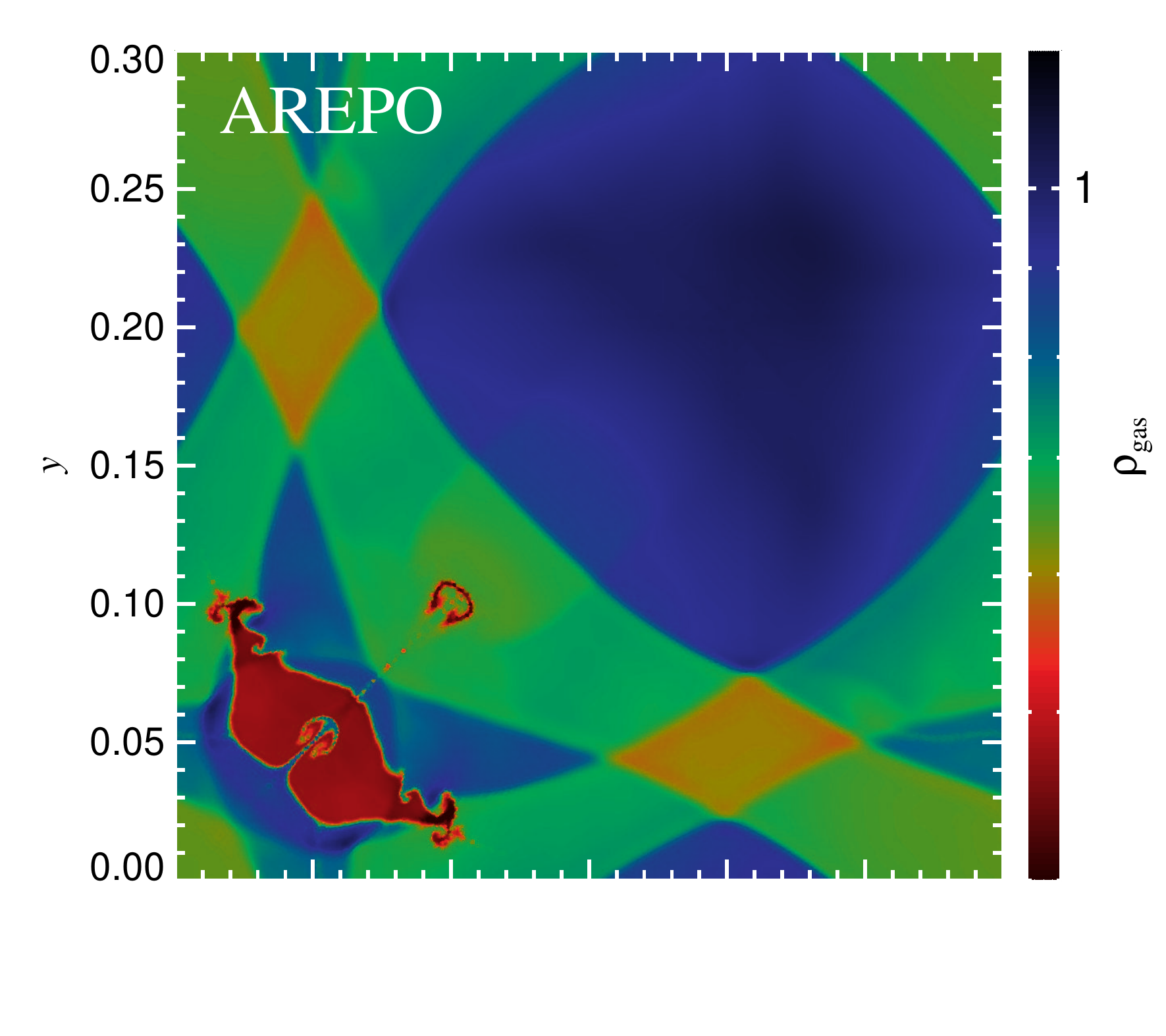}}
\vspace{-1cm}
\hbox{
\includegraphics[width=7.5truecm,height=6.6truecm]{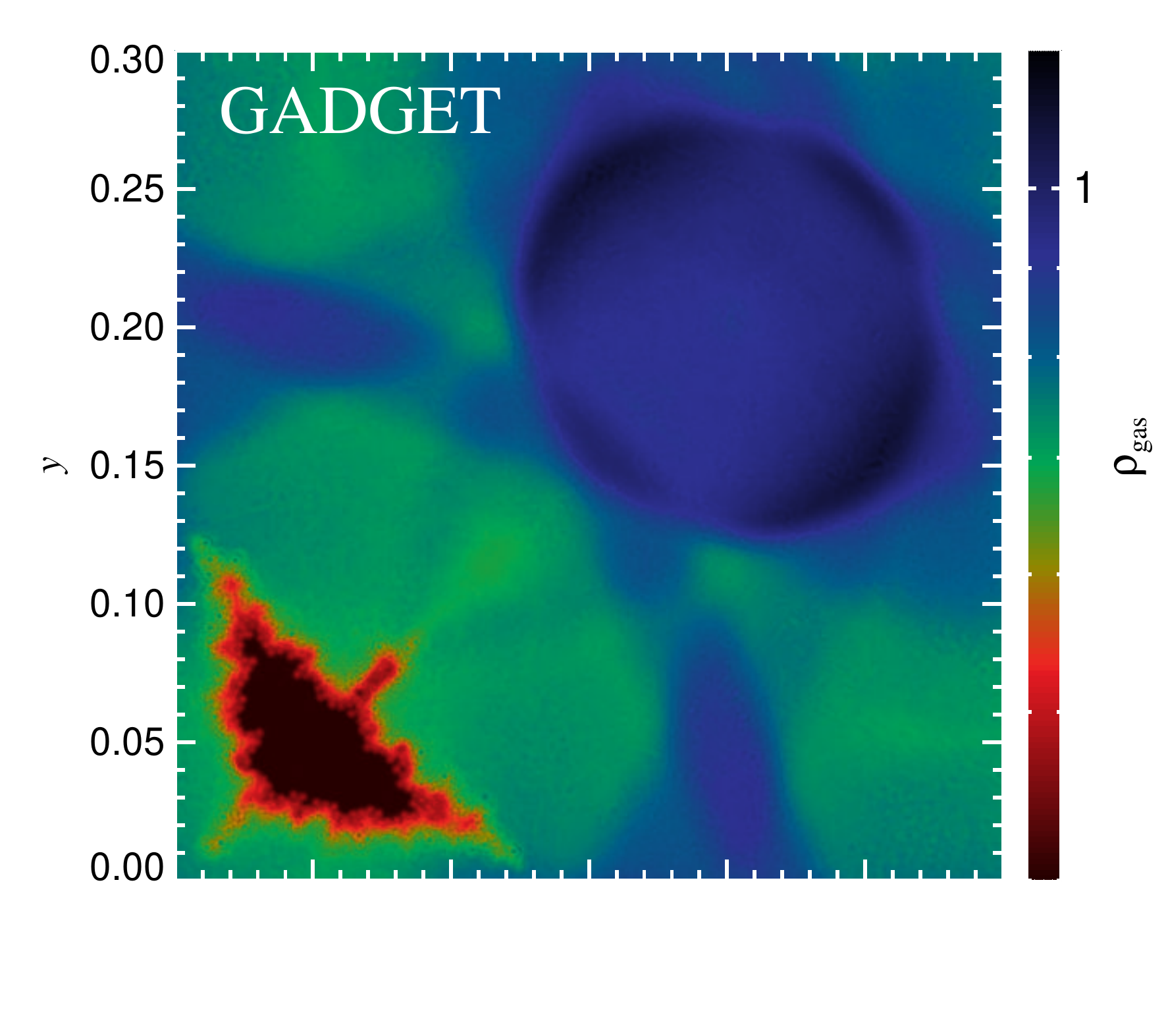}
\hspace{0.3cm}
\includegraphics[width=7.5truecm,height=6.6truecm]{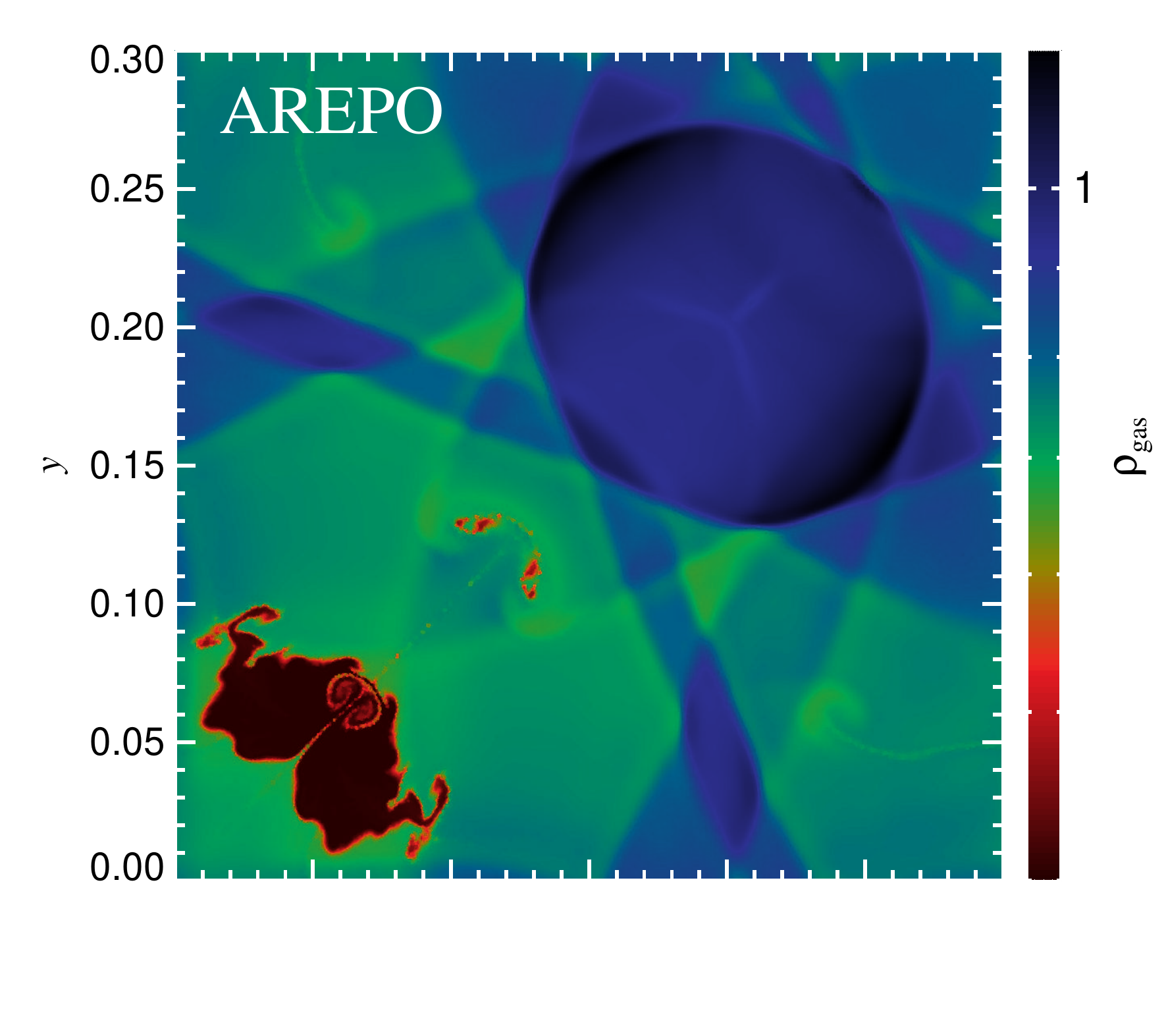}}
\vspace{-1cm}
\hbox{
\includegraphics[width=7.5truecm,height=6.6truecm]{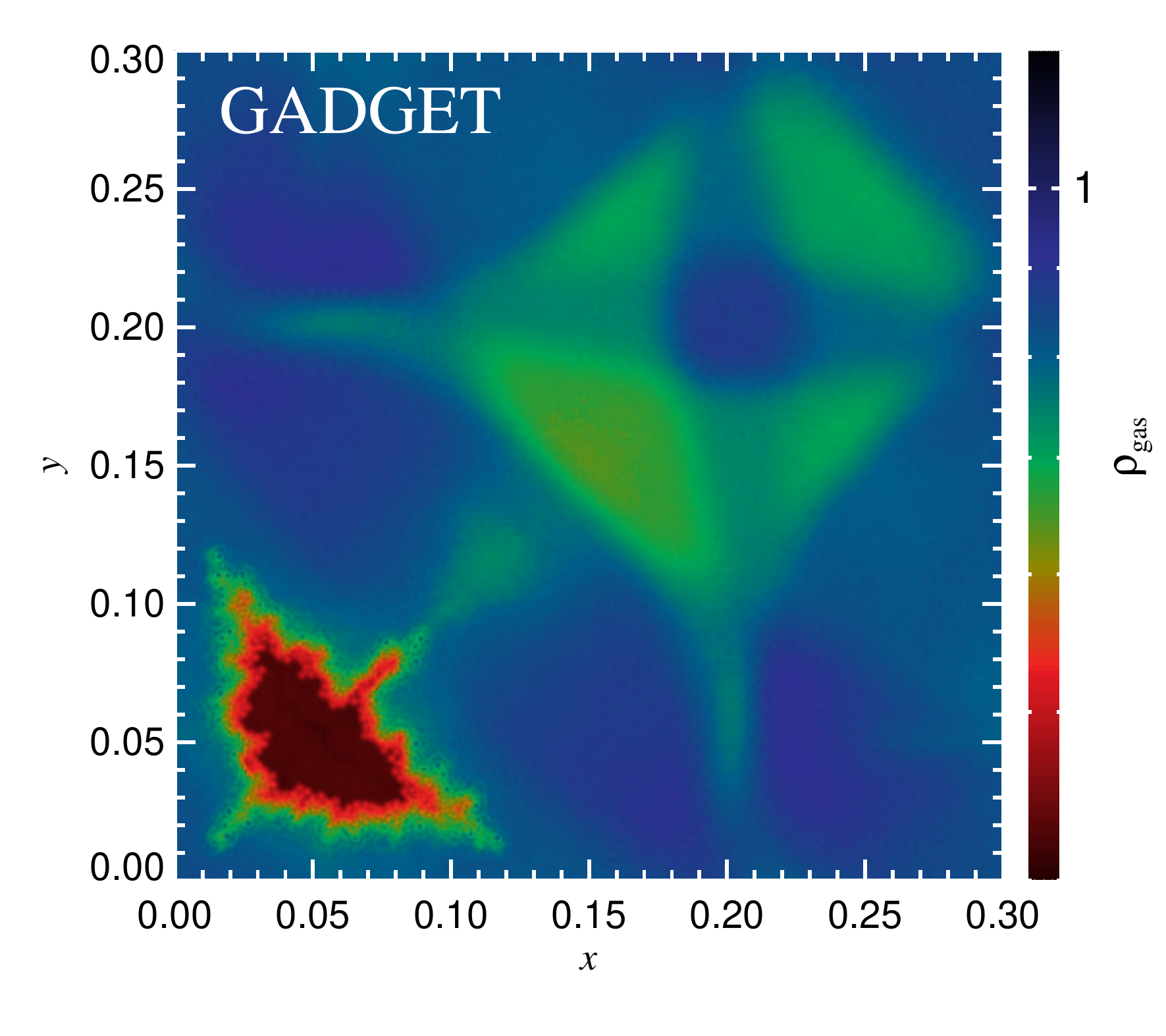}
\hspace{0.3cm}
\includegraphics[width=7.5truecm,height=6.6truecm]{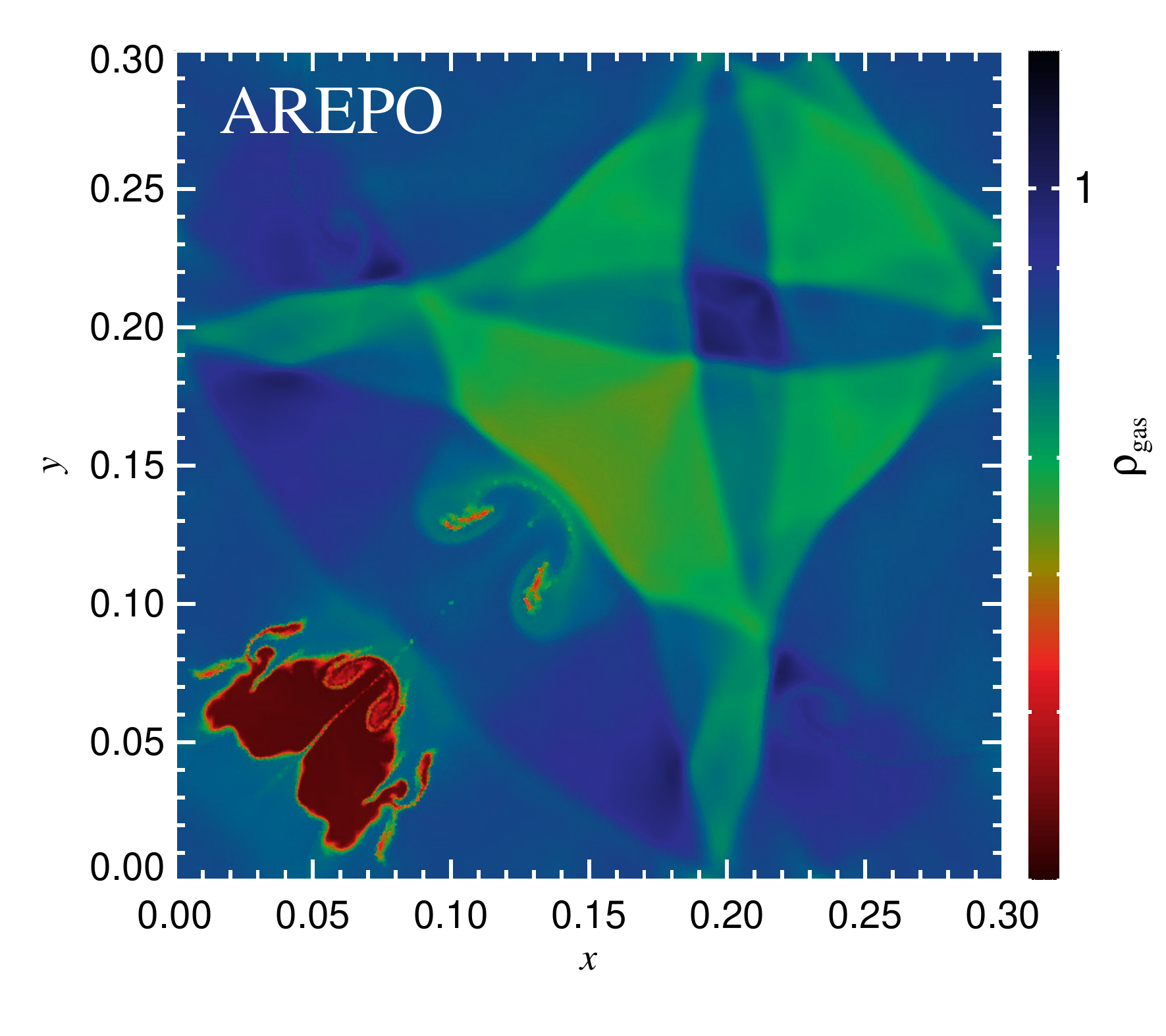}}}}
\vspace{-0.5cm}
\caption{Implosion test in 2D at times $t = 0.1$, $0.3$, $0.5$ and
  $0.7$. Left-hand panels: {\small GADGET} ($\alpha= 1.0$, $N_{\rm
    ngb} = 22$). Right-hand panels: {\small AREPO} moving-mesh. For
  each row, the density scale is the same in the left-hand and
  right-hand panels, covering the following density range: $\rho_{\rm gas}
  = 1.2-0.4$.}
\label{Fig2DShock}
\end{figure*}

\begin{figure*}\centerline{\vbox{\hbox{
\includegraphics[width=7.5truecm,height=6.6truecm]{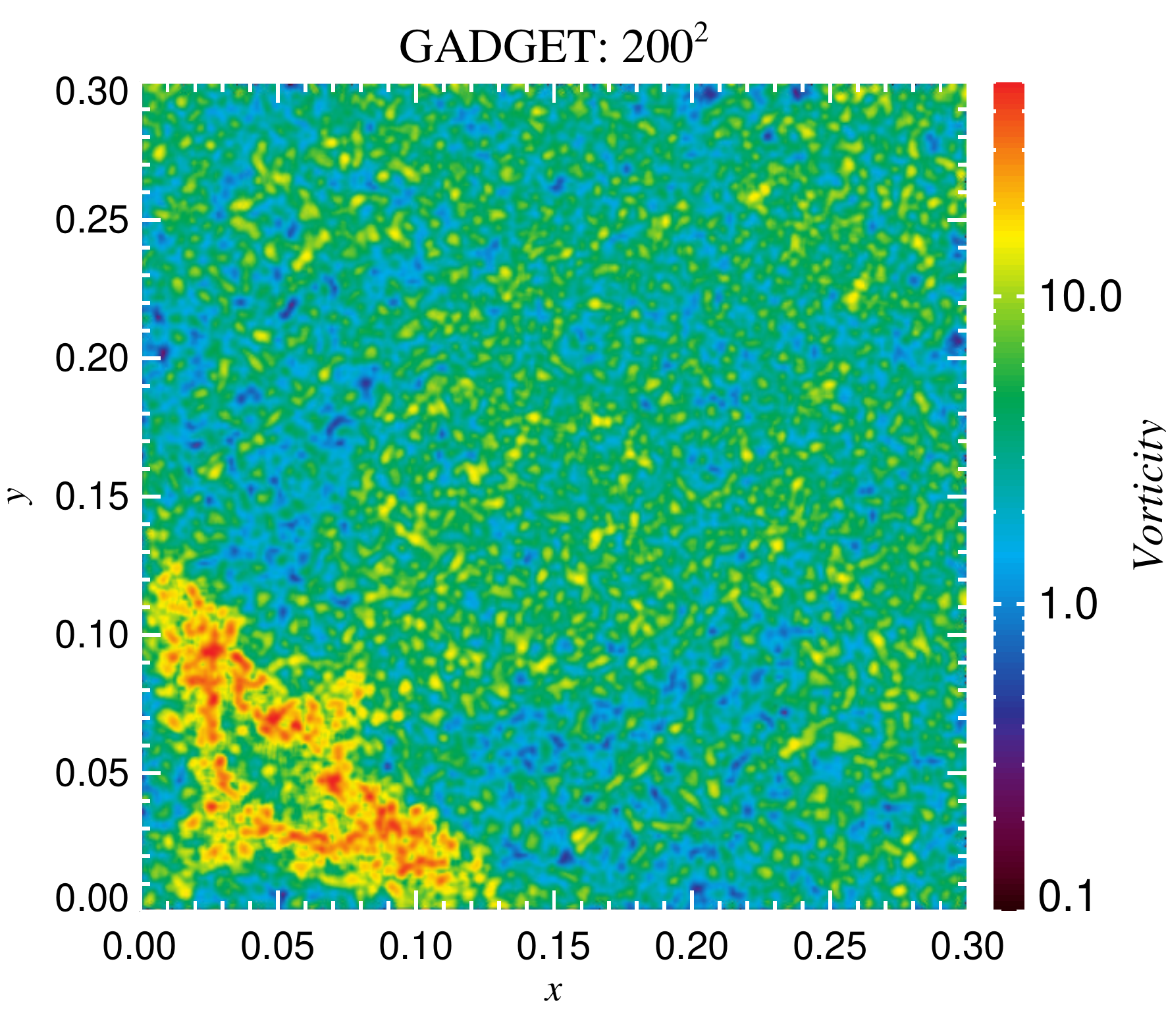}
\hspace{0.3cm}
\includegraphics[width=7.5truecm,height=6.6truecm]{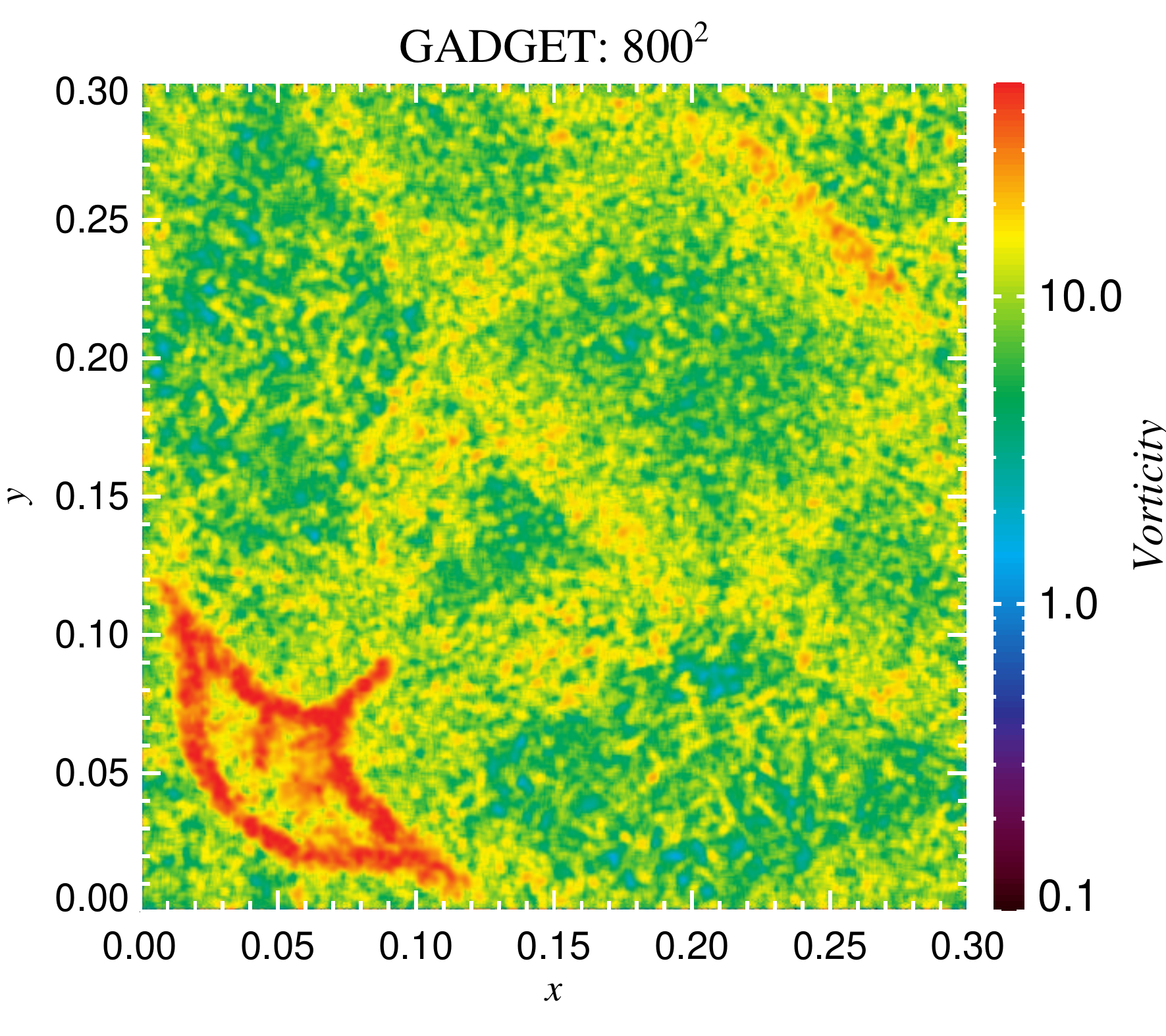}}
\hbox{
\includegraphics[width=7.5truecm,height=6.6truecm]{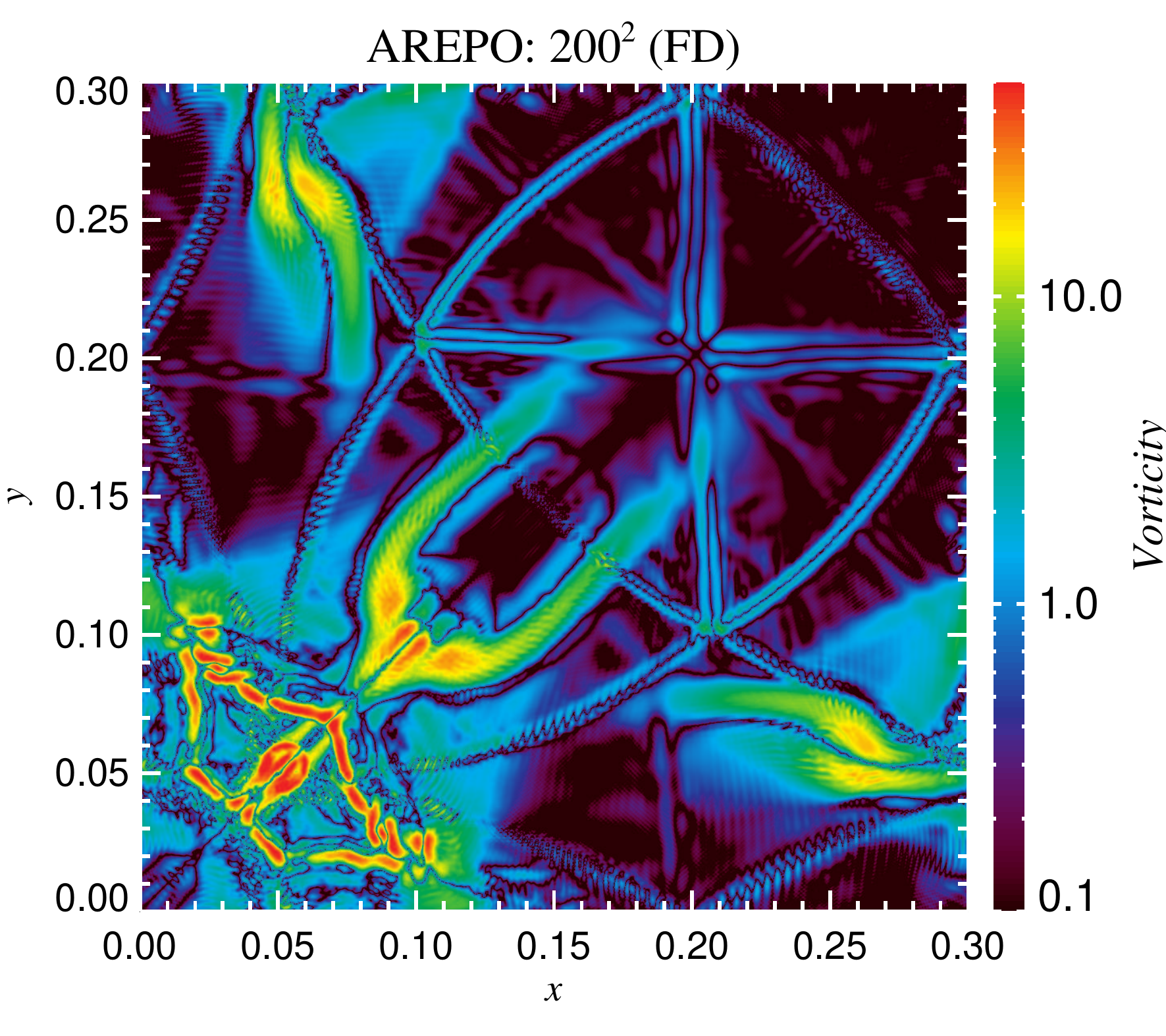}
\hspace{0.3cm}
\includegraphics[width=7.5truecm,height=6.6truecm]{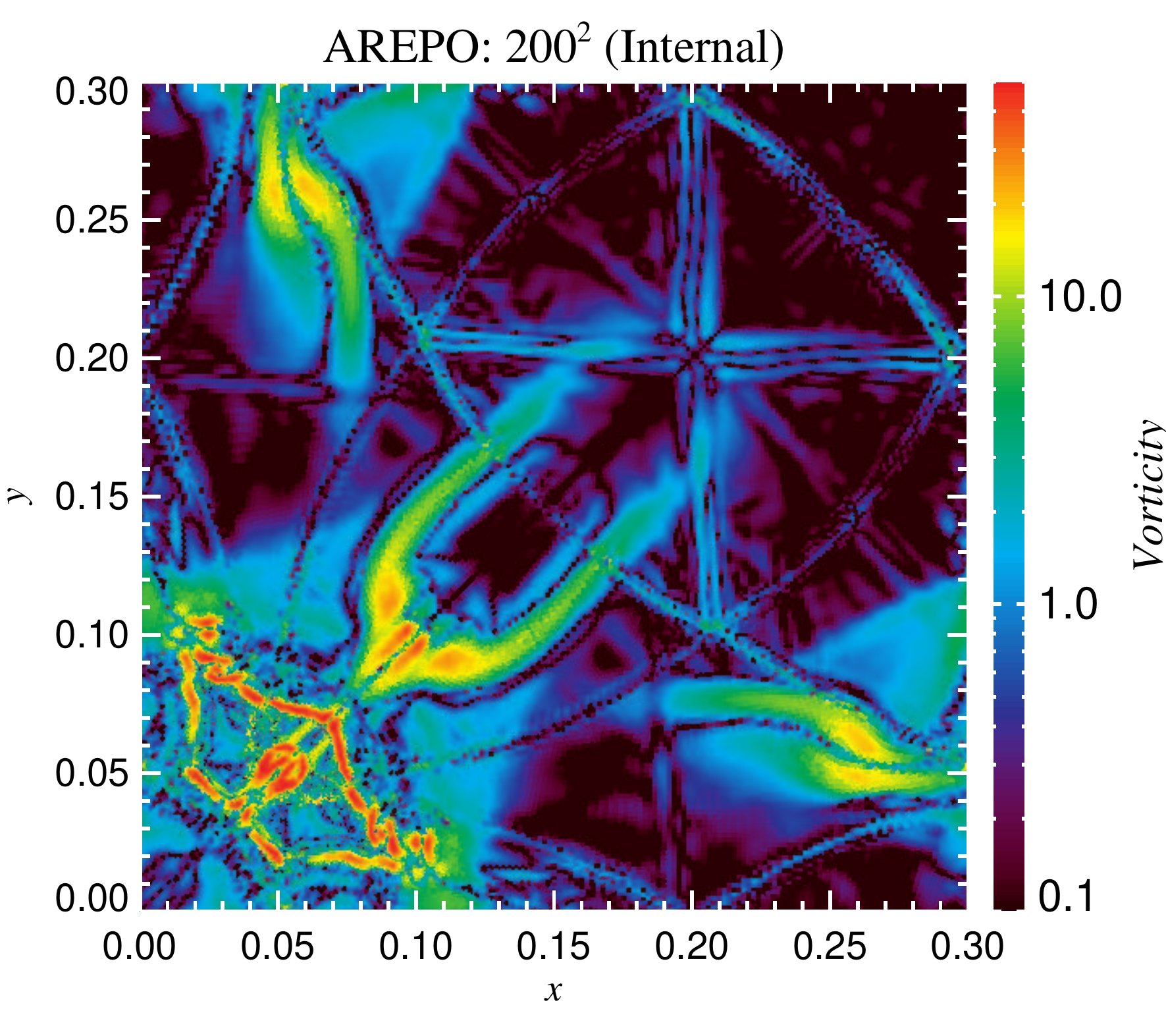}}}}
\caption{Vorticity maps of the implosion test in 2D at time $t =
  0.3$. Upper panels: {\small GADGET} runs at two different
  resolutions with $N_{\rm x} = N_{\rm y} = 200$ (left-hand panel) and
  $N_{\rm x} = N_{\rm y} = 800$ (right-hand panel). For both
  resolutions, vorticity maps are computed by finite differencing the
  velocity field. Lower panels: vorticity maps of the {\small AREPO}
  run with $N_{\rm x} = N_{\rm y} = 200$ computed by finite
  differencing the velocity field (left-hand panel), and by
  calculating vorticity in the code based on a discretized
  curl-operator directly applied to the Voronoi cells (right-hand
  panel).}
\label{Fig2DShockCurl}
\end{figure*}

While one-dimensional shock tube tests are an essential basic benchmark for
hydrodynamical code performance, they are far less demanding than
multi-dimensional flow where complex interactions of non-planar shocks may
occur, as is the case in realistic structure formation simulations. Such tests
have however not been examined widely in the SPH literature thus far. We hence
perform the so-called ``implosion test'' in two dimensions
\citep{Hui1999}\footnote{See also ``Comparison of Several difference schemes
  on 1D and 2D Test problems for the Euler equations'' by Liska,~R. and
  Wendroff,~B., on {\sf
    http://www-troja.fjfi.cvut.cz/$\sim$liska/CompareEuler/compare8/} and {\sf
    http://www.astro.princeton.edu/$\sim$jstone/Athena/tests/}.}. To set up
this test problem, we select a computational domain $L_{\rm x} = L_{\rm y}=
0.3$ with $N_{\rm x} = N_{\rm y} = 200$ particles or cells, respectively (we
have also run higher resolution versions with $N_{\rm x} = N_{\rm y} = 400$
and $800$). The initial pressure and density are $P = 1.0$ and $\rho = 1.0$
for $x + y > 0.15$, while $P=0.14$ and $\rho = 0.125$ otherwise. The adiabatic
index is $\gamma = 1.4$ and the fluid is initially at rest. We performed this
test with {\small GADGET} adopting an artificial viscosity of $\alpha = 1.0$
and $N_{\rm ngb} = 22$ smoothing neighbours (suitable for this 2D
  test). While previously this test has been considered using reflective
boundary conditions, we here adopt periodic boundary conditions due to their
more straightforward implementation in SPH codes.

In Figure~\ref{Fig2DShock}, we show density maps for simulations
performed with {\small GADGET} (left-hand panels) and with {\small
  AREPO} (right-hand panels), at times $t = 0.1$, $0.3$, $0.5$ and
$0.7$ for $N_{\rm x} = N_{\rm y} = 200$. The complex, evolving gas
density structure is caused by the continuous interaction of shocks
throughout the computational domain. Initially, due to the
discontinuity in density and pressure along $x + y = 0.15$, a planar
mild shock front develops perpendicular to the $x = y$ diagonal
traveling towards the origin. Given that we have adopted periodic
boundary conditions, the gas will interact on all four sides of the
simulated box, and in particular, interacting shock waves in the lower
left-hand corner result in the formation of a narrow jet along the $x
= y$ diagonal (which is clearly visible only in the {\small AREPO}
run). As the traveling shocks accelerate the fluid in the regions of
density discontinuity a Richtmyer-Meshkov instability develops, as
manifested by ``mushroom-like'' features in the gas distribution.

While the global gas density distribution in Figure~\ref{Fig2DShock}
agrees reasonably well in {\small GADGET} and {\small AREPO} runs
(i.e. with respect to the shape of the regions with different
densities and the magnitude of the density differences), there are
several significant differences: {\it i)} shocks and contact
discontinuities are much sharper in {\small AREPO}, in agreement with
our findings in Section~\ref{1DShock}; {\it ii)} the density
distribution in {\small GADGET} appears not only smoothed out, but it
is also noisier, as evidenced by the graininess of the density maps,
which is caused by intrinsic noise in multidimensional flows in SPH
\citep{SpringelRev2010}; {\it iii)} most strikingly, perhaps, is the
quite different appearance of the low density gas in the bottom left
corner -- a narrow, extended, dense jet along the diagonal is largely
absent in {\small GADGET} (due to the broadening of the contact
discontinuity) and Richtmyer-Meshkov instabilities along
discontinuities are suppressed. We have verified that higher
resolution simulations with {\small GADGET} ($N_{\rm x} = N_{\rm y} =
400$ and $800$) result in sharper shocks and contact discontinuities
(with a somewhat feeble jet forming), but they cannot satisfactorily cure
the absence of well-developed Richtmyer-Meshkov instabilities. This is
in agreement with previous studies \citep[e.g.][]{Agertz2007,
  SpringelRev2010} indicating fundamental limitations of the SPH
method in its widely used form. We also note that the moving mesh in
{\small AREPO} does an excellent job of maintaining symmetry across
the $x = y$ diagonal. Furthermore, the level of numerical diffusion of
discontinuities is very low, as indicated by the narrowness and length
of the jet.

In addition to the density fields it is also instructive to analyze
the vorticity distribution in the implosion test, as shown in
Figure~\ref{Fig2DShockCurl}. To construct vorticity maps we either
{\it i)} first compute spatially adaptive velocity field maps for each
component on a uniform grid (in our case only $x$ and $y$ components)
and then finite difference them to obtain the curl or {\it ii)} we
compute spatially adaptive vorticity maps, where the curl has been
evaluated directly in the code. For {\small GADGET} runs, vorticity
maps computed with method {\it i)} are shown in the upper panels of
Figure~\ref{Fig2DShockCurl} for $N_{\rm x} = N_{\rm y} = 200$ and
$N_{\rm x} = N_{\rm y} = 800$. Using method {\it ii)} vorticity maps
look essentially the same. In the case of {\small AREPO} (shown in the
lower panels for the same run with $N_{\rm x} = N_{\rm y} = 200$) the
vorticity maps are also very similar when computed with method {\it
  i)} or {\it ii)}. However, regardless of the exact details of the
vorticity map generation there are marked differences in the vorticity
field between {\small GADGET} and {\small AREPO}. For the $N_{\rm x} =
N_{\rm y} = 200$ run, the vorticity map in {\small GADGET} is largely
featureless and noisy, with artificial suppression of vorticity
generation at locations where surfaces of constant density and of
constant pressure are not aligned (i.e. baroclinic source term
$\propto \vec \nabla \rho \times \vec \nabla p$). Strikingly, even in
the case of the high resolution {\small GADGET} run with $N_{\rm x} =
N_{\rm y} = 800$ (upper right-hand panel) vorticity generation in
  the regions with large baroclinic term is largely suppressed
compared to the {\small AREPO} simulation, even though the latter has
a much smaller number of resolution elements. It is also worth
  noting that the overall magnitude of the vorticity is relatively
  high in {\small GADGET} with respect to {\small AREPO} in the
  regions which should be characterized by having low vorticity. This
  is due to the noise alluded to above, visible both in the density
  and vorticity maps.  This noise is inherently present in
  multidimensional SPH simulations. It is caused by inaccurate
  pressure gradient estimates and is particularly evident in the
  subsonic regime \citep[for further details see
  e.g.][]{SpringelRev2010, Bauer2011}. The jitter in gas velocities
  caused by the noisy gradient estimates introduces an artificial
  vorticity 'floor' throughout the simulated volume of the box. The
poor treatment of vorticity in {\small GADGET} has also direct
consequences for the effectiveness of the widely used Balsara switch
\citep{Balsara1995} for artificial viscosity which relies on
evaluation of velocity divergence and curl.
        
Hence, from the results of the ``implosion test'' we conclude that
while {\small GADGET} can accurately capture the main fluid properties
in the case of dynamically interacting shocks with complicated
geometries, there are systematic biases in detailed aspects,
related to the development of fluid instabilities and to the level of
fluid mixing at different entropies. These unwanted features lead to
the suppression of angular momentum transport by vortices and to the
damping of turbulence in the wake of curved shocks \citep[see
also][]{Bauer2011}.

\subsection{Dynamical fluid instabilities}\label{Instabilities}

\subsubsection{Blob experiment}\label{Blob}

\begin{figure*}\centerline{\vbox{\hbox{
\includegraphics[width=9.truecm,height=4.5truecm]{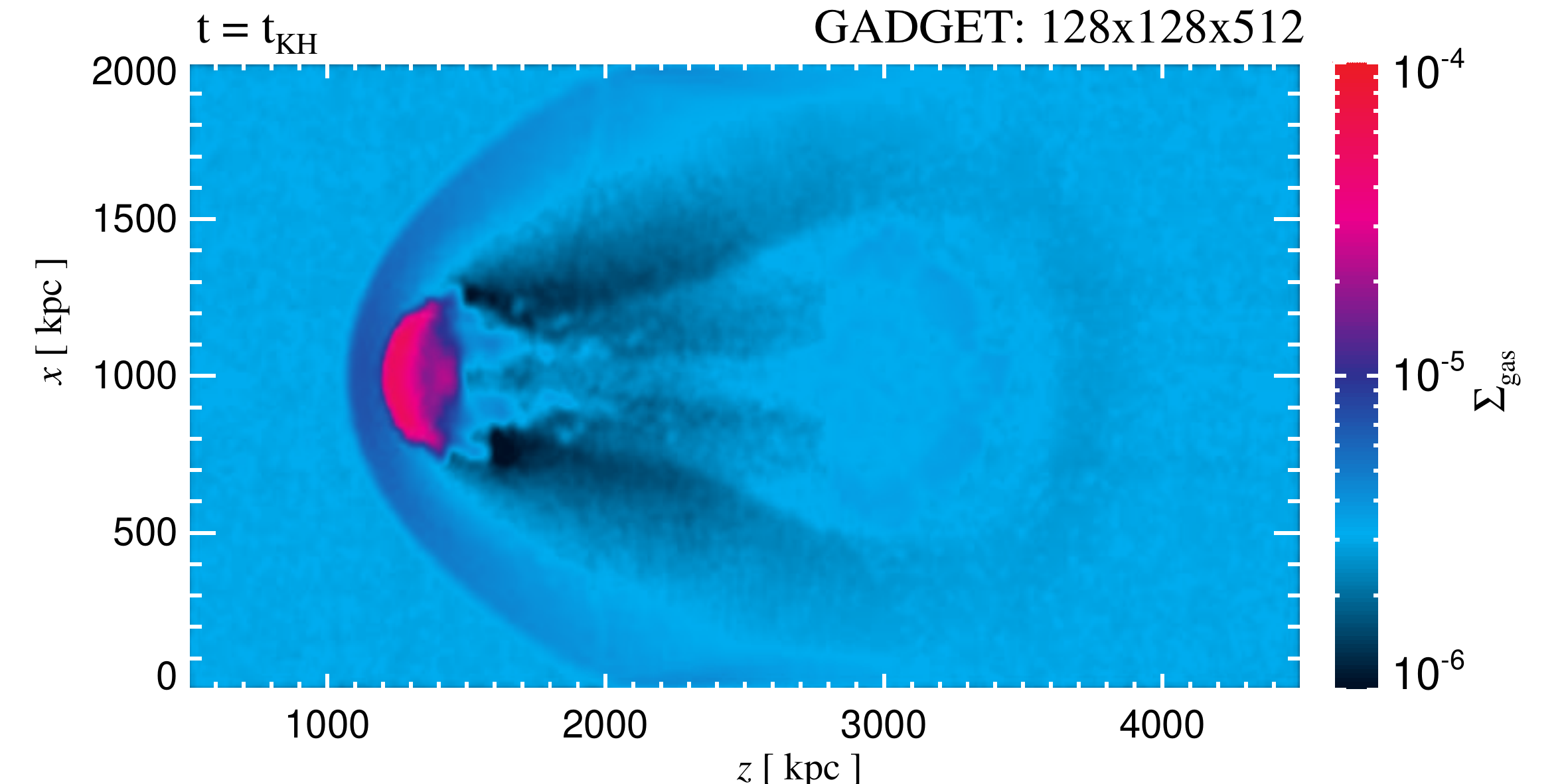}
\hspace{-0.35cm}
\includegraphics[width=9truecm,height=4.5truecm]{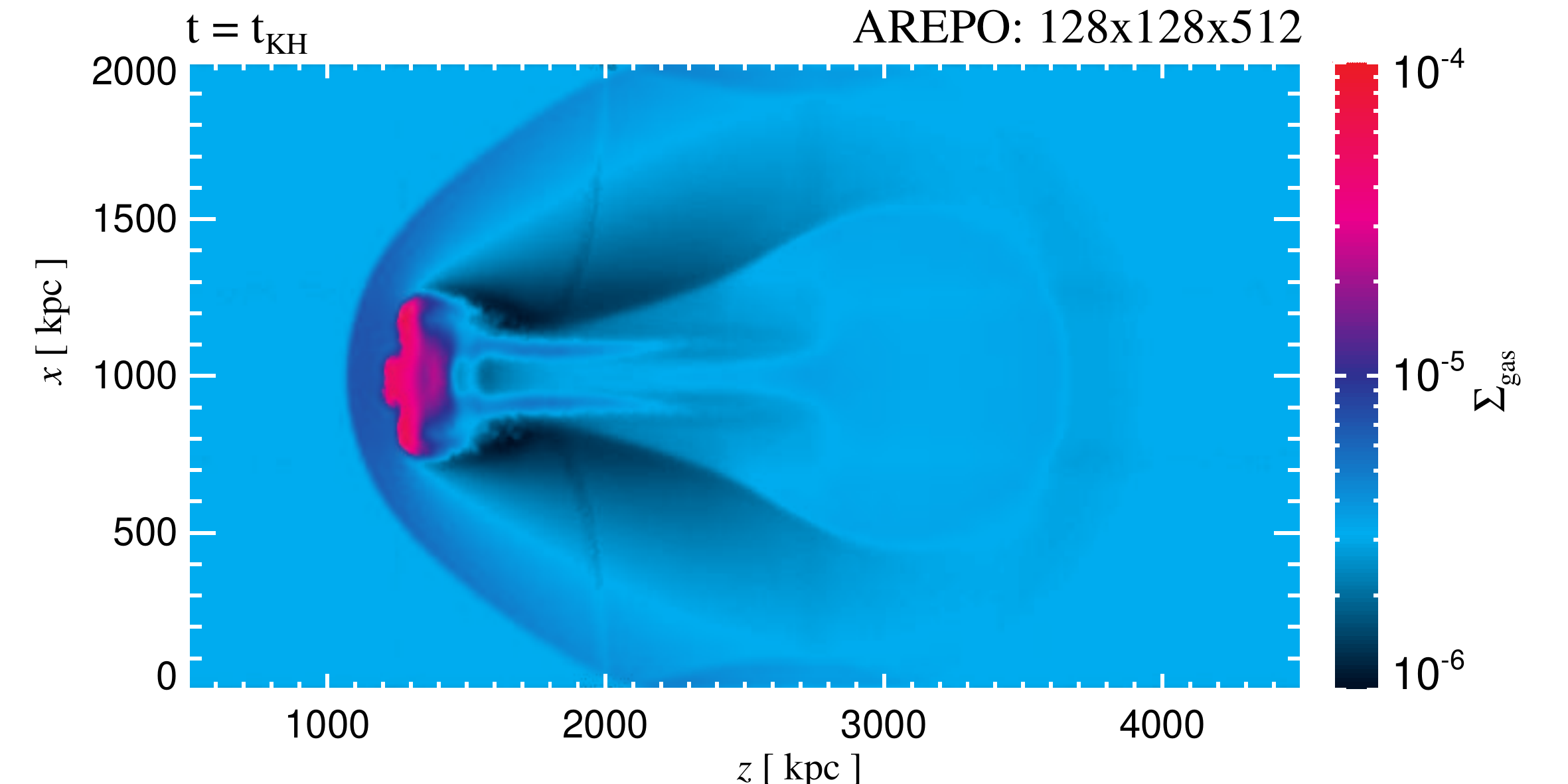}}
\hbox{
\includegraphics[width=9.truecm,height=4.5truecm]{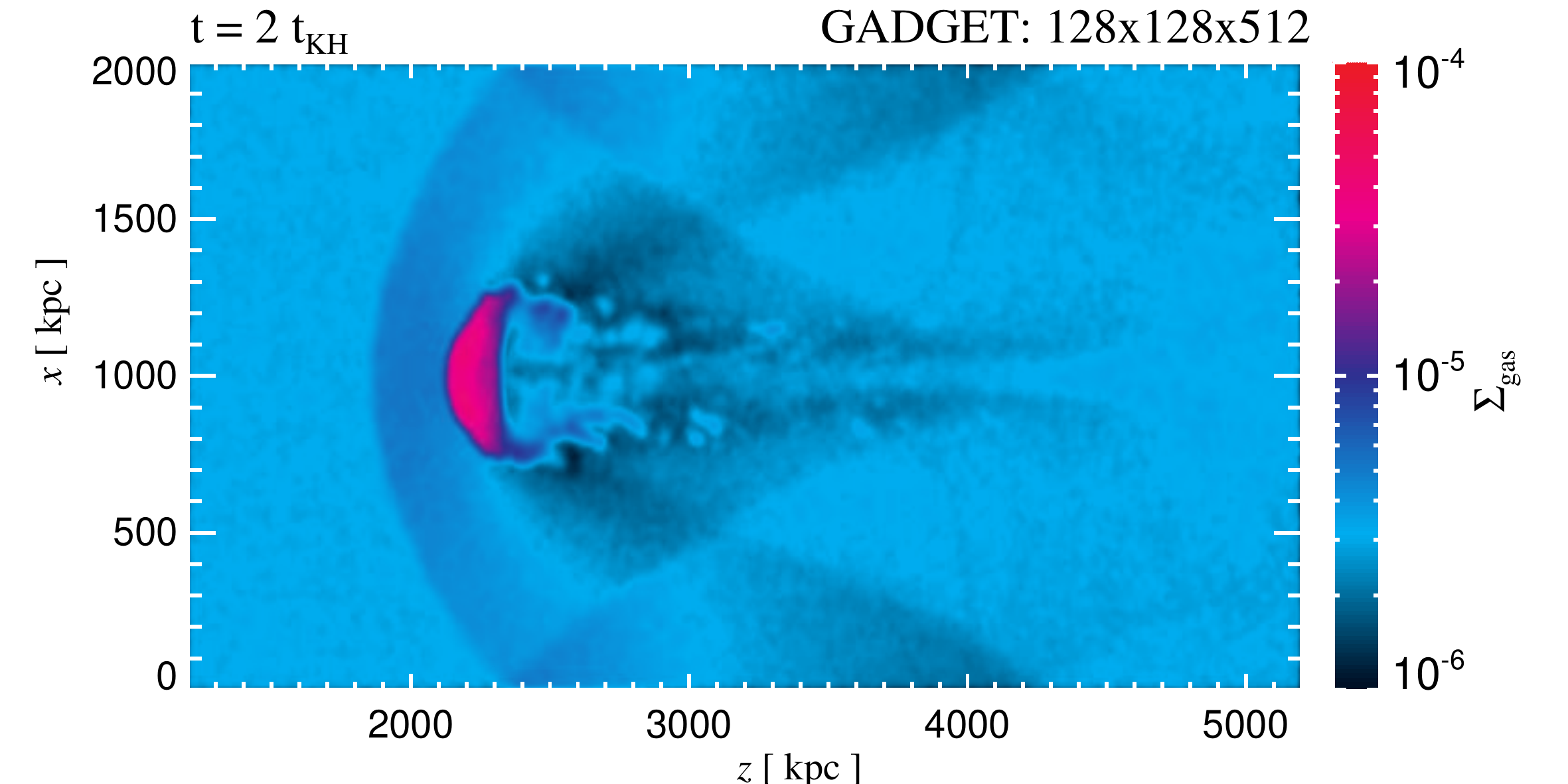}
\hspace{-0.35cm}
\includegraphics[width=9truecm,height=4.5truecm]{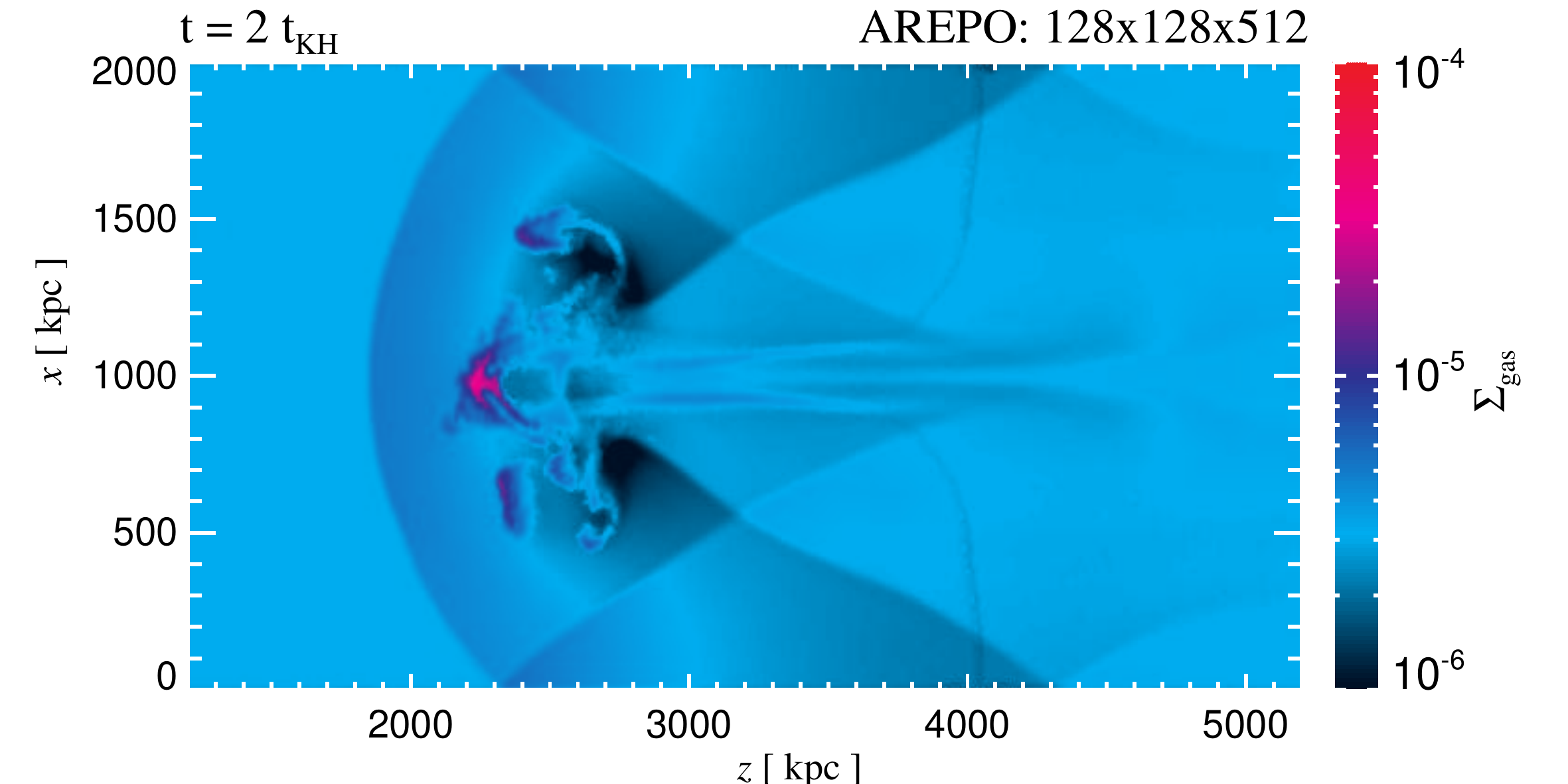}}
\hbox{
\includegraphics[width=9.truecm,height=4.5truecm]{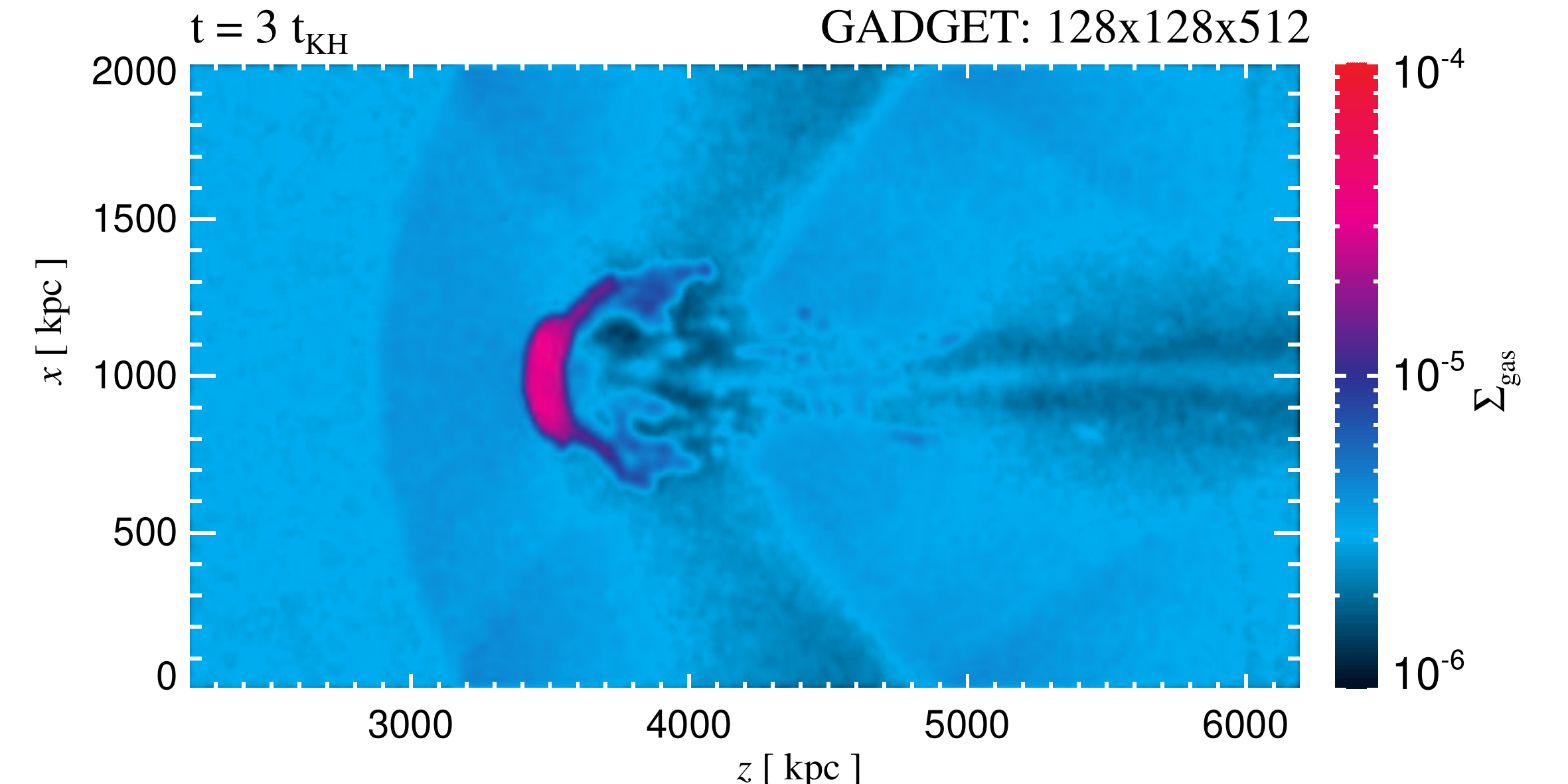}
\hspace{-0.35cm}
\includegraphics[width=9truecm,height=4.5truecm]{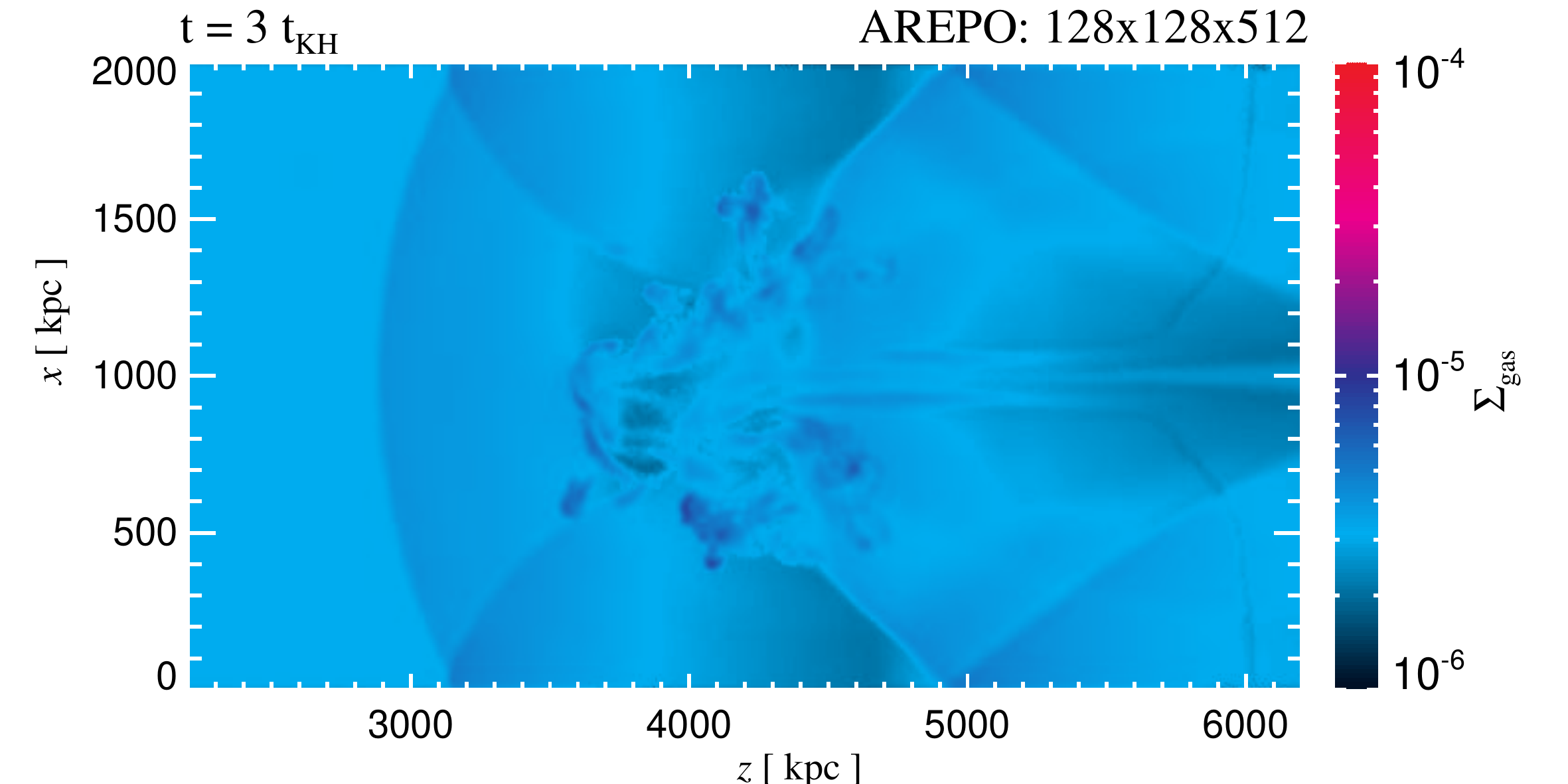}}}}
\caption{Projected surface density maps in units of $[{\rm
    M_{\odot}}{\rm kpc}^{-2}]$ at times $t = t_{\rm KH}$ (top row), $t
  = 2\, t_{\rm KH}$ (middle row) and $t = 3\, t_{\rm KH}$ (bottom row)
  for {\small GADGET} (left-hand panels; $\alpha= 1.0$, $N_{\rm ngb} =
  64$) and {\small AREPO} (right-hand panels; moving mesh). The
  thickness of the slices is $\Delta y = 100\,{\rm kpc}$, and they are
  centred on $y_{\rm c} = 1000\,{\rm kpc}$. Dynamical fluid
  instabilities are responsible for blob shredding on the
  characteristic Kelvin-Helmholtz timescale in the case of {\small
    AREPO}, but they are artificially suppressed in the run with
  {\small GADGET}, prolonging the survival time of the blob.}
\label{FigBlob}
\end{figure*}

\begin{figure*}\centerline{\vbox{\hbox{
\includegraphics[width=9.truecm,height=4.5truecm]{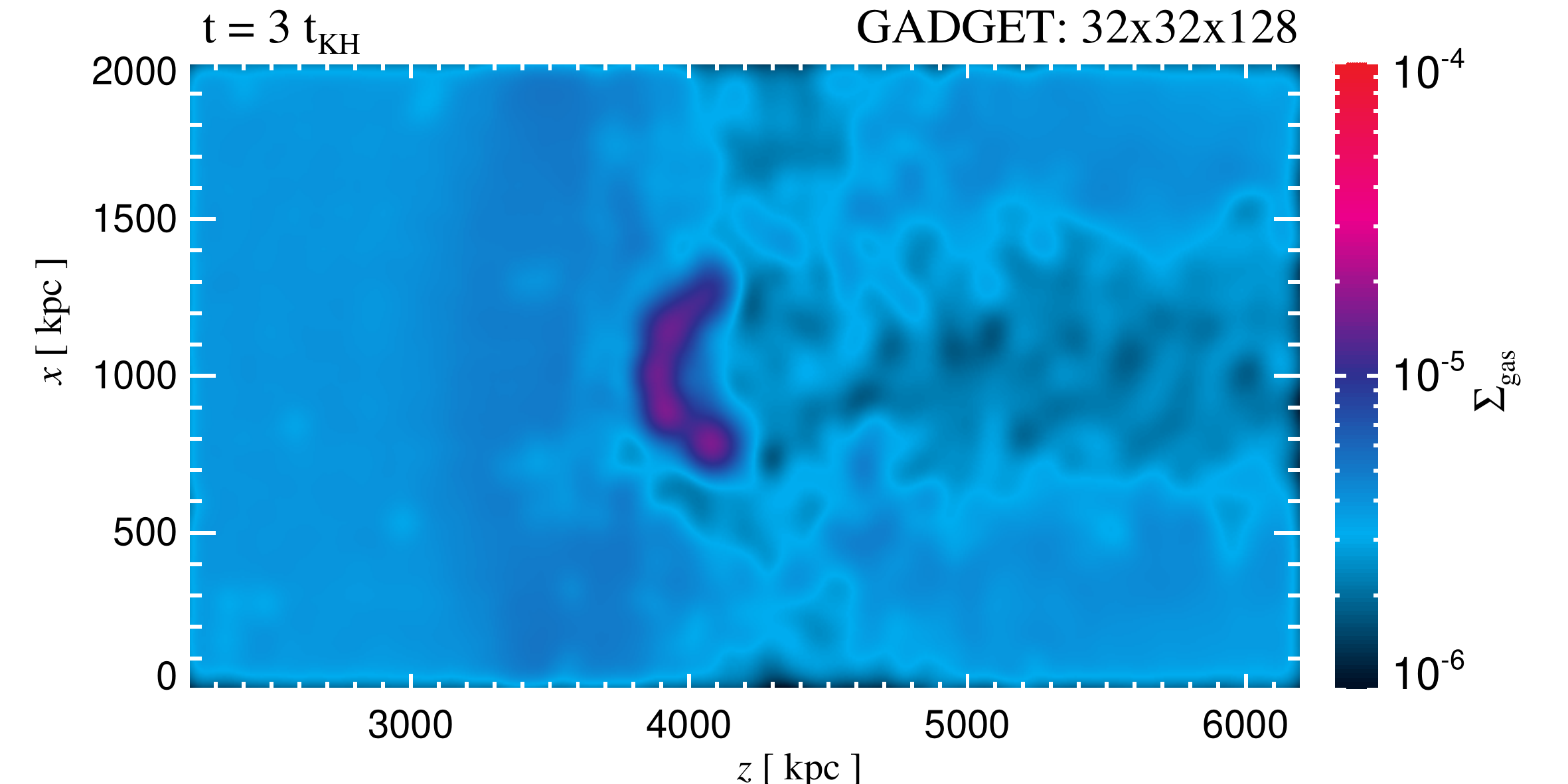}
\hspace{-0.35cm}
\includegraphics[width=9truecm,height=4.5truecm]{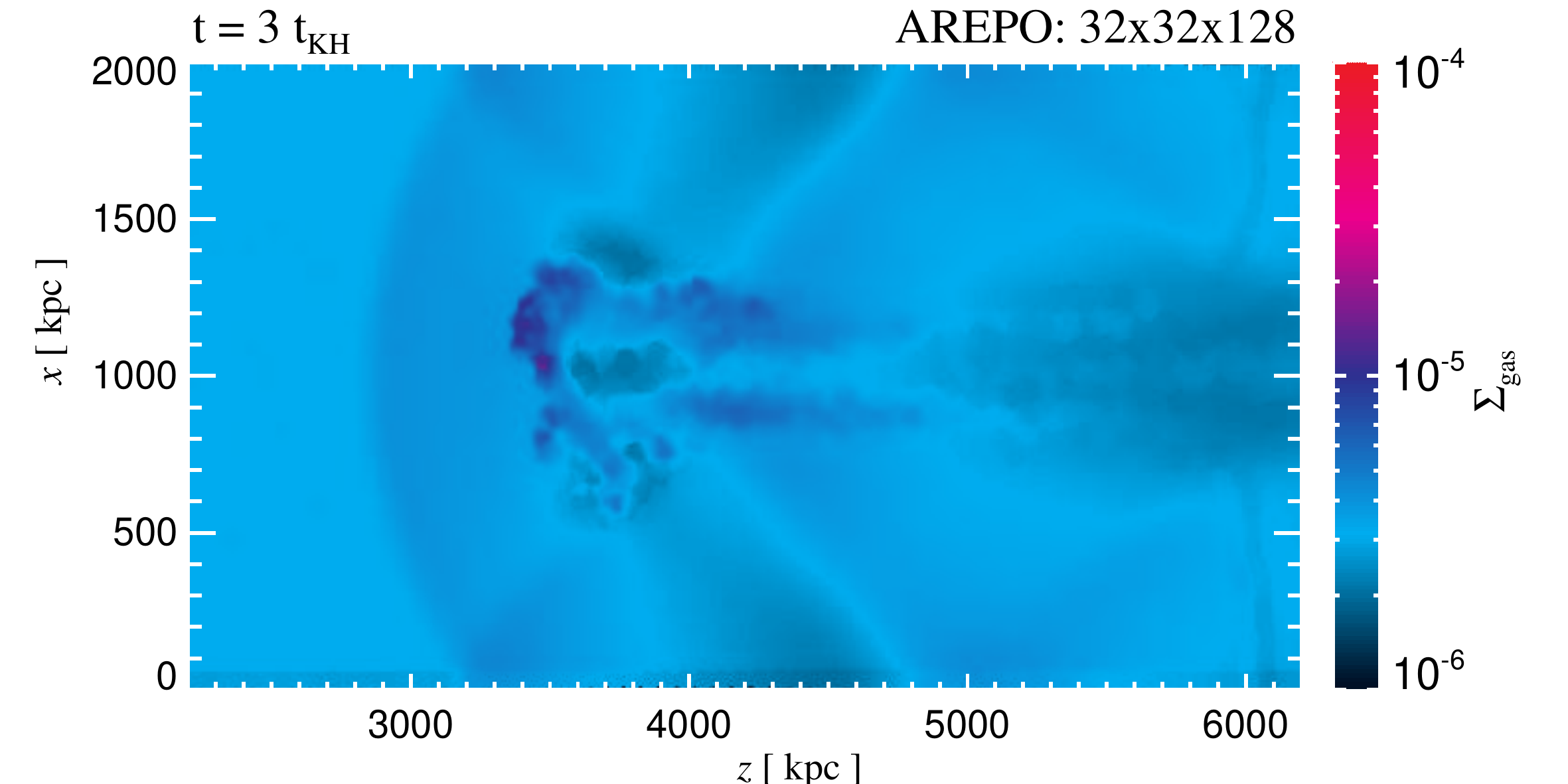}}
\hbox{
\includegraphics[width=9.truecm,height=4.5truecm]{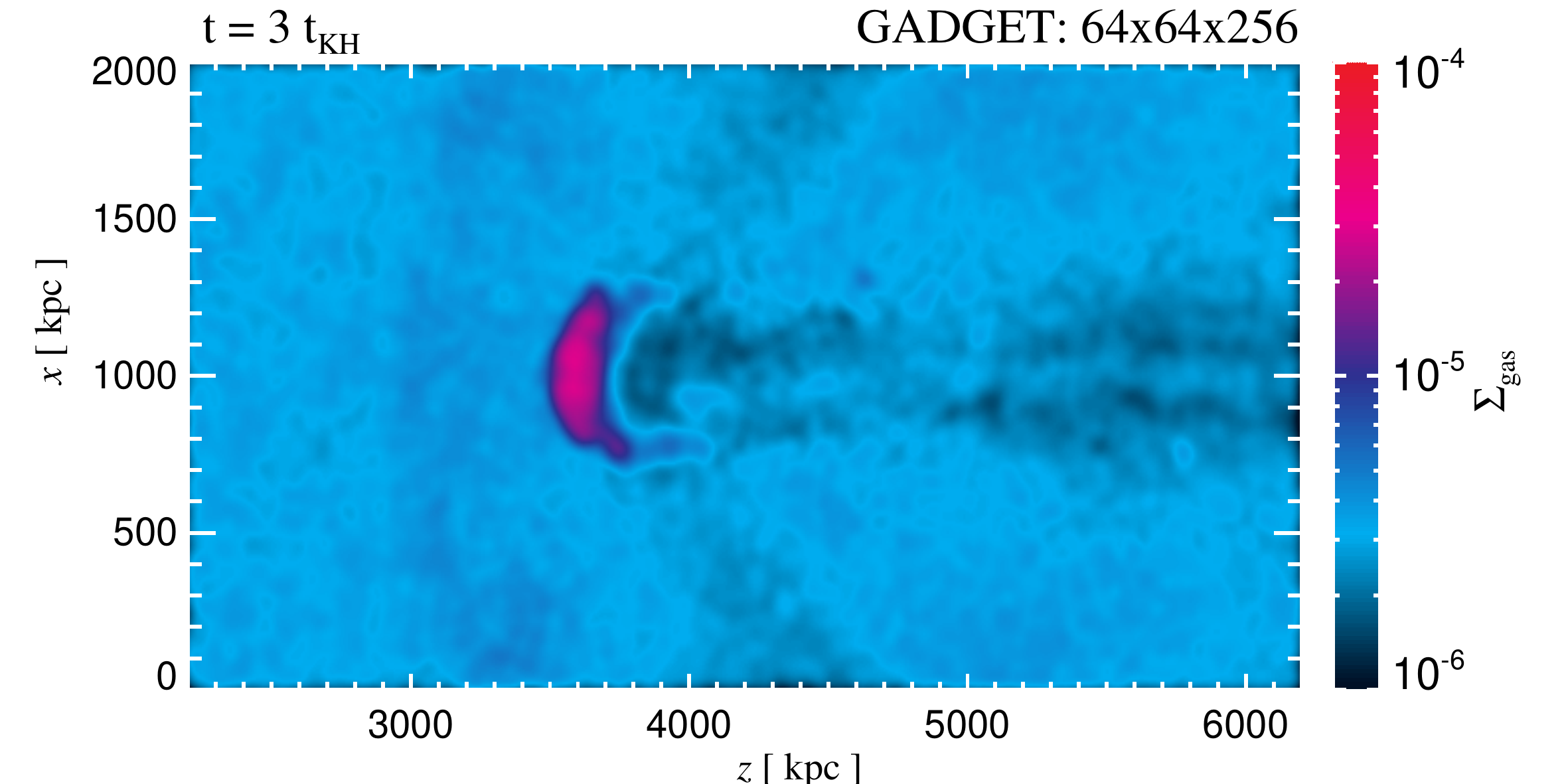}
\hspace{-0.35cm}
\includegraphics[width=9truecm,height=4.5truecm]{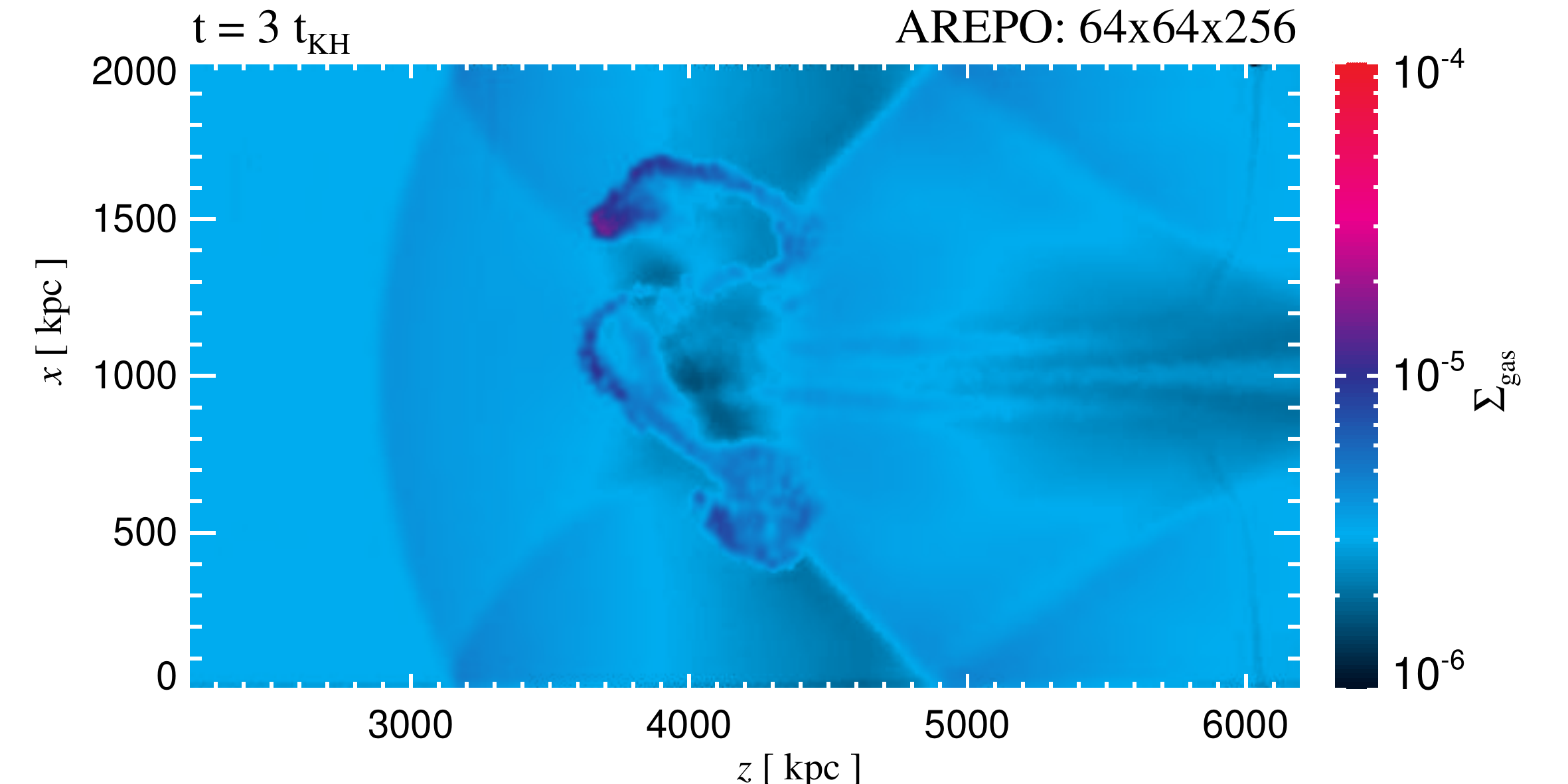}}}}
\caption{Projected surface density maps in units of $[{\rm
    M_{\odot}}{\rm kpc}^{-2}]$ at $t = 3\, t_{\rm KH}$ for the low
  (top row) and intermediate resolution simulation (bottom row) with
  {\small GADGET} and {\small AREPO}. The thickness of the slices is
  $\Delta y = 100\,{\rm kpc}$ and they are centred on $y_{\rm c} =
  1000\,{\rm kpc}$. Even at our lowest resolution, {\small AREPO}
  captures the dynamical evolution of the cold blob much more
  accurately.}
\label{FigBlobLR}
\end{figure*}

Our findings from the implosion test (Section~\ref{2DShock}) indicate
that there are fundamental differences in the treatment of fluid
instabilities between {\small GADGET} and {\small AREPO}. To
investigate this issue in detail, we first consider the so-called
``blob'' test as proposed by \citet{Agertz2007}, which has been
analyzed with many different codes by now \citep{Agertz2007, Hess2010,
  Murante2011}. The idea behind this test is to simulate the evolution
of a dense cold blob immersed in a hot windtunnel, mimicking in a
simplified way the motion of a satellite galaxy through the
intracluster medium \citep[see also][]{Hess2011}. If the relative
velocity between the blob and the surrounding medium is supersonic, a
bow shock will develop in front of the blob.  Additionally, dynamical
instabilities (mostly of Kelvin-Helmholtz and Rayleigh-Taylor type)
will grow in the subsonic part of the flow between the bow shock and
the surface of the blob \citep[see][for a detailed
description]{Agertz2007}. These instabilities will greatly influence
the evolution of the blob, leading to its eventual disintegration on
the characteristic Kelvin-Helmholtz timescale set by the initial
conditions.

To simulate this problem, we adopt the initial conditions used in the
original \citet{Agertz2007} paper, which are publicly
available\footnote{\sf http://www.astrosim.net/}. The simulated domain
consists of a periodic box with extensions $L_{\rm x} = 2000\,{\rm
  kpc}$, $L_{\rm y} = 2000\,{\rm kpc}$, $L_{\rm z} = 8000\,{\rm kpc}$,
and the blob is initially placed at $x_{\rm c} = y_{\rm c} = z_{\rm c}
= 1000\,{\rm kpc}$. The blob has a radius $R_{\rm blob} = 197\,{\rm
  kpc}$, and it is ten times colder and denser than the surrounding
medium, such that pressure equilibrium is ensured. The density and
temperature of the external medium is $\rho_{\rm medium} = 3.13 \times
10^{-8}{\rm M_{\odot}}{\rm kpc}^{-3}$ and $T_{\rm medium} =
10^{7}\,{\rm K}$, respectively, and it is moving with a constant
velocity $v_{\rm medium} = 1000\, {\rm km \, s}^{-1}$ along the
$z$-axis. The adiabatic index is set to $\gamma = 5/3$. Following
\cite{Agertz2007}, we define the characteristic Kelvin-Helmholtz
timescale as $t_{\rm KH} = 1.6\, t_{\rm cr}$, where $t_{\rm cr}$ is
the blob crushing time defined as $t_{\rm cr} = 2 R_{\rm blob}
(\rho_{\rm blob} / \rho_{\rm medium})^{1/2} / v_{\rm medium}$. For
these initial conditions, the characteristic Kelvin-Helmholtz
timescale is $t_{\rm KH} \sim 1.98\,{\rm Gyr}$.

We have performed the blob test at three different resolutions with {\small GADGET} and
{\small AREPO}. The resolutions used are: $32 \times 32 \times 128$, $64 \times 64
\times 256$, and $128 \times 128 \times 512$. The two higher resolutions
correspond exactly to the resolution of the simulations used in
\citet{Agertz2007}, while we constructed the initial conditions for the lowest
resolution run by sub-sampling. 

In Figure~\ref{FigBlob} we show projected surface density maps of a thin slice
($\Delta y = 100\,{\rm kpc}$) centred around the blob position. Left-hand
panels illustrate {\small GADGET} runs at times $t = t_{\rm KH}$, $t = 2\,
t_{\rm KH}$ and $t = 3\, t_{\rm KH}$, while the right-hand panels are for the
simulations with {\small AREPO} in the moving mesh mode. The position and the
shape of the bow shock are rather similar between {\small GADGET} and {\small
  AREPO} runs, especially at early times, but the shock is broader and less
crisp in {\small GADGET}. In the run with {\small GADGET}, the blob acquires a
cap-like appearance caused by the internal shock which compresses it, and it
undergoes continuous ablation due to the low pressure region which forms in
the wake of the blob \citep{Agertz2007}. While initially the blob evolution is
similar in {\small AREPO}, after $t = t_{\rm KH}$ well developed
Kelvin-Helmholtz and Rayleigh-Taylor instabilities lead to an efficient
shredding of the blob, mixing it with the external medium. In
Figure~\ref{FigBlobLR} we also show projected surface density maps centred
around the blob position for the $32 \times 32 \times 128$ and $64 \times 64
\times 256$ runs at $t = 3\, t_{\rm KH}$.

\begin{figure}\centerline{
\includegraphics[width=8.truecm,height=8truecm]{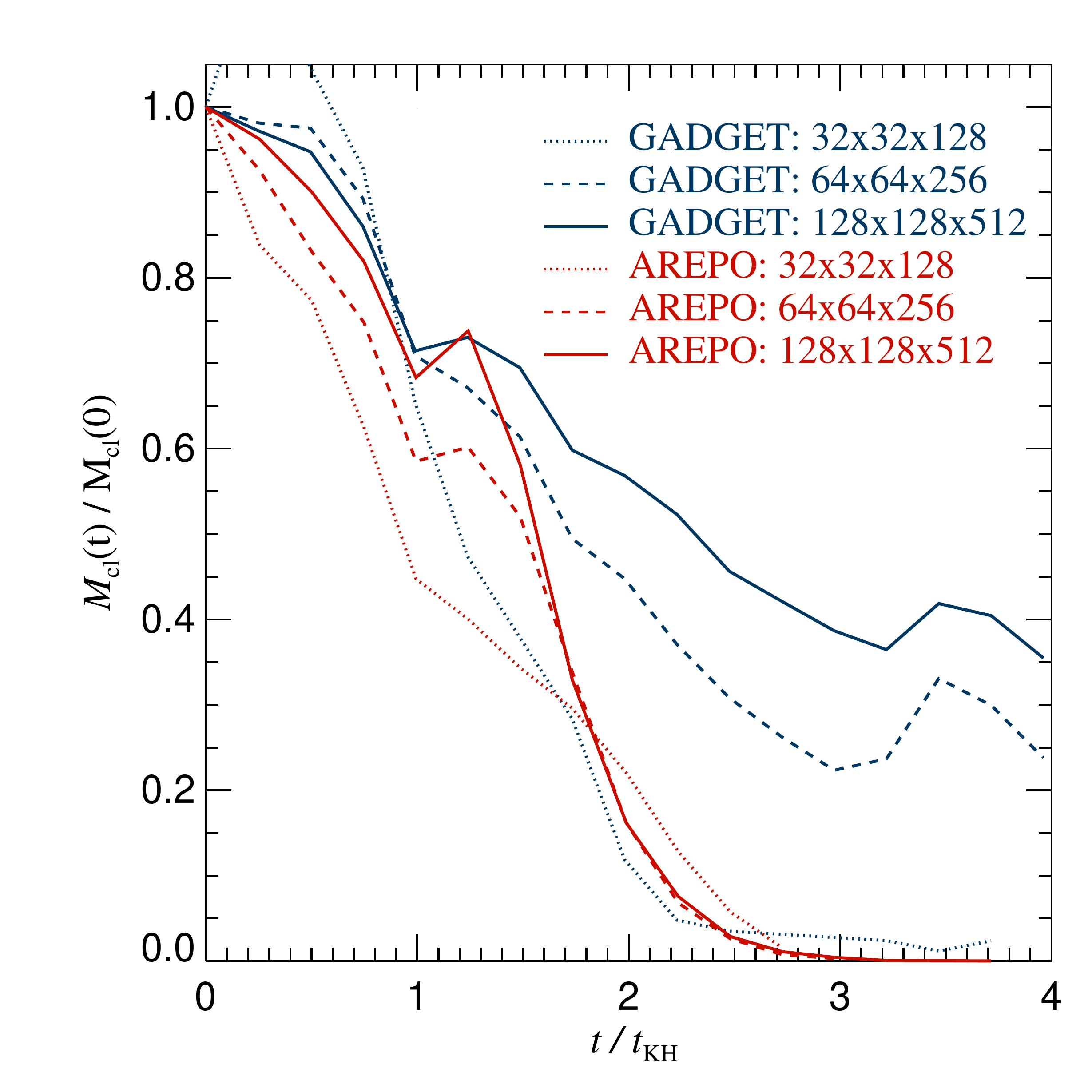}}
\caption{The remaining blob mass fraction as a function of time in
  units of $t_{\rm KH}$. {\small GADGET} and {\small AREPO} results at
  three different resolutions are shown, as indicated in the legend.}
\label{FigBlobMfrac}
\end{figure}

We quantify the time evolution of the blob in Figure~\ref{FigBlobMfrac}. Here
we show the remaining blob mass fraction as a function of time, where the
material associated with the blob is selected such as to satisfy: $T < 0.9\,
T_{\rm medium}$ and $\rho > 0.64\, \rho_{\rm blob}$, as has been done in the
previous studies. We compare {\small GADGET} and moving-mesh {\small AREPO}
simulations, performed at three different resolutions (as indicated on the
legend). While for our highest resolution runs there is a broad agreement
between {\small GADGET} and {\small AREPO} for $t < t_{\rm KH}$, blob mass
fractions are systematically different afterwards. This is exactly when the
transition in the mass loss rate from the ablation-dominated to the fluid
instability dominated regime occurs. In the latter regime, {\small AREPO}
clearly delivers a physically more trustworthy solution. The moving-mesh
{\small AREPO} simulation agrees qualitatively very well with the outcome of
Eulerian grid codes studied in \cite{Agertz2007}. However, as noted by
  \citet{Springel2011}, there is a small but systematic difference with the
  moving-mesh code delivering slightly higher remaining blob mass for $t > 1.5
  \times t_{\rm KH}$. As expected, the {\small GADGET} simulation results
agree well with the findings of \citet{Agertz2007} for the other SPH
codes. This indicates that the inaccuracies we find for {\small GADGET} are
inherent to the standard SPH method, and have prompted a number of works
  suggesting possible improvements to the SPH method \citep[e.g.][see also
    Section~\ref{OtherSPH}]{Price2008, Wadsley2008, Hess2010, Saitoh2012}.

\subsection{Shocks and fluid instabilities in isolated halo models}\label{NonRadHaloes}

\subsubsection{Isolated haloes in hydrostatic equilibrium}\label{IsolatedHaloEquil}     

To explore more directly the impact of shocks and fluid instabilities
in the context of structure formation, we have devised a number of
idealized test problems involving isolated halo models. Here we
briefly describe how we set up and verify hydrostatic equilibrium
configuration of the initial conditions, which form the backbone for a
series of numerical experiments discussed in
Sections~\ref{NonRadHaloes} and \ref{RadHaloes}.

The initial conditions are constructed by populating static or live
dark matter potentials with gas particles, whose positions are drawn
randomly. For the dark matter distribution we assume a Hernquist
profile \citep{Hernquist1990} so that we can easily generate
self-consistent models when we use live haloes. The gas density
profile traces the dark matter at large radii but is slightly softened
in the centre, i.e.  \be \rho_{\rm gas} (r) = \frac{M_{\rm vir}}{(2
  \pi a^3) (x + x_{0}) (x + 1)^3}, \ee where $M_{\rm vir}$ is the
virial mass of the system, $a$ is the Hernquist scale length
parameter, $x = r/a$, and $x_{0} = 0.01$ is the softening scale length
parameter for the gaseous halo. For the simulations without any net
rotation, the initial gas velocities are set to zero, while for the
simulation with non-vanishing angular momentum we assign gas
velocities such that the halo is characterized by the dimensionless
spin parameter \be \lambda \, = \,\frac{J|E|^{1/2}}{GM_{\rm
    vir}^{5/2}}, \ee where $J$ represents the angular momentum, $E$ is
the total energy of the halo, and we assume solid body rotation.

For validation purposes we evolve isolated haloes with {\small GADGET} and
{\small AREPO} for $2.45\,{\rm Gyr}$ with gas self-gravity and no radiative
losses. These test runs confirm that the gas is in very good hydrostatic
equilibrium within the dark matter potential. The differences in gas density,
temperature and entropy between {\small GADGET} and {\small AREPO} are within
a few percent throughout the whole halo at the final time in the case of
static dark matter haloes (see Figure~\ref{A1} in
Appendix~\ref{AppIsolatedHaloEquil}; for live haloes see
Section~\ref{N-bodyHeating}). 

We also observe that even though the initial gas velocities are zero
(in the case with $\lambda = 0$) some small gas velocities develop
over time ($\sigma_{\rm gas, 3D} \sim 30\,{\rm km\,s^{-1}}$). This is
primarily caused by the initial Poisson sampling of gas positions,
which implies an initial state that is not perfectly relaxed.  While
this numerical artifact could be avoided by explicitly relaxing the
initial conditions, we note that these residual gas velocities do not
have any significant bearing on our results: the total gas kinetic
energy $E_{\rm kin}$ is less than $ 0.005$ of either the total
potential or the total internal gas energy at the final time, as shown in
Figure~\ref{A2} in Appendix~\ref{AppIsolatedHaloEquil}. It is,
however, interesting to note that while the total $\sigma_{\rm gas}$
is very similar between {\small GADGET} and {\small AREPO}, there are
some systematic differences in the radial profiles of $\sigma_{\rm
  gas}$, which are caused by dissipation of gas motions on different
spatial scales (see bottom panel of Figure~\ref{A1}). 
 
\subsubsection{Radial collapse of cold gas in a static dark matter halo}\label{ColdInflow}

\begin{figure}\centerline{
\includegraphics[width=8.3truecm,height=20.truecm]{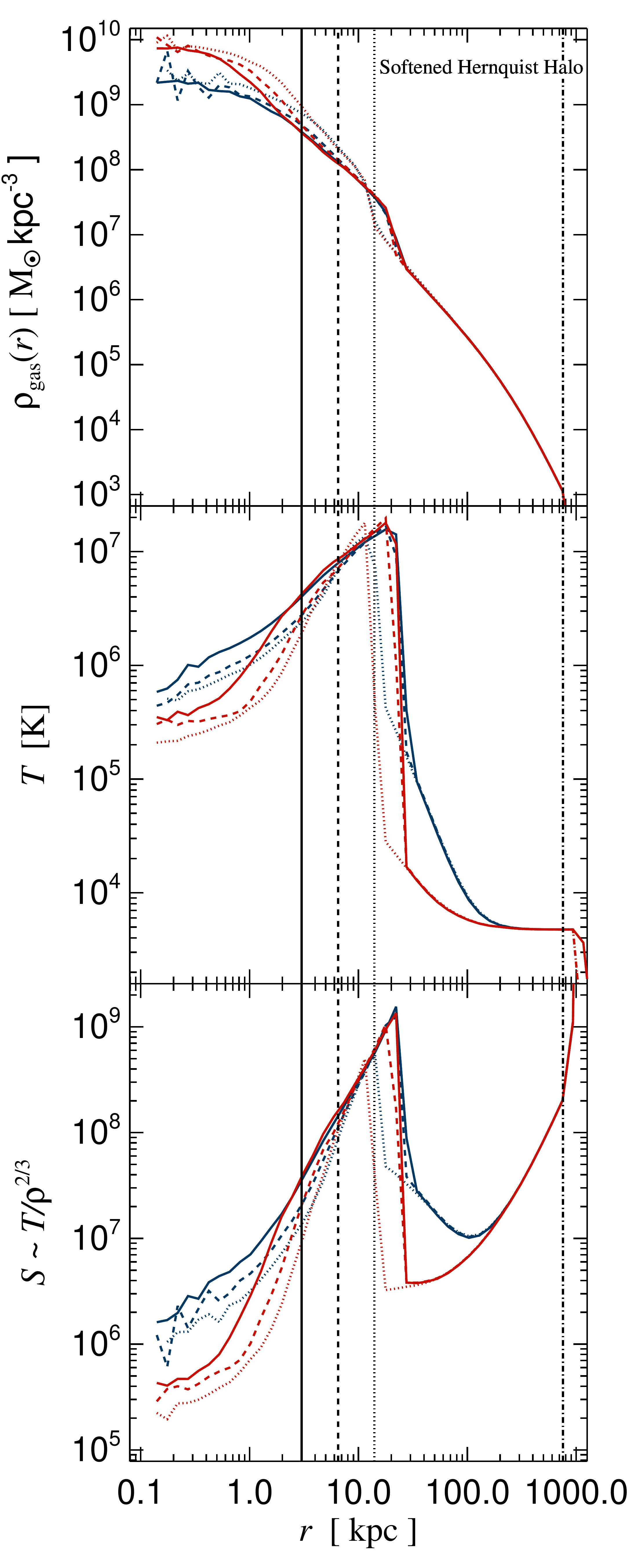}}
\caption{Radial profiles of gas density, temperature and entropy at
  time $t = 0.1\,{\rm Gyr}$ for {\small GADGET} (blue lines) and for
  {\small AREPO} (red lines). For each code we show three different
  resolution runs: $N_{\rm gas} = 10^4$ and $r_{\rm soft} = 14.0
  \,{\rm kpc}$ (dotted lines), $N_{\rm gas} = 10^5$ and $r_{\rm soft}
  = 6.5 \,{\rm kpc}$ (dashed lines) and $N_{\rm gas} = 10^6$ and
  $r_{\rm soft} = 3.0 \,{\rm kpc}$ (continuous lines). Vertical black
  lines with the same style indicate the softening scales of the dark
  matter potential (note that this does not correspond strictly to the
  spatial resolution limit because the gas is not
  self-gravitating). The black vertical dot-dashed line denotes the
  virial radius of the system.}
\label{FigColdInflow}
\end{figure}

\begin{figure}\centerline{
\includegraphics[width=8.3truecm,height=6.0truecm]{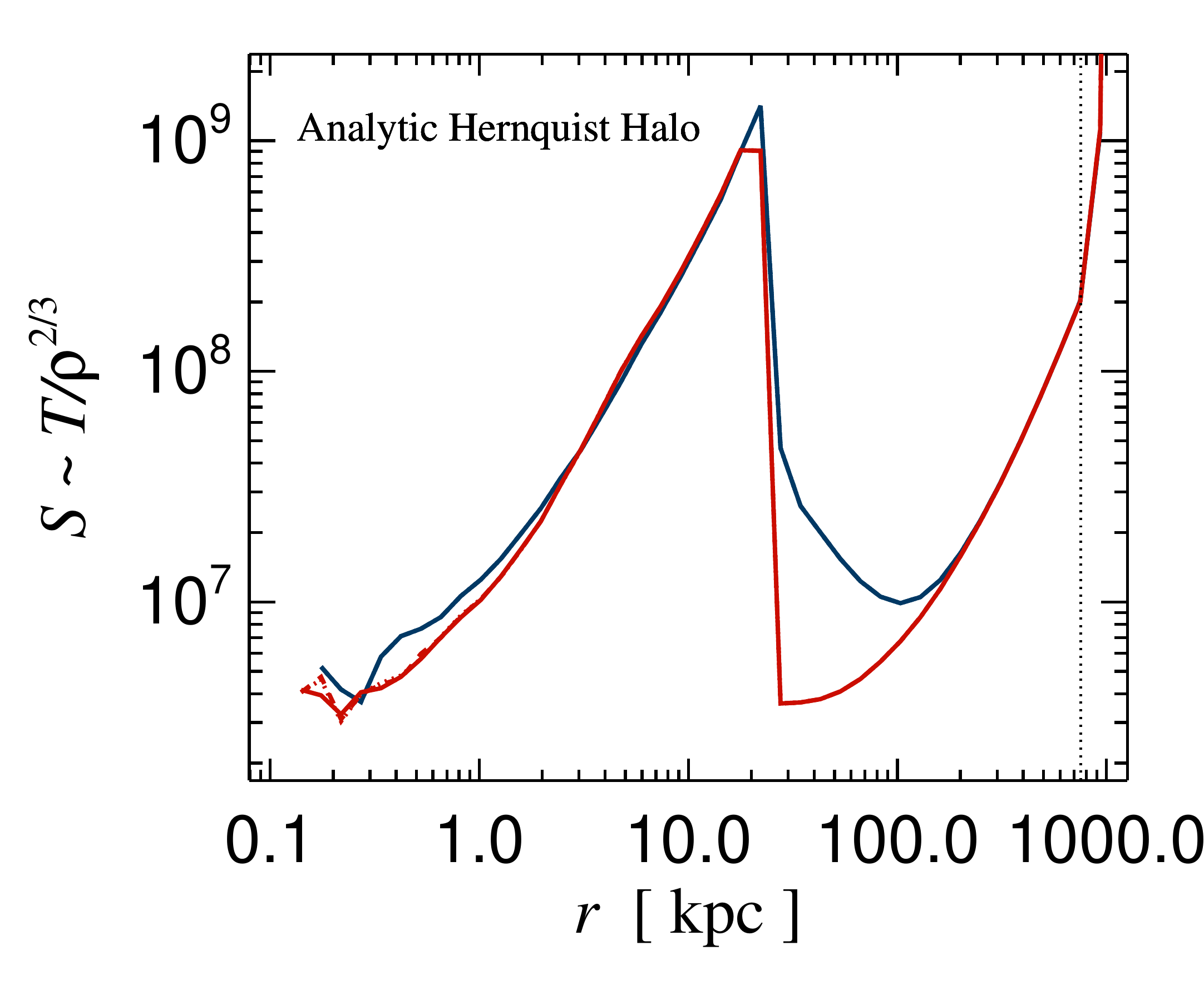}}
\caption{Radial profiles of gas entropy at time $t = 0.1\,{\rm Gyr}$
  for {\small GADGET} (blue lines) and for {\small AREPO} (red
  lines). For each code we show three different resolution runs:
  $N_{\rm gas} = 10^4$ (dotted lines), $N_{\rm gas} = 10^5$ (dashed
  lines) and $N_{\rm gas} = 10^6$ (continuous lines), which are
  largely overlapping. These runs have been computed by adopting an
  analytic Hernquist dark matter halo without any softening. The black
  vertical dotted line denotes the virial radius of the system.}
\label{FigColdInflowNoSm}
\end{figure}

We now analyze how differences in shock capturing between {\small
  GADGET} and {\small AREPO} affect the radial infall of gas in dark
matter haloes, a problem of direct cosmological interest. For this
purpose we intentionally adopt a set-up as simple as possible in order
to isolate differences between the codes driven by the shock
treatment only. We initially set up gas in hydrostatic equilibrium and at
rest within a static Hernquist potential with mass $M_{\rm vir} =
10^{14} {\rm M_{\odot}}$, scale length $a = 176 \,{\rm kpc}$, and a
gas fraction of $f_{\rm gas} = 0.17$. We consider both an analytic
gravitational potential and a potential with a centrally softened
core\footnote{We modify the analytic Hernquist potential by convolving
  it with the spline softened potential in the centre, as we describe
  in detail in Section~\ref{N-bodyHeating}.}. We introduce a small
modification in the codes such that gas self-gravity is switched off
-- the gas only feels the static dark matter potential and
hydrodynamical forces in a purely non-radiative regime. We then
artificially reduce the internal energies of gas particles/cells so as
to displace the gas from the equilibrium solution. The newly assigned
gas temperature is $4.7\times10^3\,{\rm K}$, equal for all resolution
elements. This test is hence analogous to the well-studied Evrard
collapse \citep{Evrard1988}, but it is even simpler in nature because
we intentionally neglect gas self-gravity.

As a consequence of the dramatic reduction in its temperature, the gas
will suddenly lose pressure support and radially free-fall towards the
centre. As the gas collapses, a radial shock develops in the centre,
steepening while it propagates outwards and ploughing trough the
remainder of the outer material which is still falling in. The Mach
number of the shock varies over the range $\sim 3 - 8$, well matched
to the 1D shock tube test problem described in
Section~\ref{1DShock}. Finally, as the shock propagates beyond the
virial radius of the halo, the gas will reach a new hydrostatic
equilibrium solution within the static dark matter potential. In
Figure~\ref{FigColdInflow}, we show radial profiles of gas density,
temperature and entropy computed at time $t = 0.1\,{\rm Gyr}$ after
the start of the gas collapse for dark matter haloes with softened
potentials. For each code ({\small GADGET}, $N_{\rm gb} = 64$,
$\alpha=1.0$: blue lines; {\small AREPO}: red lines) we perform three
runs at different resolutions: $N_{\rm gas} = 10^4$ and $r_{\rm soft}
= 14.0 \,{\rm kpc}$ (dotted lines), $N_{\rm gas} = 10^5$ and $r_{\rm
  soft} = 6.5 \,{\rm kpc}$ (dashed lines) and $N_{\rm gas} = 10^6$ and
$r_{\rm soft} = 3.0 \,{\rm kpc}$ (continuous lines). Note that the
$r_{\rm soft}$ values indicate the spatial scale over which we smooth
the analytic dark matter potential. They do not necessarily represent
the minimum spatial resolution of these simulations, given that the
gas is not self-gravitating and that the dark matter halo is
rigid. The smoothing lengths of gas particles in the central region
are of the order of $r_{\rm hsml} \sim 1 \,{\rm kpc}$ in {\small
  GADGET} for the lowest resolution run with $N_{\rm gas} = 10^4$,
while the typical central cell sizes in {\small AREPO} are a few
hundred ${\rm pc}$.

From Figure~\ref{FigColdInflow} it can be seen that there are
systematic differences in the gas properties predicted by the
simulations with {\small GADGET} and {\small AREPO}. In particular,
the gas entropy distribution is broader in {\small GADGET} both in
pre- and post-shock gas in all three simulations, while for the two
lower resolution runs there is a slight mismatch in the exact position
of the shock front and in its strength \citep[similar to the findings
for the Evrard collapse in][]{Springel2010}, which is minimized in the
case of $N_{\rm gas} = 10^6$ particles. Note that the differences in
shock front position between different resolution simulations for a
given code are not driven by resolution effects but by different
spatial softening of the central potential, which affects the gas
collapse in the innermost regions. We have checked this explicitly by
performing runs with $N_{\rm gas} = 10^4, 10^5$ and $10^6$ for both
codes, but this time simulating cold gas collapse within an analytic
Hernquist potential, as shown in Figure~\ref{FigColdInflowNoSm}. In
this case the shock properties for different resolution runs are
almost identical for a given code, indicating that the simulation with
$N_{\rm gas} = 10^4$ resolution elements is in principle sufficient
for capturing the shock position accurately. Nonetheless, regardless
of the resolution, differences between {\small GADGET} and {\small
  AREPO} in the pre- and post-shock gas persist also in the case of
analytic Hernquist potentials.

The differences in entropy content of pre- and post-shock gas in
{\small GADGET} and {\small AREPO} are in part due to the large shock
broadening in SPH, as discussed in Sections~\ref{1DShock} and
\ref{2DShock}. In fact, if we adopt $N_{\rm ngb} = 32$ (which is
considered the minimum number still permissable in three-dimensional
simulations) instead of $N_{\rm ngb} = 64$, the gas entropy profile in
{\small GADGET} becomes less broad, but is still not as sharp as in
{\small AREPO}, indicating that simulations with a larger number of
particles are needed in {\small GADGET} than in {\small AREPO} to
recover shock features with the same accuracy.

Moreover, there is another numerical effect leading to spatially
different entropy generation in {\small GADGET}: in the converging
subsonic part of the flow, artificial viscosity (as implemented in
{\small GADGET}) leads to artificial dissipation which increases the
entropy in the pre-shock gas. While this feature is clearly visible in
Figure 40 of \citet{Springel2010} for the case of the Evrard collapse,
here we see that it also enlarges the central entropy in {\small
  GADGET}. The reason for this is the following: as soon as the gas is
brought out of equilibrium and starts free-falling towards the centre,
the gas entropy will be boosted in the central region due to an active
artificial viscosity in the converging flow, creating an entropy bump
that extends up to several ${\rm kpc}$ (or even several tens of ${\rm
  kpc}$ in the case of potentials with large cores) away from the
centre, even though the shock has not fully formed yet at this
point. This artificial entropy generation is much smaller in
{\small AREPO}. As the shock forms and propagates outwards, it will
lead to additional physical dissipation of much higher magnitude,
bringing the entropy profiles of {\small GADGET} and {\small AREPO}
into better agreement. Interestingly, in the case of the Evrard
collapse, the initial difference in central entropy profiles is
minimized with time due to the gas self-gravity (as we checked
explicitly by running a simulation with exactly the same gas
configuration but with gas self-gravity and without static dark matter
potential), while it persists in our test runs even when the new
equilibrium solution of the system is reached. The central entropy is
higher in {\small GADGET} by a factor $\sim 1.2$ and $\sim 1.5$ for
analytic and softened potentials, respectively. Thus it follows that
differences in the gas properties due to different spatial dissipation
of kinetic energy in {\small GADGET} and {\small AREPO} can be
aggravated in the case of non self-gravitating gas.

Note, however, that the difference in the central entropy profiles
between the two codes goes in the opposite direction to what is found
in non-radiative simulations of hierarchically forming galaxy
clusters, where the central entropy is higher in mesh-based
calculations. This indicates that accretion shocks during cosmological
structure formation do not seem to be the likely cause of this central
entropy discrepancy.
 
\subsubsection{Infall of two gaseous spheres in a static dark matter halo}\label{Collision}

\begin{figure*}\centerline{\vbox{
\hbox{
\includegraphics[width=4.8truecm]{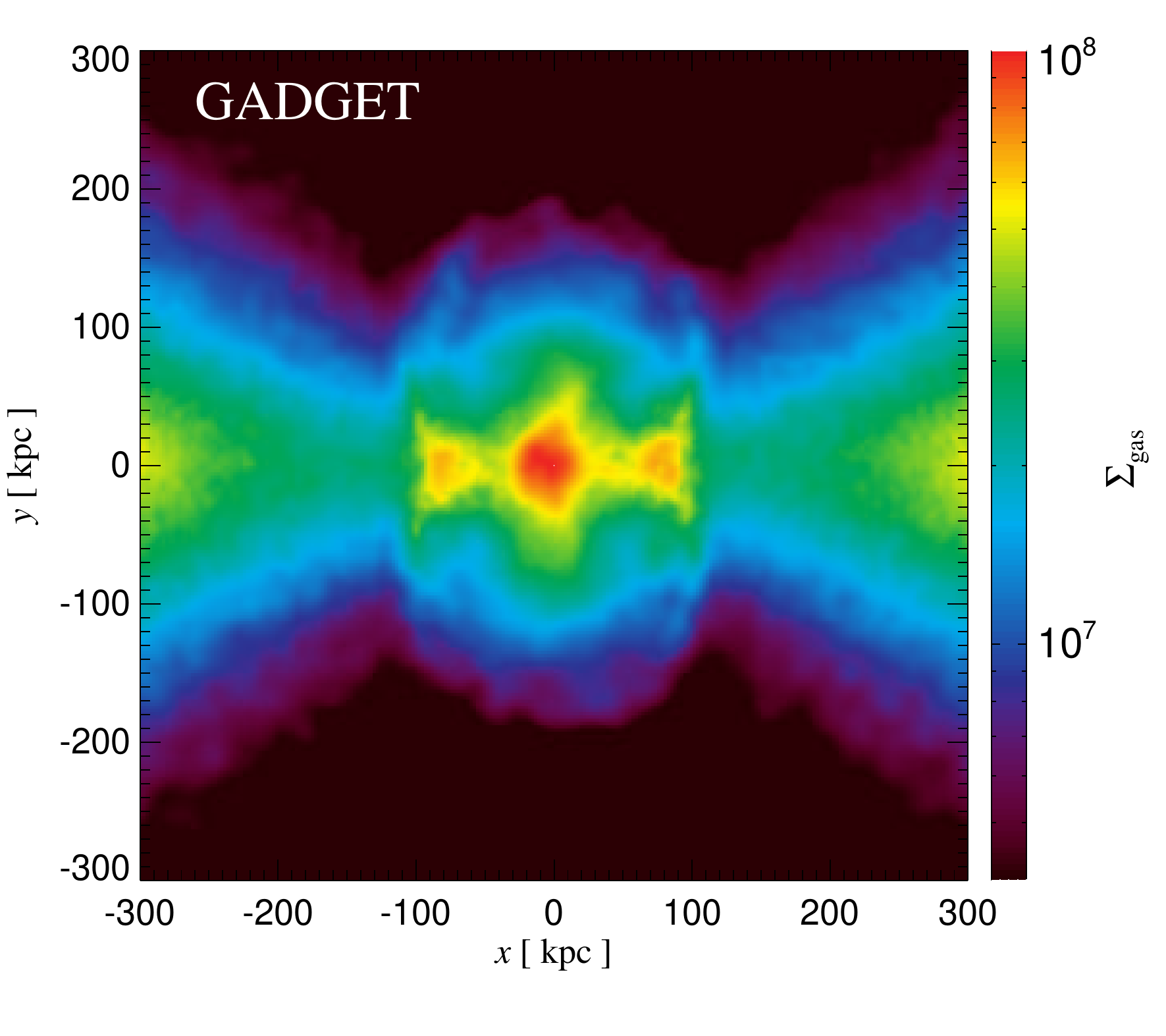}
\hspace{-0.5cm}
\includegraphics[width=4.8truecm]{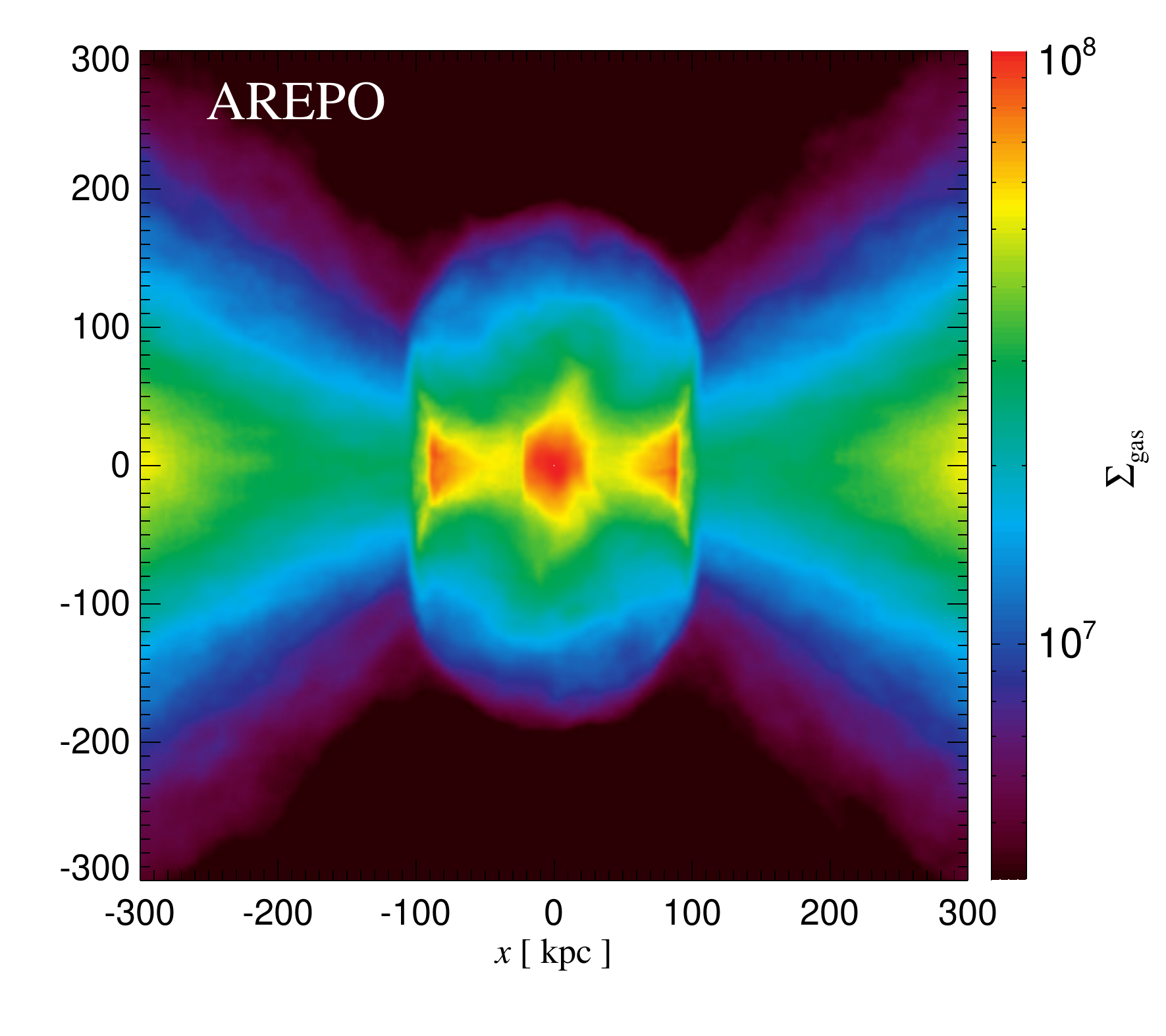}
\includegraphics[width=4.8truecm]{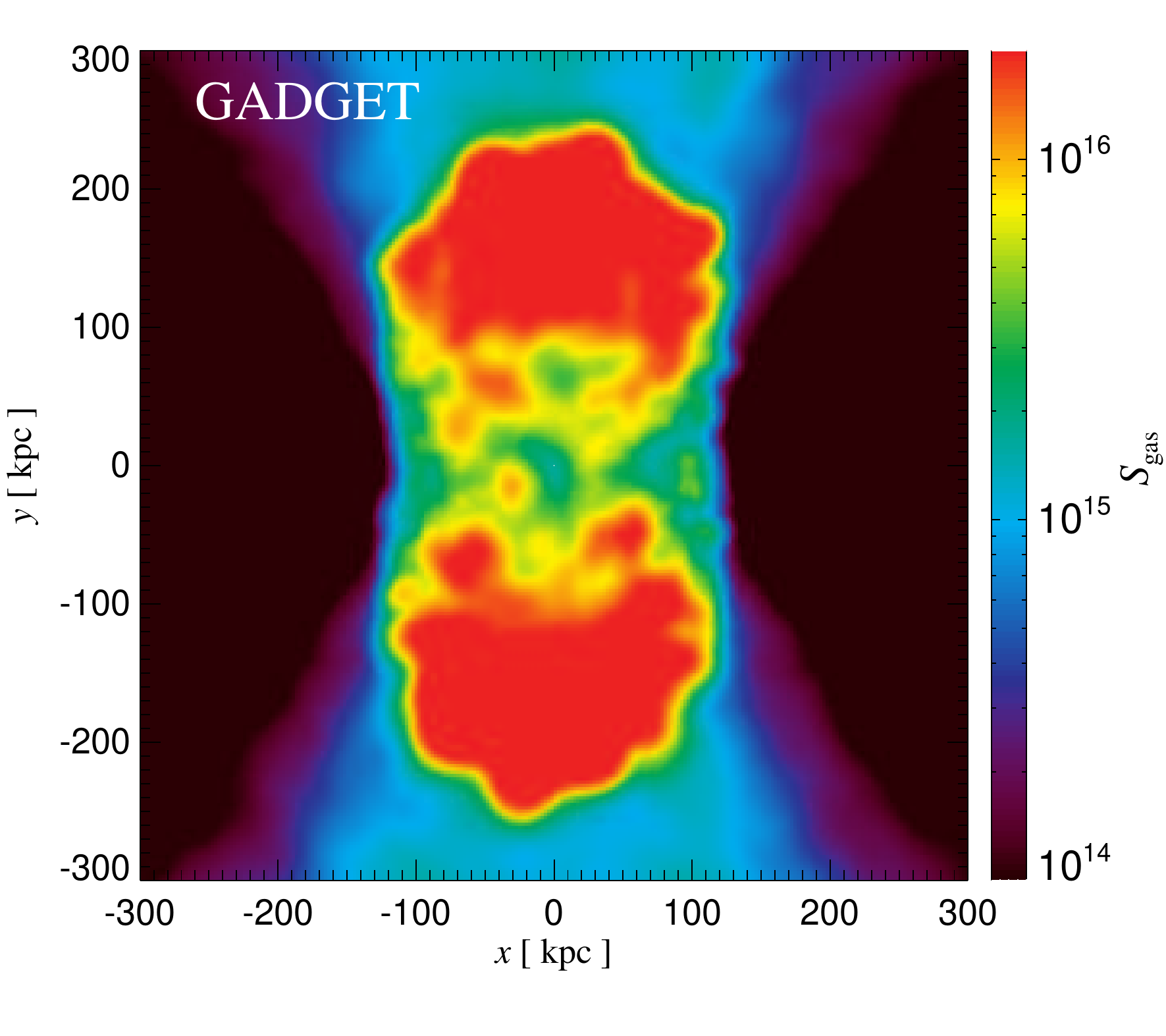}
\hspace{-0.5cm}
\includegraphics[width=4.8truecm]{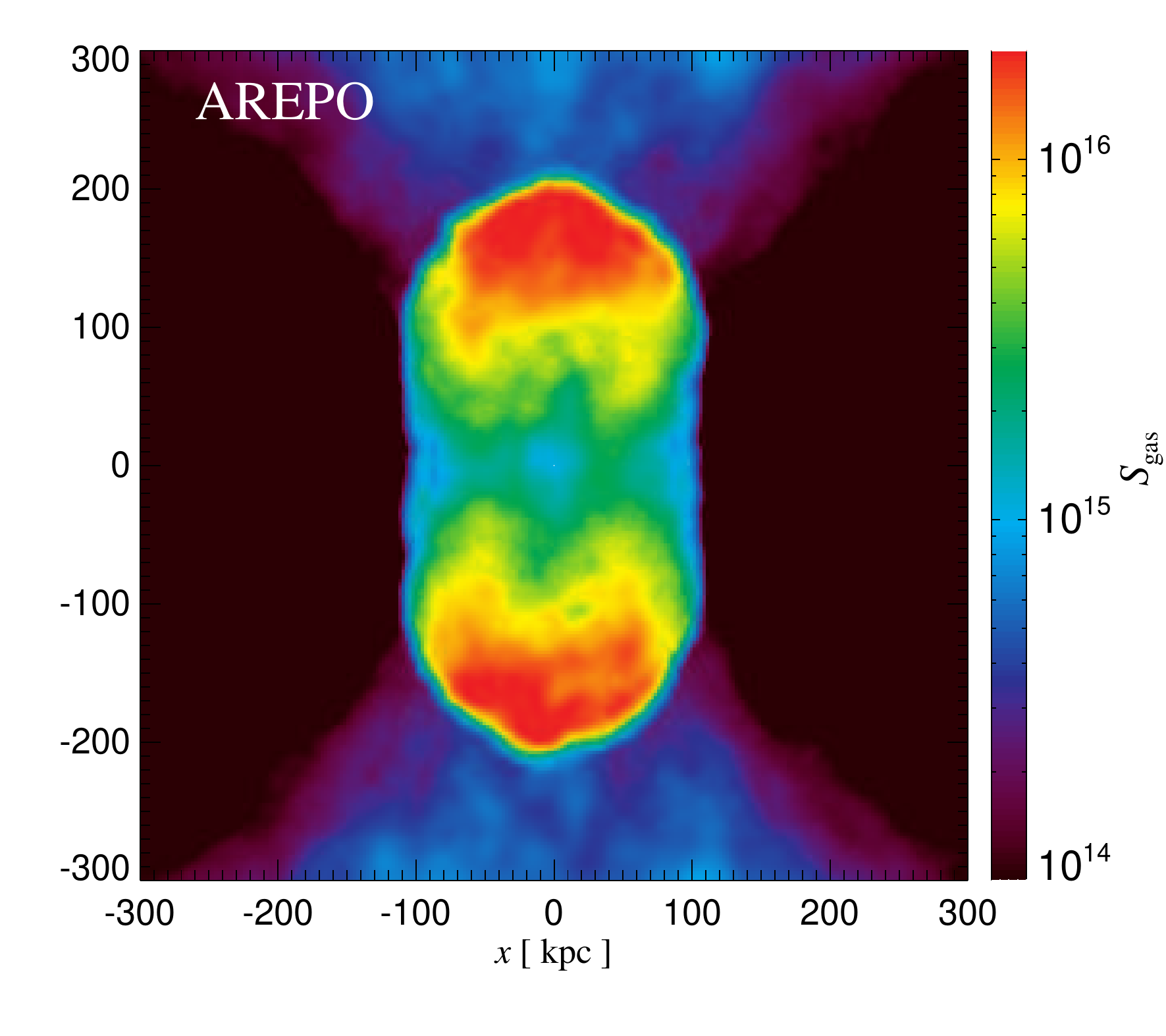}
}
\vspace{-0.25cm}
\hbox{
\includegraphics[width=4.8truecm]{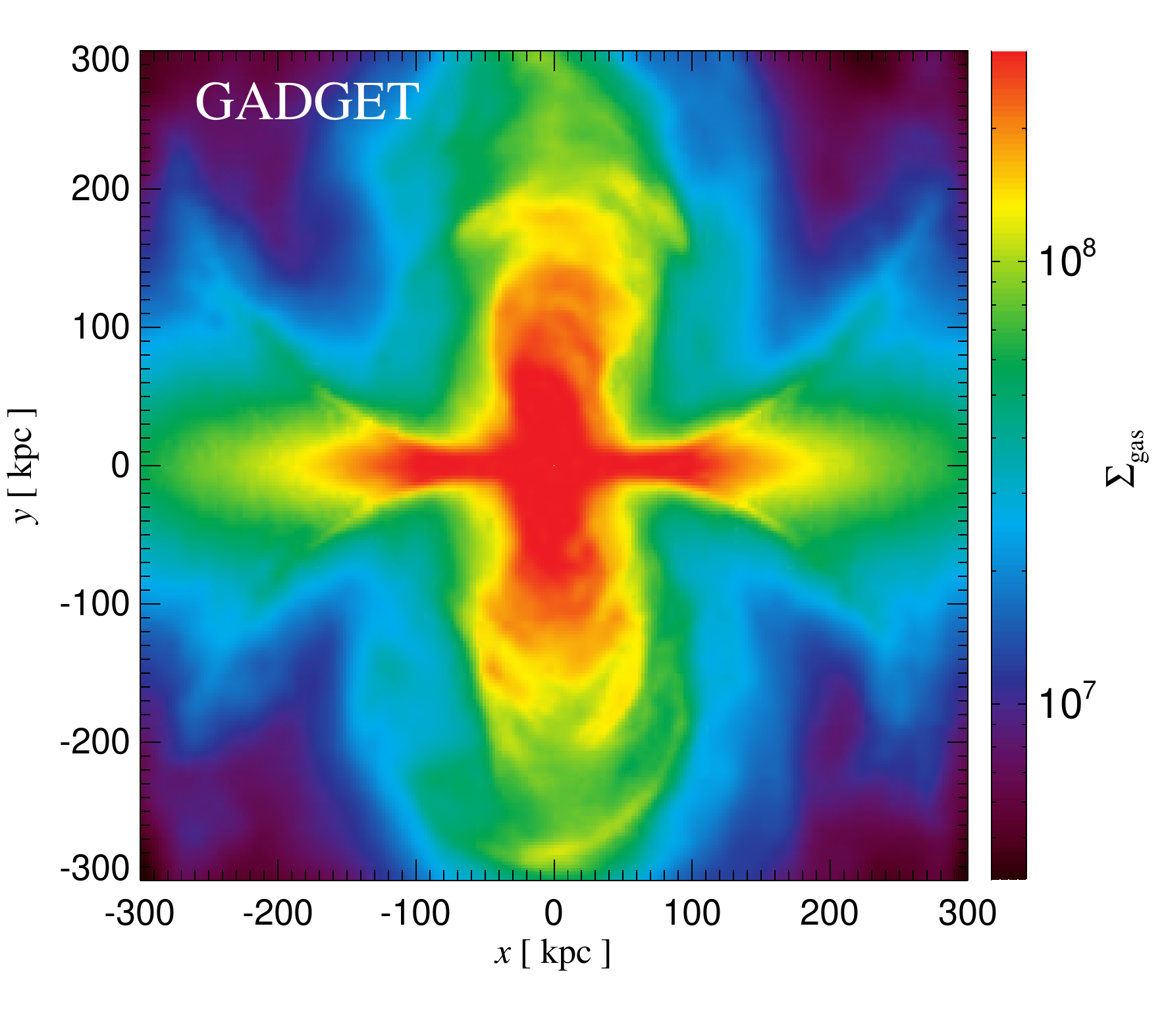}
\hspace{-0.5cm}
\includegraphics[width=4.8truecm]{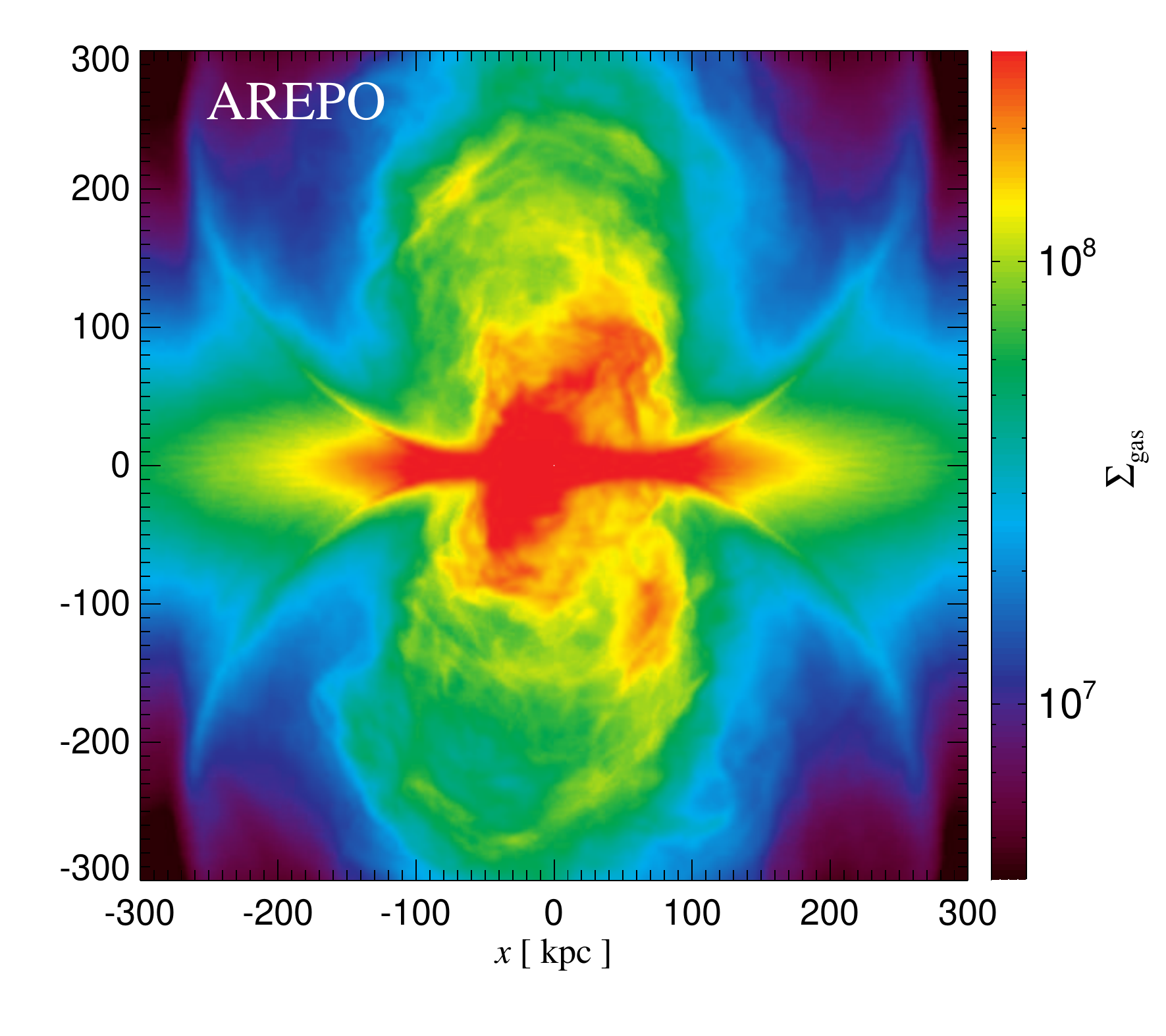}
\includegraphics[width=4.8truecm]{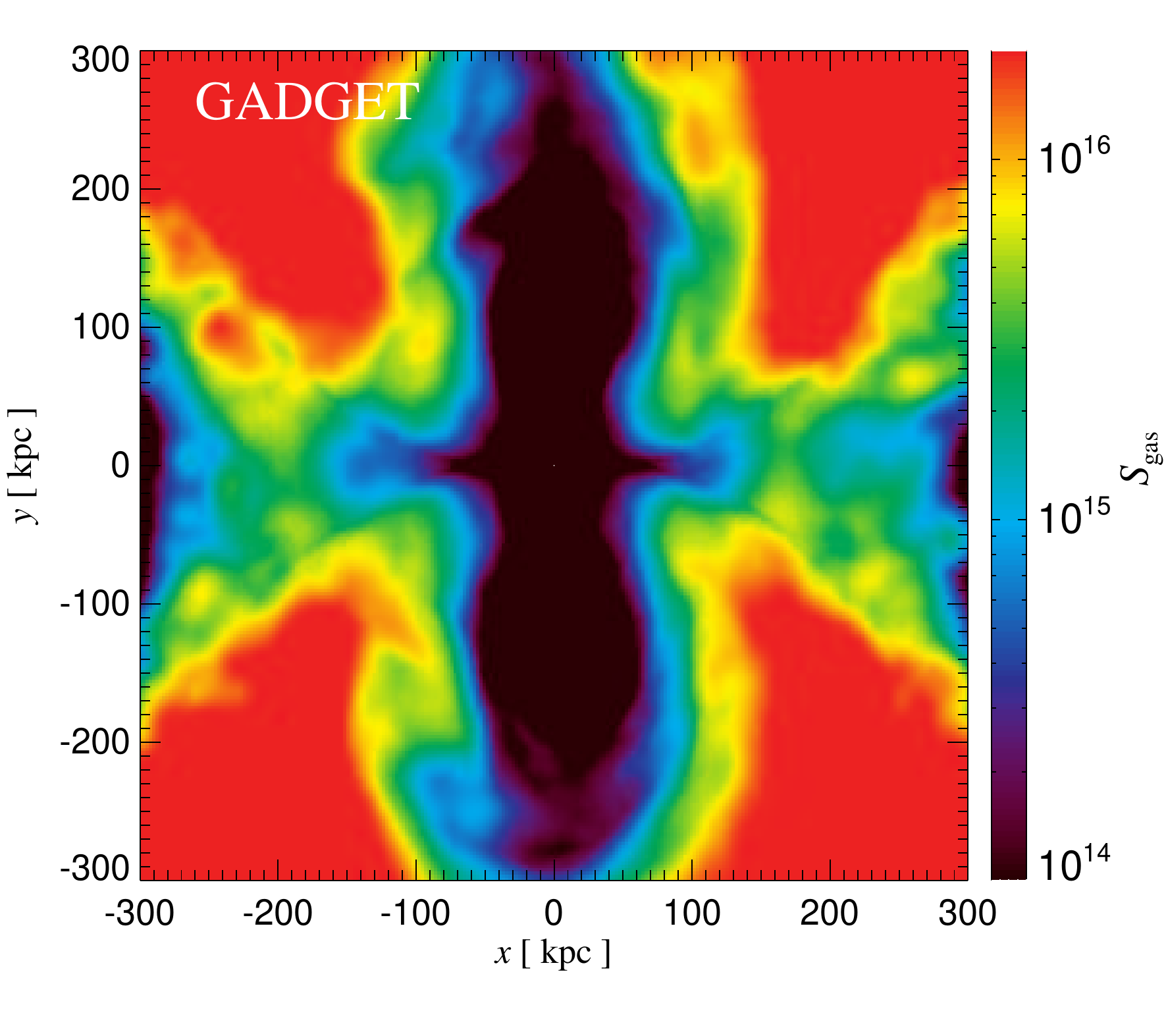}
\hspace{-0.5cm}
\includegraphics[width=4.8truecm]{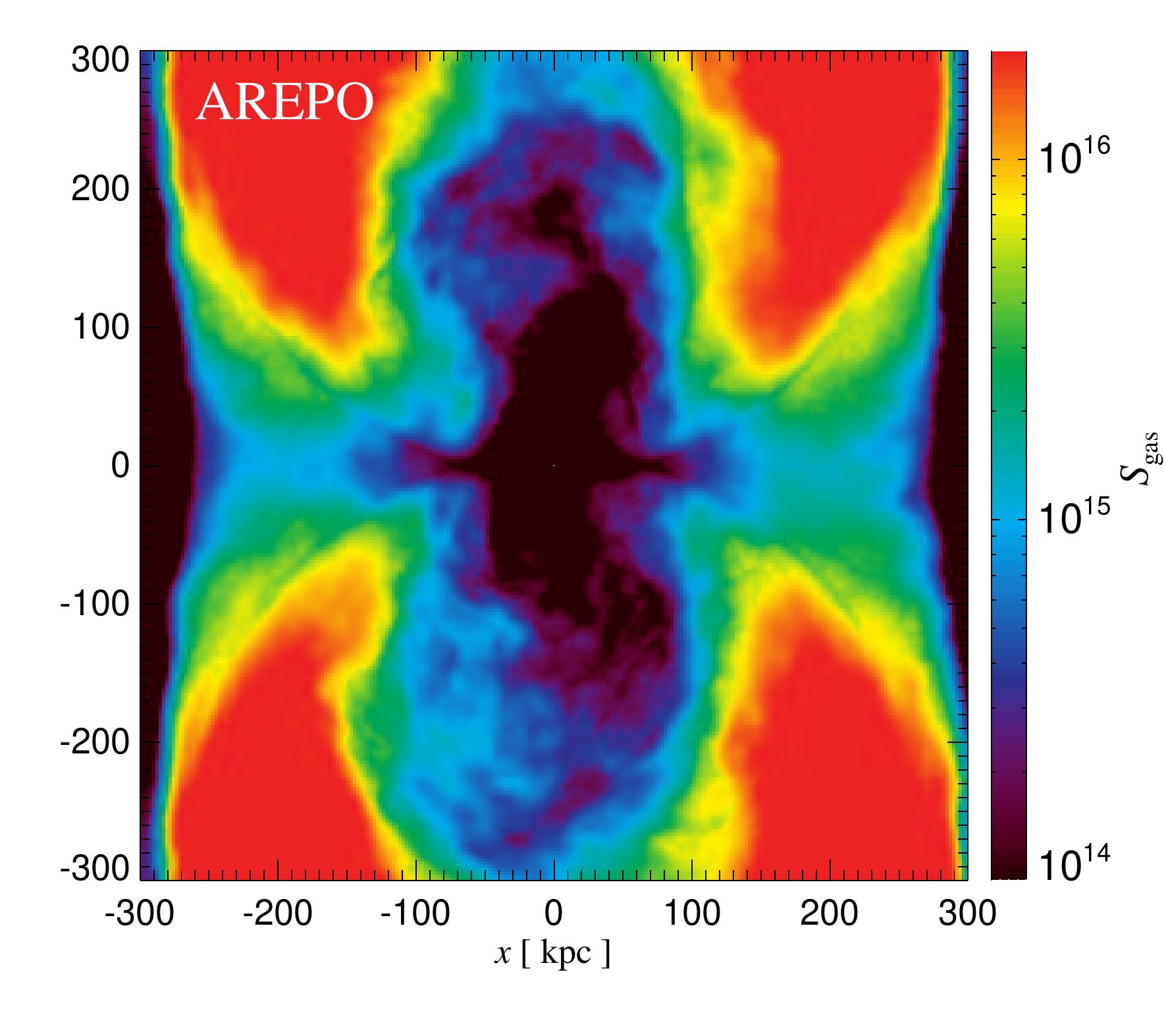}
}
\vspace{-0.25cm}
\hbox{
\includegraphics[width=4.8truecm]{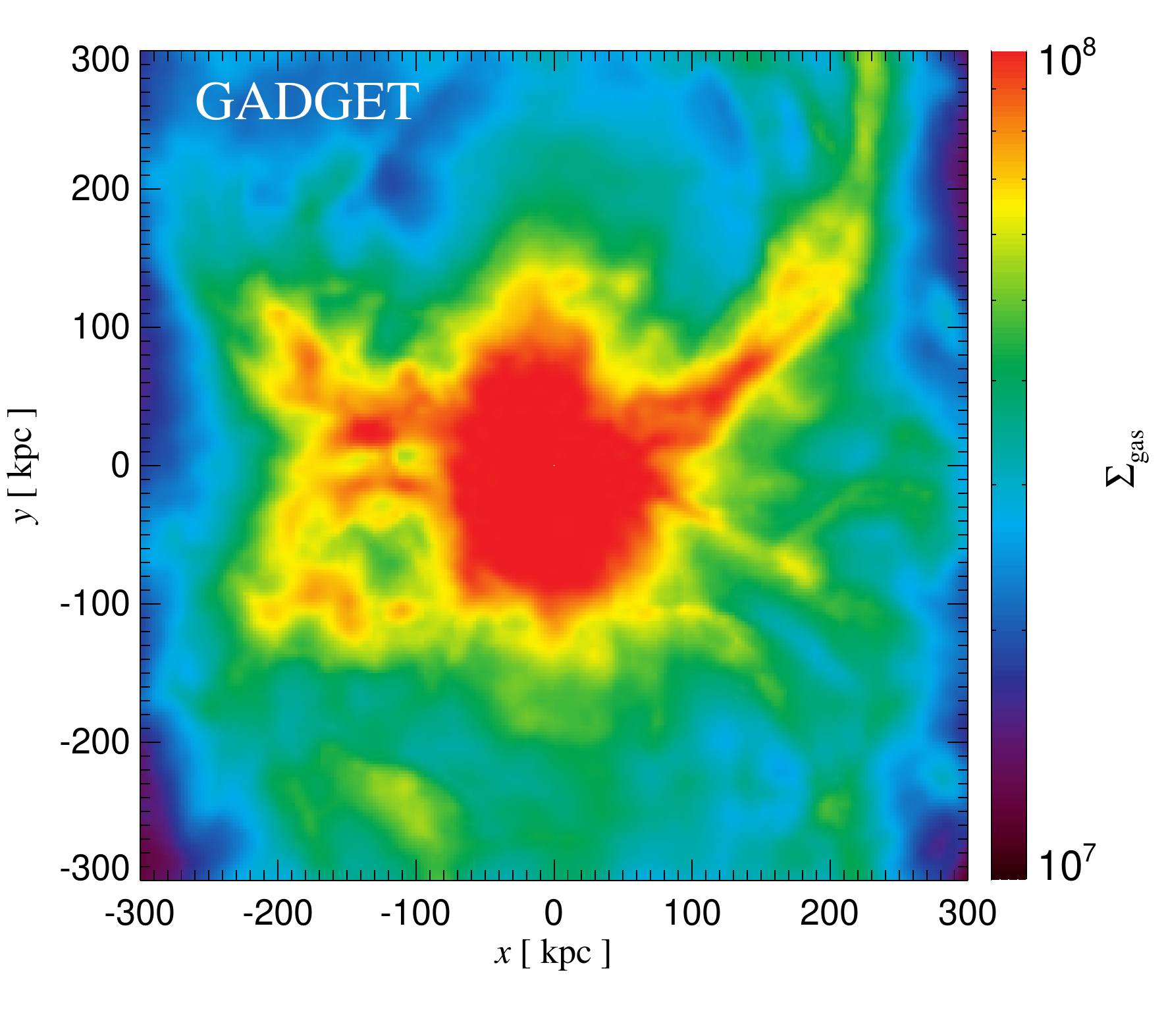}
\hspace{-0.5cm}
\includegraphics[width=4.8truecm]{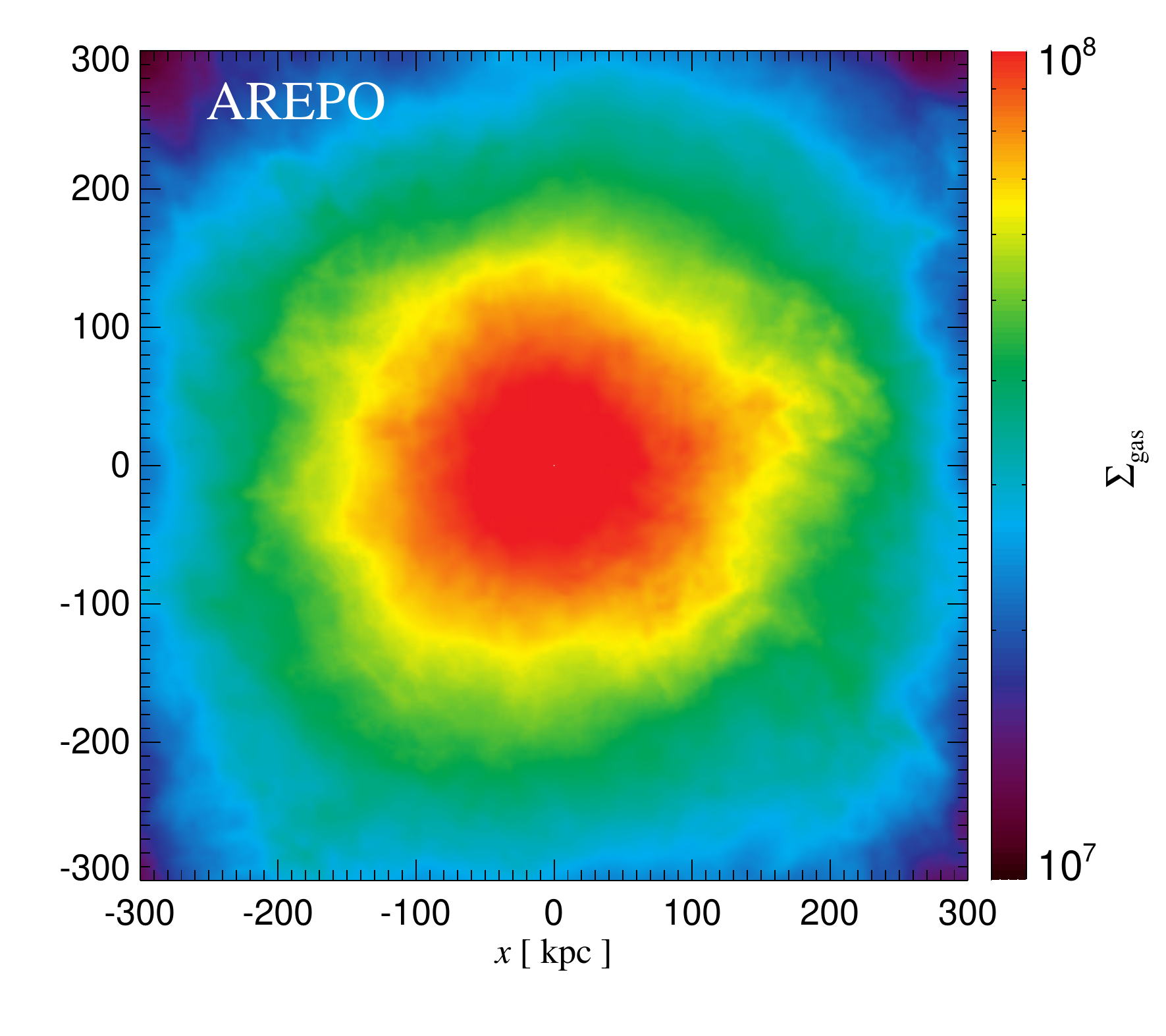}
\includegraphics[width=4.8truecm]{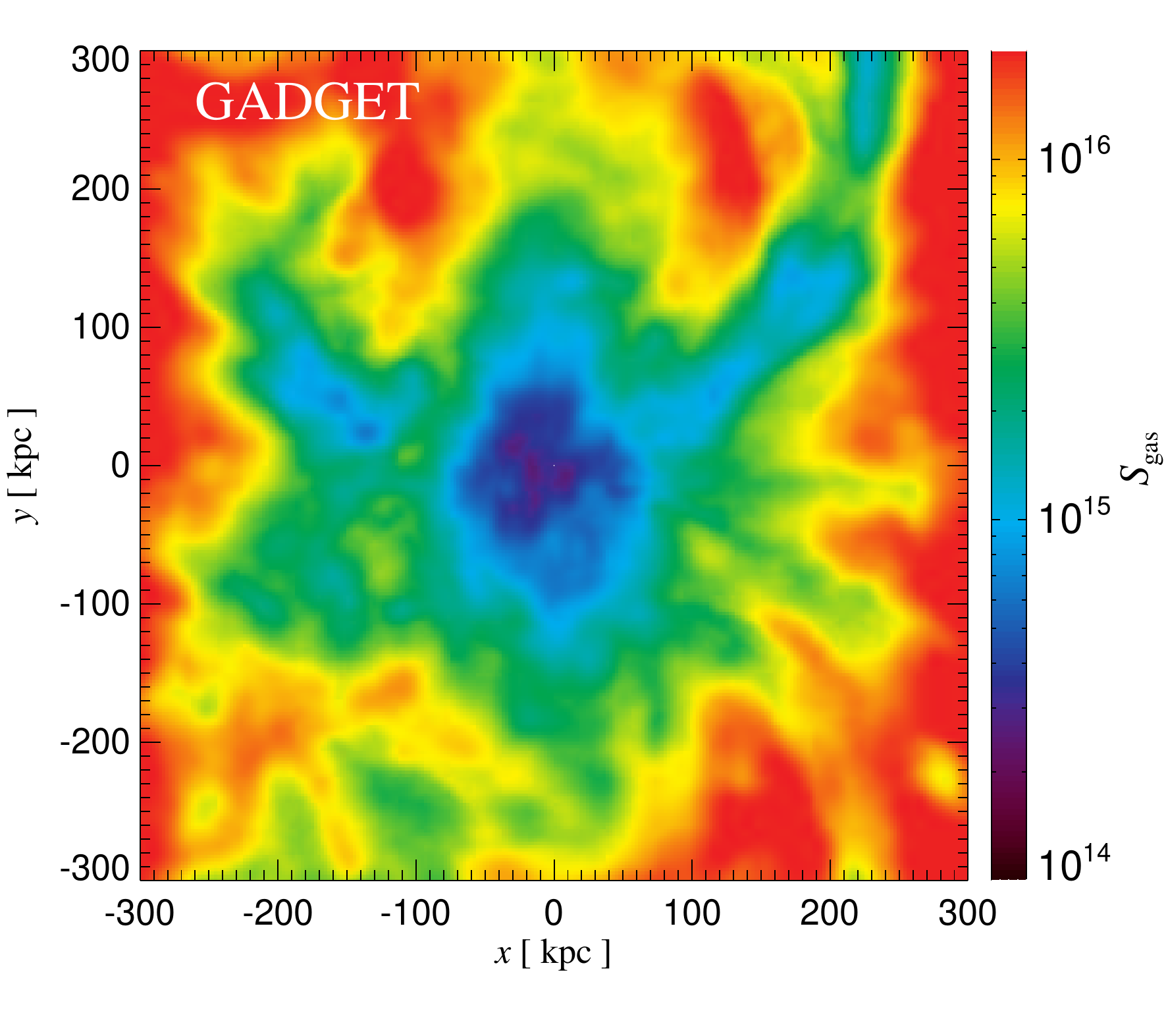}
\hspace{-0.5cm}
\includegraphics[width=4.8truecm]{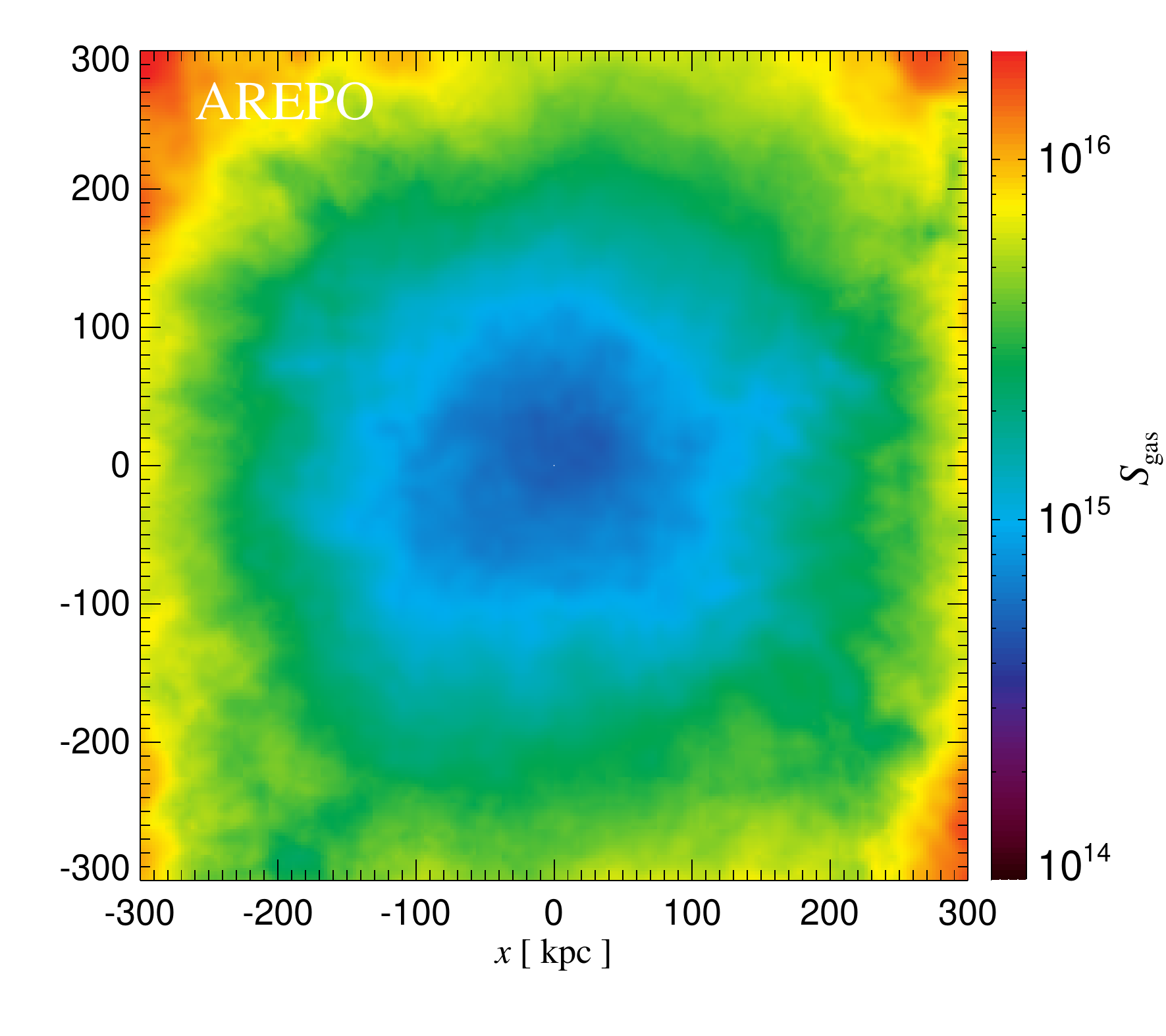}
}
\vspace{-0.25cm}
\hbox{
\includegraphics[width=4.8truecm]{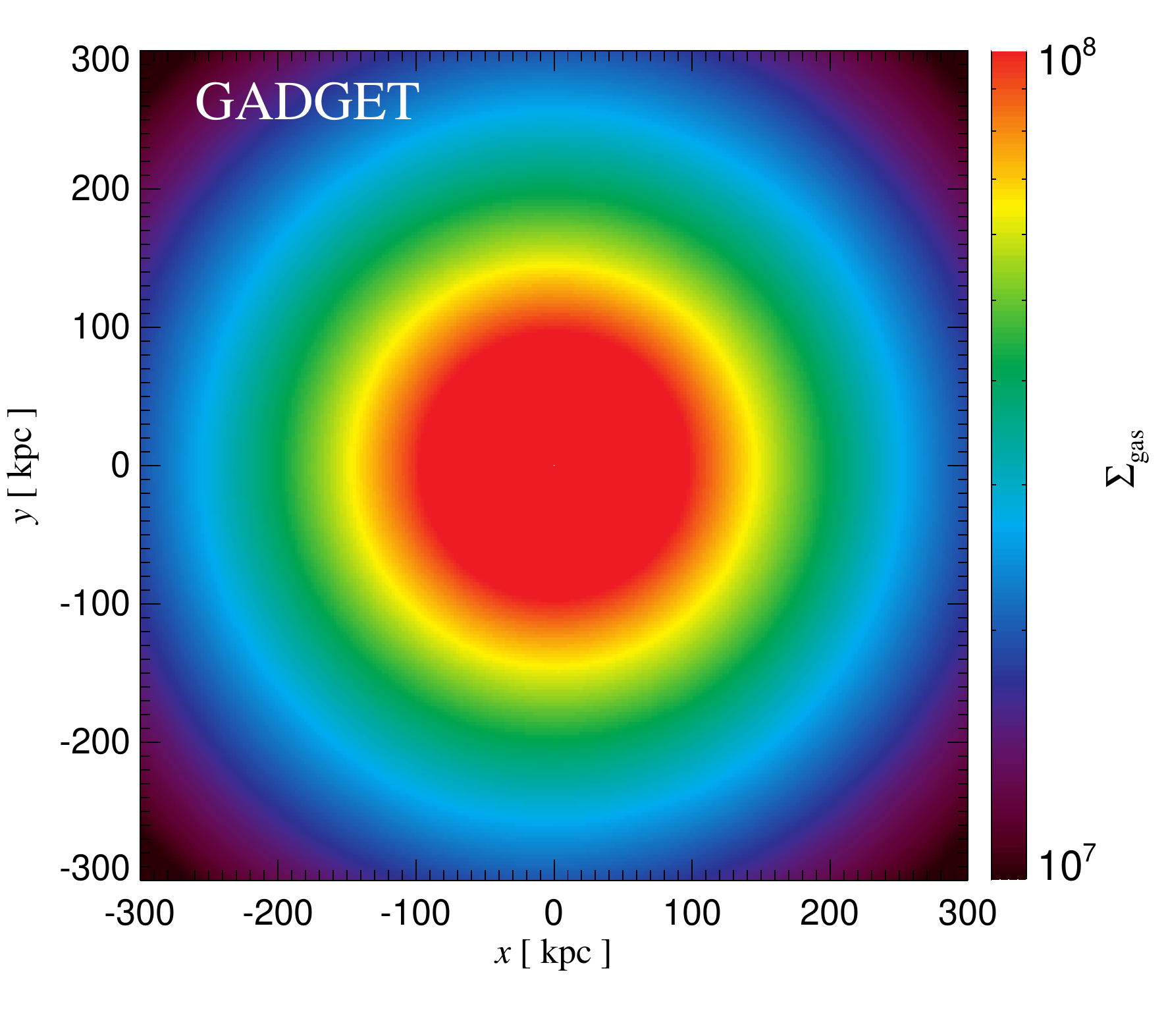}
\hspace{-0.5cm}
\includegraphics[width=4.8truecm]{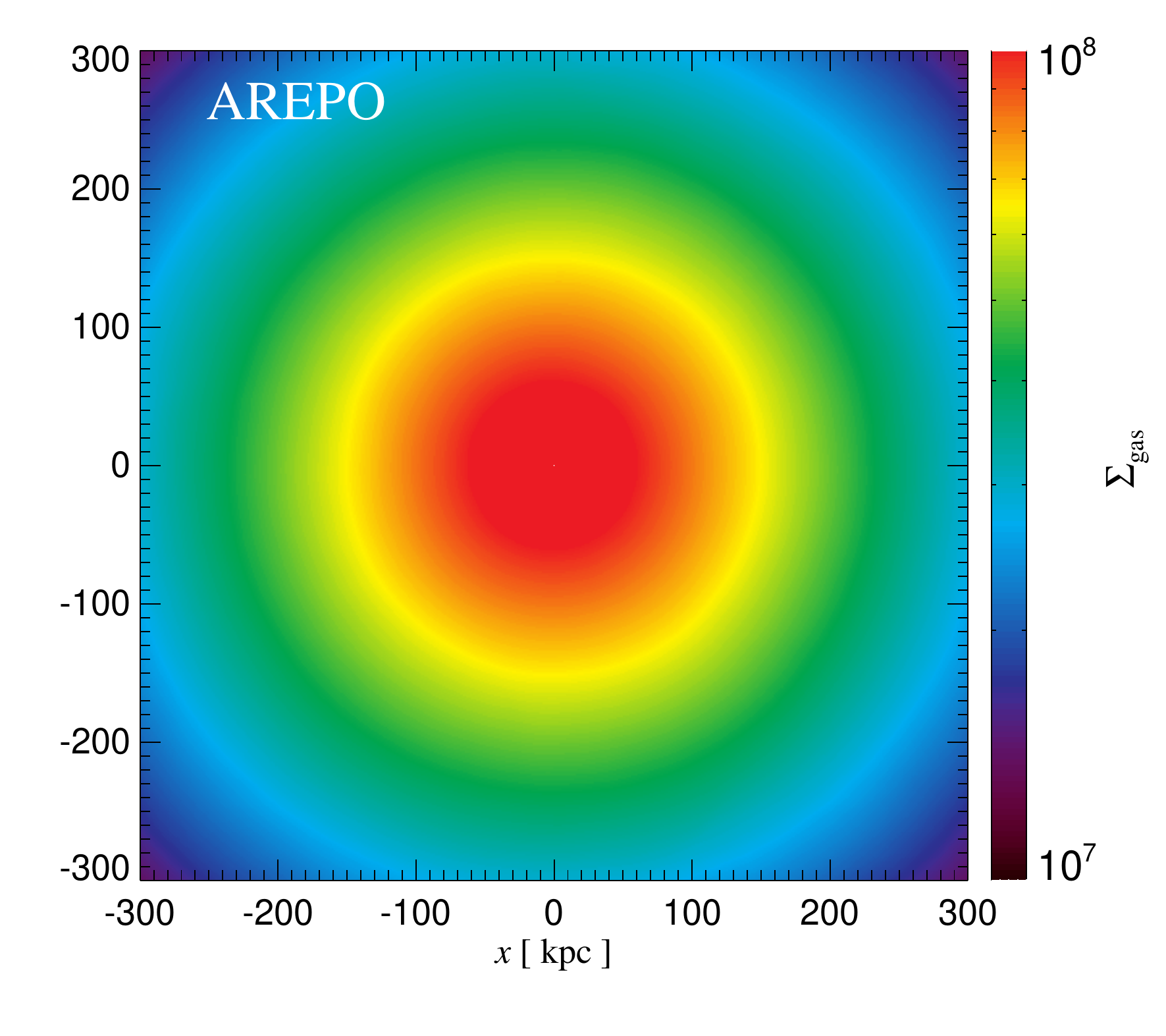}
\includegraphics[width=4.8truecm]{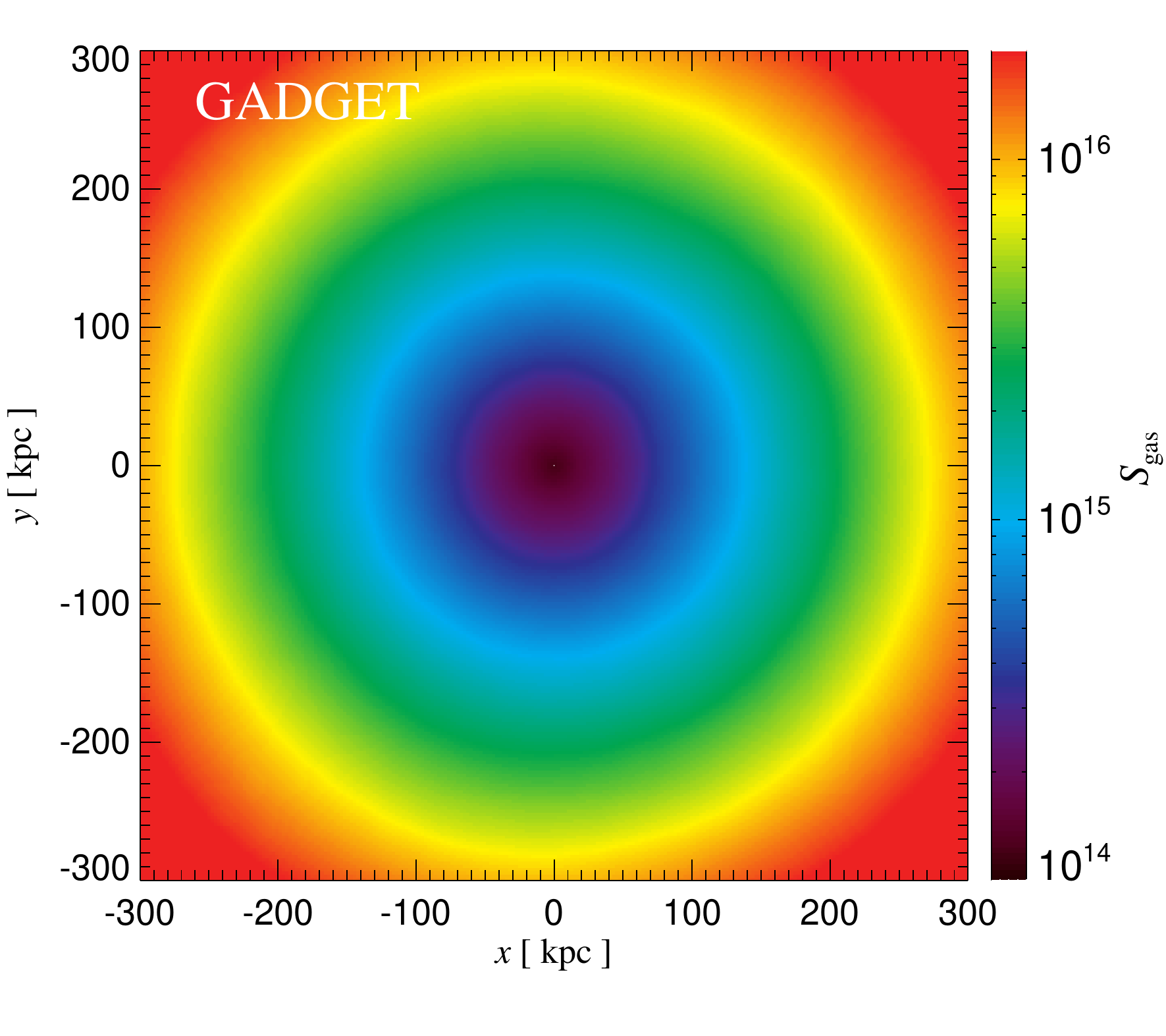}
\hspace{-0.5cm}
\includegraphics[width=4.8truecm]{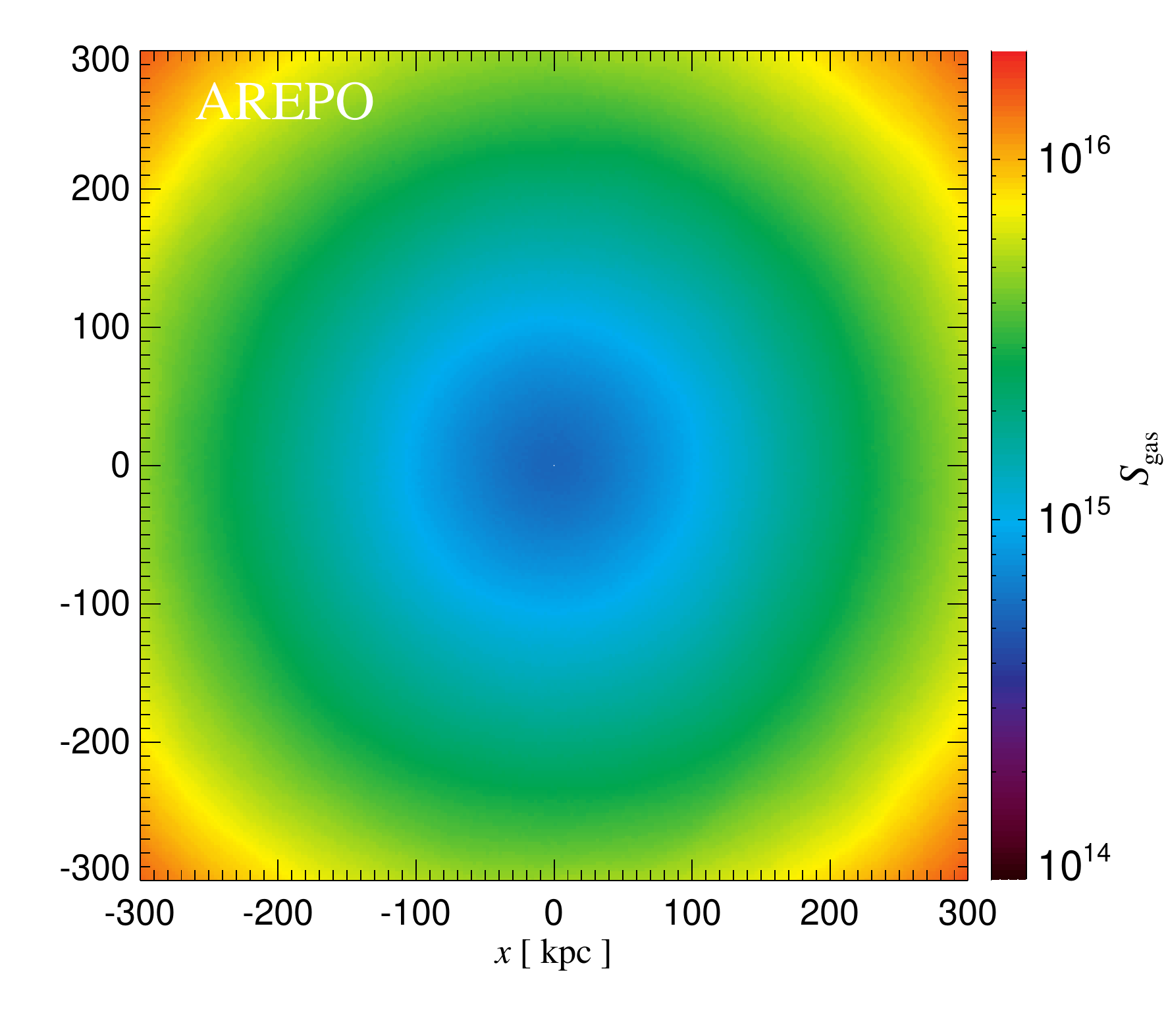}
}}}
\caption{Projected surface density maps (first two columns; in units of $[{\rm
      M_{\odot}}{\rm kpc}^{-2}]$) and mass-weighted entropy maps (last two
  columns; in internal units) for {\small GADGET} and {\small AREPO}
  simulations, showing the collision of two gaseous spheres with $N_{\rm gas}
  = 2 \times 10^6$ at times $t = 0.54\,{\rm Gyr}$ (first row), $t = 1.2\,{\rm
    Gyr}$ (second row), $t = 2.9\,{\rm Gyr}$ (third row), and $t = 15\,{\rm
    Gyr}$ (fourth row). The plotted spatial domain is $0.6 \times 0.6 \times
  1$Mpc.}
\label{FigCollision}
\end{figure*}

We now further increase the complexity of the problem by considering two
gaseous spheres instead of one, collapsing into one common static dark matter
halo placed in-between the two spheres. Each sphere has an
initial spatial displacement from the centre of the halo. This test is similar
in spirit to a number of previous works which analyzed collisions of two
galaxy clusters in isolation \citep[see e.g.][]{Ricker2001, Ritchie2002,
  McCarthy2007, Springel2007, Mitchell2009, ZuHone2011}, but here we devise a
cleaner set-up in order to minimize additional possible numerical effects
(e.g. gravitational N-body heating, differences due to gas self-gravity,
etc.). As in Section~\ref{ColdInflow}, we simulate a static analytic Hernquist
dark matter halo (with exactly the same parameters) but instead of one cold
gaseous sphere we generate two identical cold spheres separated by $1.2\, {\rm
  Mpc}$ along the $x$-axis, and we again neglect any radiative losses and gas
self-gravity. The gaseous spheres are constructed in the same way as in
Section~\ref{ColdInflow} i.e.~gas is first set up to be in hydrostatic
equilibrium within a static Hernquist dark matter potential, and then its
internal energy is reduced to $4.7 \times 10^3\,{\rm K}$. For each code we
perform runs with different gas particle numbers, i.e.~$N_{\rm gas} = 10^4$,
$N_{\rm gas} = 10^5$ and $N_{\rm gas} = 10^6$ per sphere.

Under the gravitational pull from the central dark matter potential the two
cold spheres collapse towards its centre and violently collide. The
interesting aspect of this problem is that radial symmetry is broken and the
gas interaction results in much more complicated shock geometries. This
  is illustrated in Figure~\ref{FigCollision}, where we show a time sequence
  of projected gas density maps (first two columns) and mass-weighted entropy
  maps (last two columns) for runs with $N_{\rm gas} = 2 \times 10^6$
  performed with the two codes. While the {\small GADGET} simulation shows
  qualitatively similar gas structures, the detailed properties substantially
  differ. 

Initially, as the spheres start to fall in, the gas is compressed in the
centre of the halo, generating a spherical overdensity, which is somewhat
broader and less peaked in {\small GADGET}, largely due to a poorer effective
spatial resolution and non-negligible artificial viscosity (see
Section~\ref{ColdInflow}). Also, during this initial stage more entropy is
produced in the central region in our SPH calculation. As more gas falls in, a
shock develops which rapidly assumes a cocoon-like geometry elongated
perpendicular to the direction of collapse (see top panels of
Figure~\ref{FigCollision}). From the gas density maps it can be seen that the
shock front is narrower and sharper in {\small AREPO}, whereas in {\small
  GADGET} it has a more splotchy-like appearance, caused by kernel
averaging. High entropy plumes propagating outwards along the $y$-axis are
clearly visible in the right-hand panels, where the central entropy in {\small
  GADGET} within $\sim 250\,{\rm kpc}$ is still higher than in the moving mesh
simulation. As the cocoon propagates against the infalling material, gas in
the very centre is pushed perpendicular to the $x$-axis, generating a dense
sheet-like region (see second row of Figure~\ref{FigCollision}). Dynamical
fluid instabilities at the boundary of this dense region induce typical
mushroom-like morphologies (cap-like in projection). However, even for our
highest resolution simulations with $N_{\rm gas} = 2 \times 10^6$ particles,
there are some marked differences between the two codes.  Mushroom-like
features (corresponding to the red-orange colours in the density maps)
originating at the very boundary of the dense central region are more coherent
in {\small GADGET}, while in {\small AREPO} they break up and mix more
efficiently with the surrounding medium. This also leads to the more efficient
mixing of different entropy gas in the very core in the moving mesh
simulation, as can be seen from the entropy maps.

\begin{figure}\centerline{
\includegraphics[width=9.truecm,height=8.truecm]{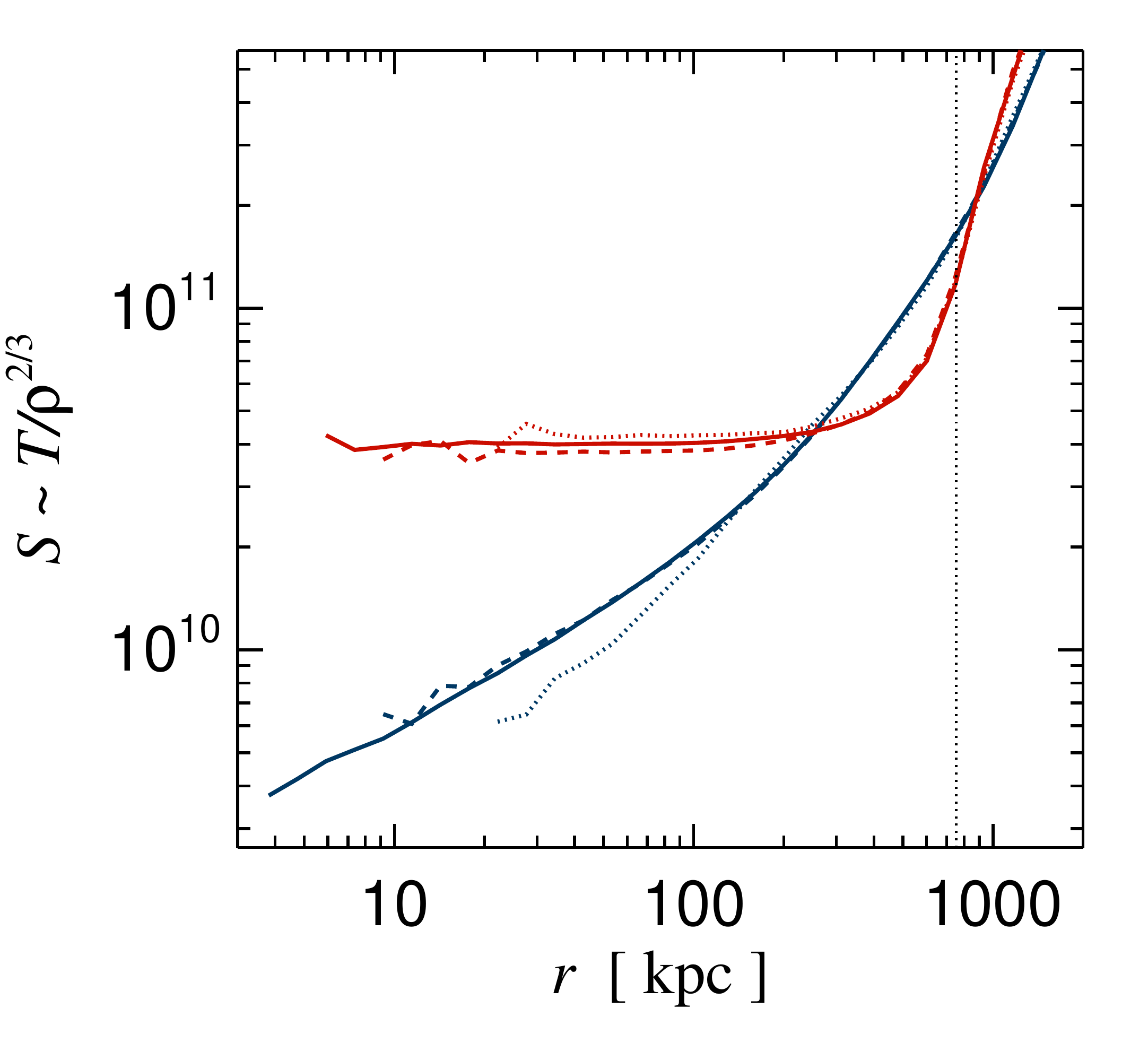}}
\caption{Radial entropy profiles at $t \sim 15\,{\rm Gyr}$ from the start of
  the simulation when the system has reached equilibrium. For each code
  ({\small GADGET}: blue lines; {\small AREPO}: red lines) three simulations
  with increasing gas particle number are shown: $N_{\rm gas} = 2 \times 10^4$
  (dotted lines), $N_{\rm gas} = 2 \times 10^5$ (dashed lines), $N_{\rm gas} =
  2 \times 10^6$ (continuous lines). The vertical dotted line indicates the
  virial radius of the underlying dark matter halo. While for $N_{\rm gas} \ge
  2 \times 10^5$ the entropy profiles seem converged for each code, they
  converge to a very different result.}
\label{FigCollision2}
\end{figure}

The differences in the fluid properties at the early stages of the simulated
system as described above are, however, not the only reason why the
thermodynamic properties of the gas are systematically discrepant between the
two codes when a new equilibrium state is reached. With time, dense shells of
gas completely disperse by mixing with the surrounding material in {\small
  AREPO}, while in the case of {\small GADGET} filaments and blobs of dense
gas survive and gradually sink back to the centre (see third row of
Figure~\ref{FigCollision}). This buoyantly-driven deposition of low entropy
material  in {\small GADGET} causes even larger differences between the final
entropy distributions, which are illustrated in Figure~\ref{FigCollision2}
(see also bottom row of Figure~\ref{FigCollision}). The radial entropy
profiles are computed  once the system has reached hydrostatic equilibrium at
time $t \sim 15\,{\rm Gyr}$ from the start of the simulation. The blue lines
denote {\small GADGET} results at three different resolutions (dotted lines:
$N_{\rm gas} = 2 \times 10^4$, dashed lines: $N_{\rm gas} = 2 \times 10^5$,
solid lines: $N_{\rm gas} = 2 \times 10^6$), while the red lines are for the
runs with {\small AREPO}. For $N_{\rm gas} \ge 2 \times 10^5$, both codes seem
to produce converged entropy profiles, but they converge to very different
results. While in {\small GADGET} the entropy profiles steadily decrease
towards the centre, in the moving mesh code a large entropy core is
produced. This systematic difference in the central entropy profiles is in
good agreement with the previous study by \citet{Mitchell2009} of idealized
major merger simulations of haloes in the non-radiative regime.

\subsubsection{Generalized blob test: non-radiative case}\label{GeneralBlob}

\begin{figure*}\centerline{\vbox{
\hbox{
\includegraphics[width=9.truecm,height=8truecm]{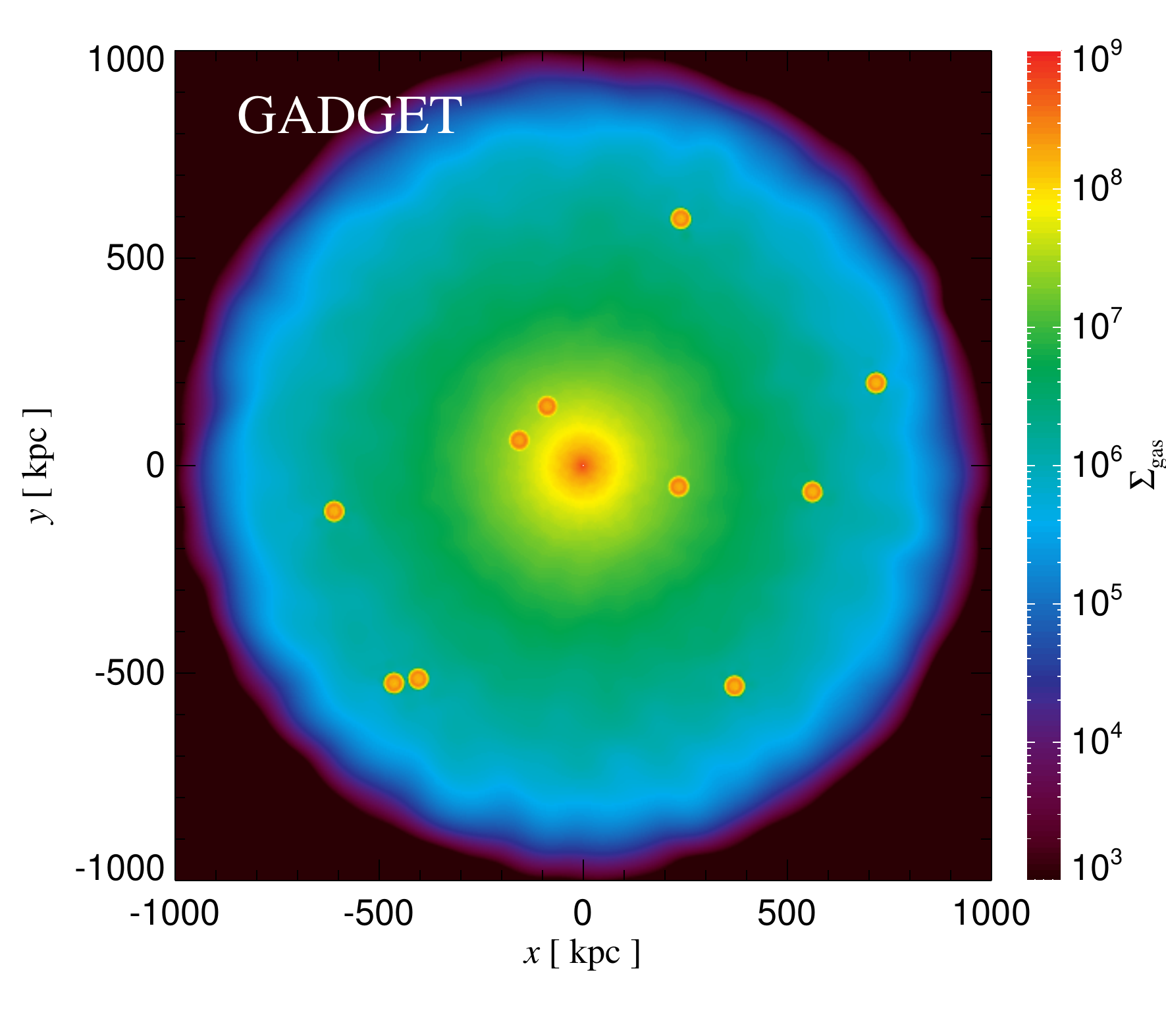}
\includegraphics[width=9.truecm,height=8truecm]{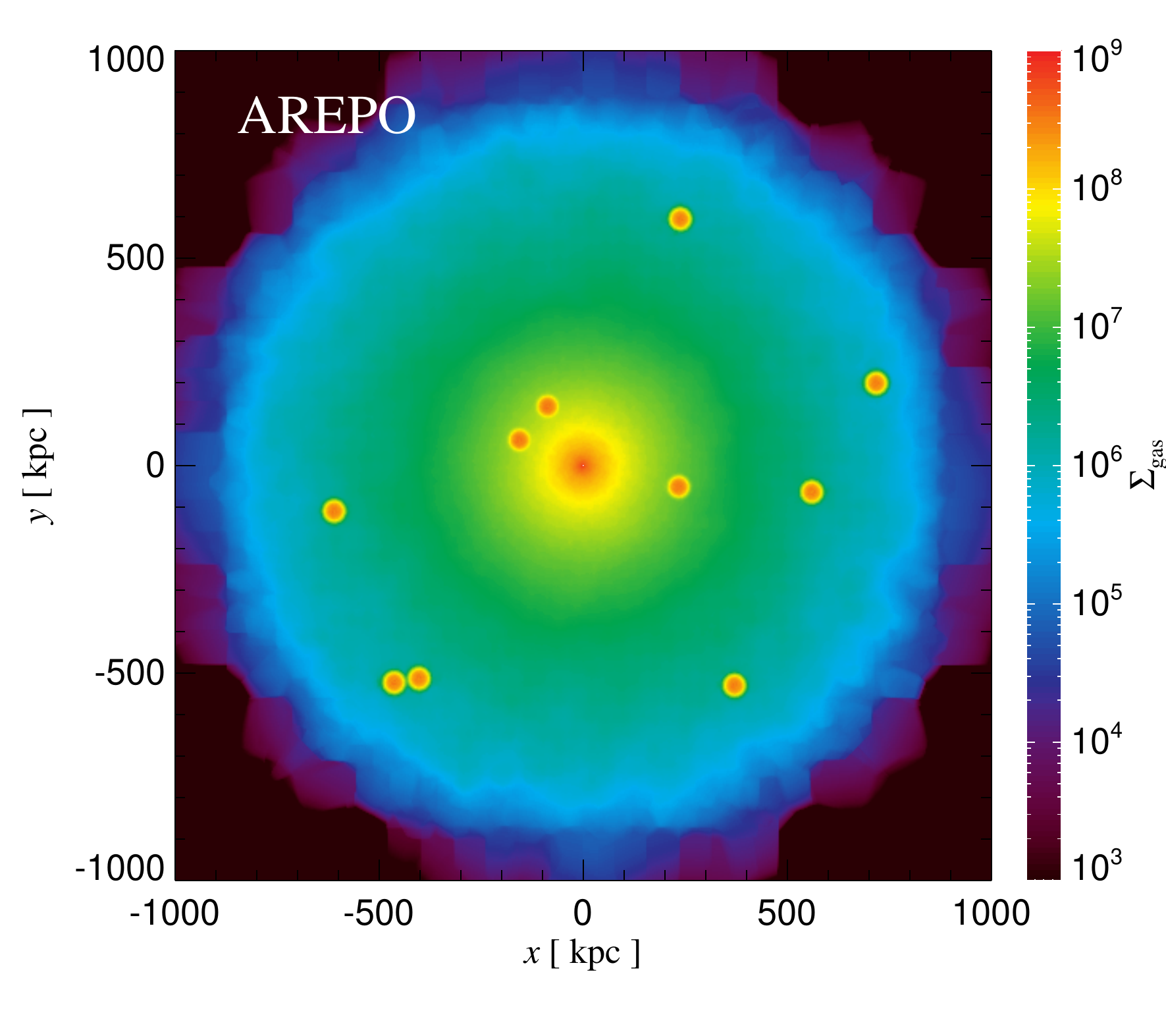}}
\vspace{-0.7cm}
\hbox{
\includegraphics[width=9.truecm,height=8truecm]{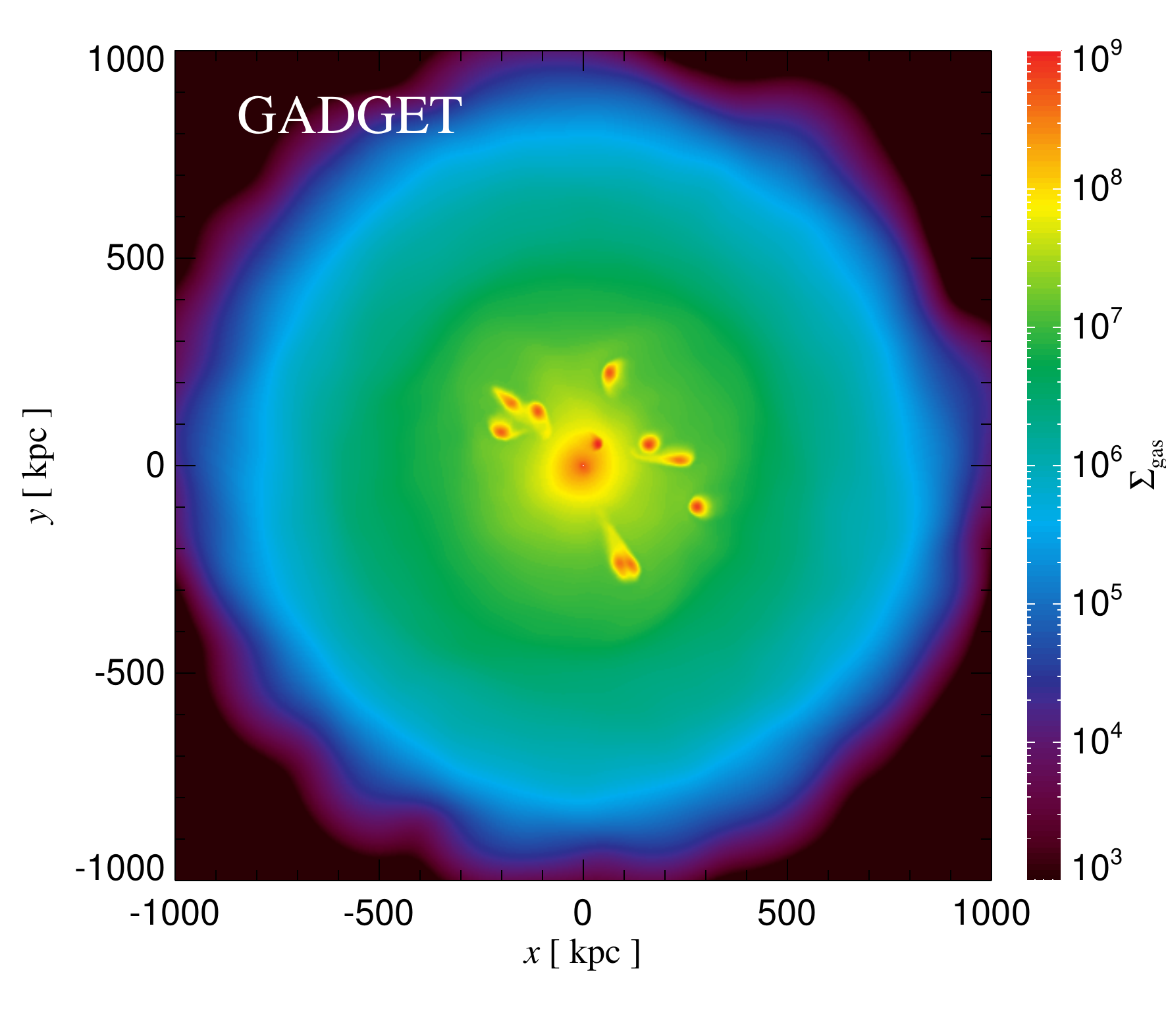}
\includegraphics[width=9.truecm,height=8truecm]{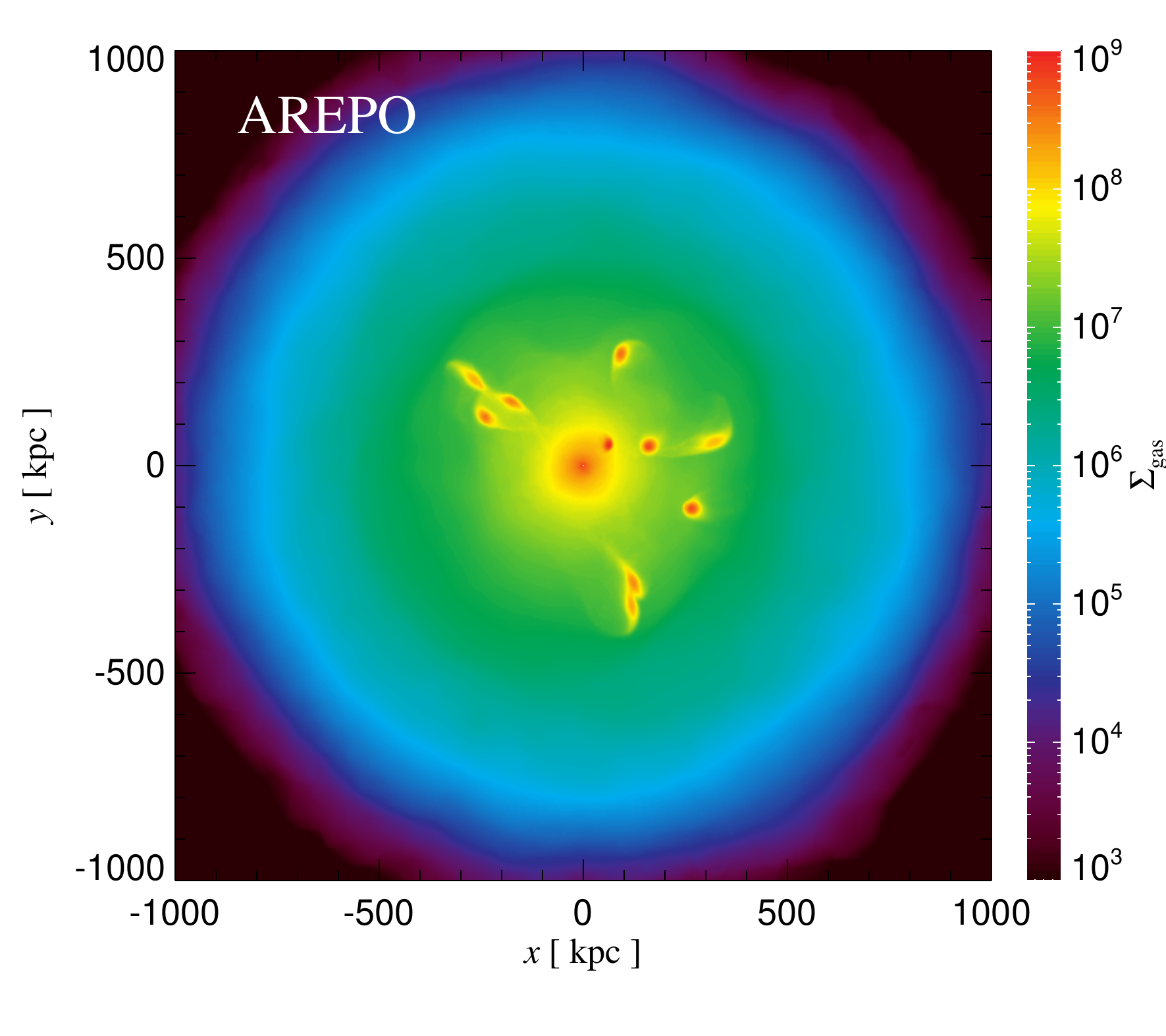}}
\vspace{-0.7cm}
\hbox{
\includegraphics[width=9.truecm,height=8truecm]{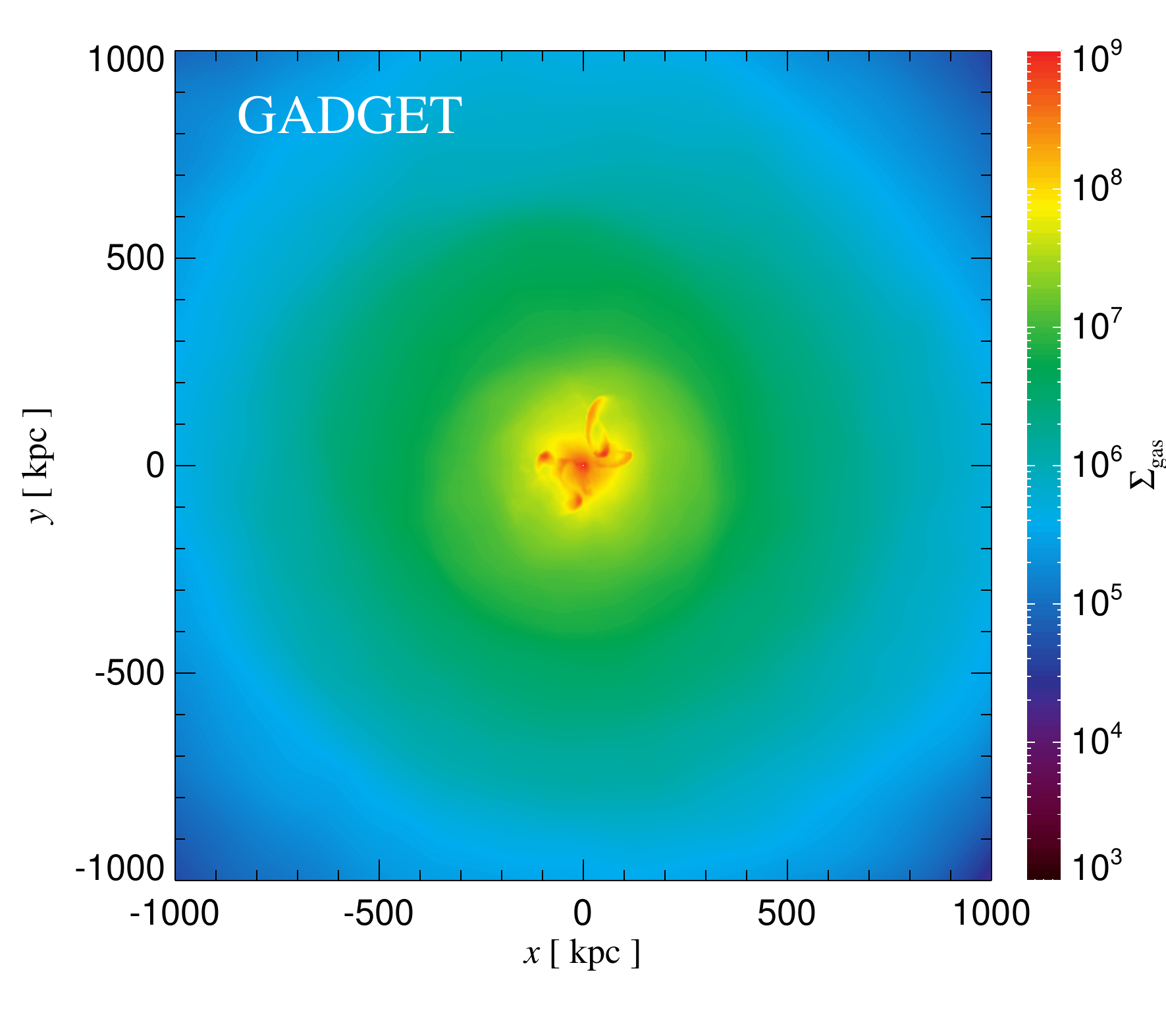}
\includegraphics[width=9.truecm,height=8truecm]{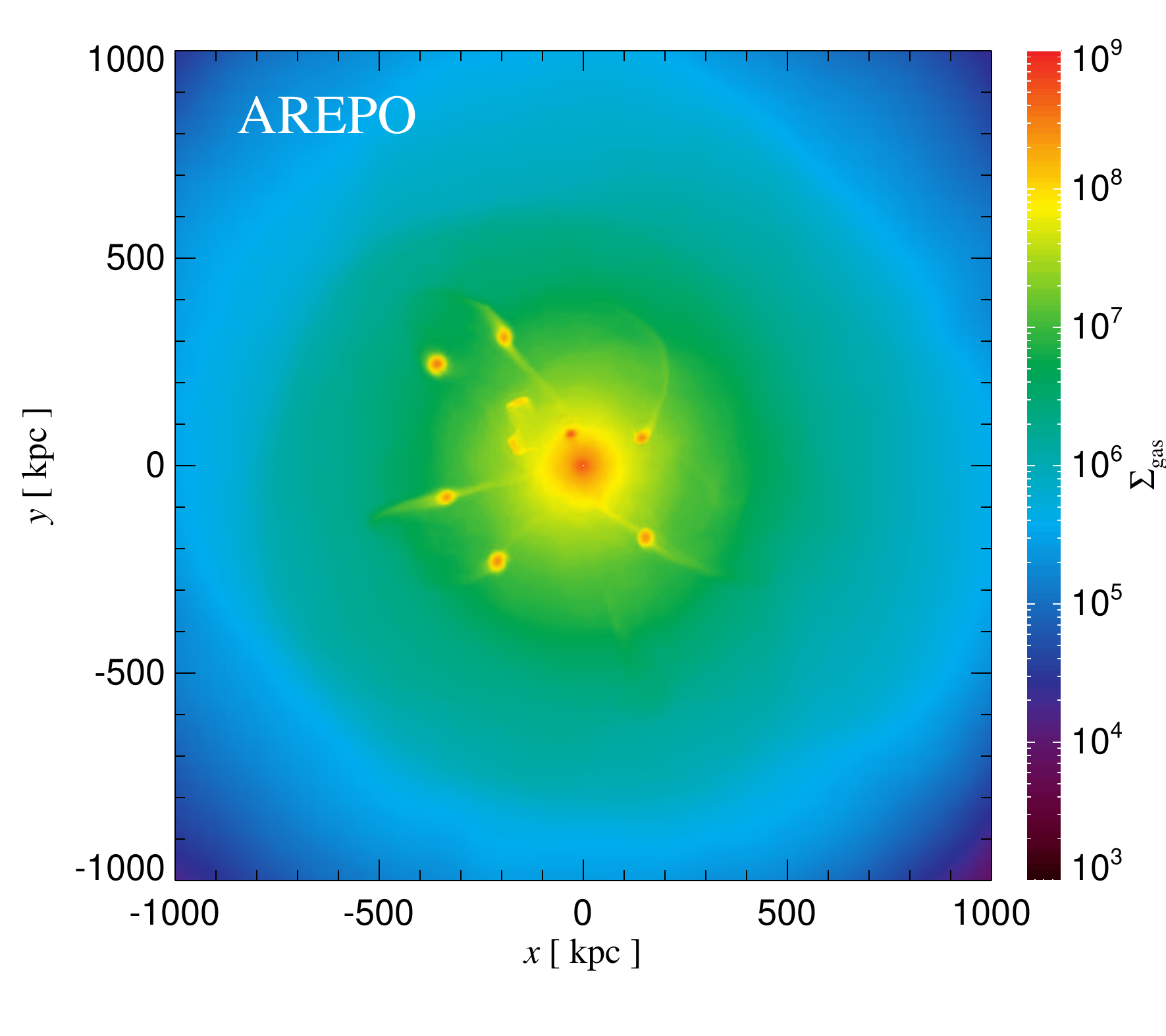}}}}
\vspace{-0.5cm}
\caption{Projected surface density maps in units of $[{\rm
    M_{\odot}}{\rm kpc}^{-2}]$ at the initial time (top panels), at $t
  = 1.37\,{\rm Gyr}$ (middle panels), and at $t = 2.65\,{\rm Gyr}$
  (bottom panels) for {\small GADGET} ($\alpha= 1.0$, $N_{\rm ngb} =
  64$) and {\small AREPO}. The plotted spatial domain is $2 \times 2
  \times 2\,{\rm Mpc}$, so as to encompass initially the whole halo
  whose virial radius is $R_{\rm 200} = 755\,{\rm kpc}$. While in the
  beginning the orbits and the morphologies of the blobs are very
  similar in the two codes, for $t \ge 1 \,{\rm Gyr}$ they begin to
  diverge, with blobs sinking to the centre faster in {\small
    GADGET}.}
\label{FigGenBlobMap}
\end{figure*}

\begin{figure*}\centerline{\hbox{
\includegraphics[width=8.5truecm,height=8.5truecm]{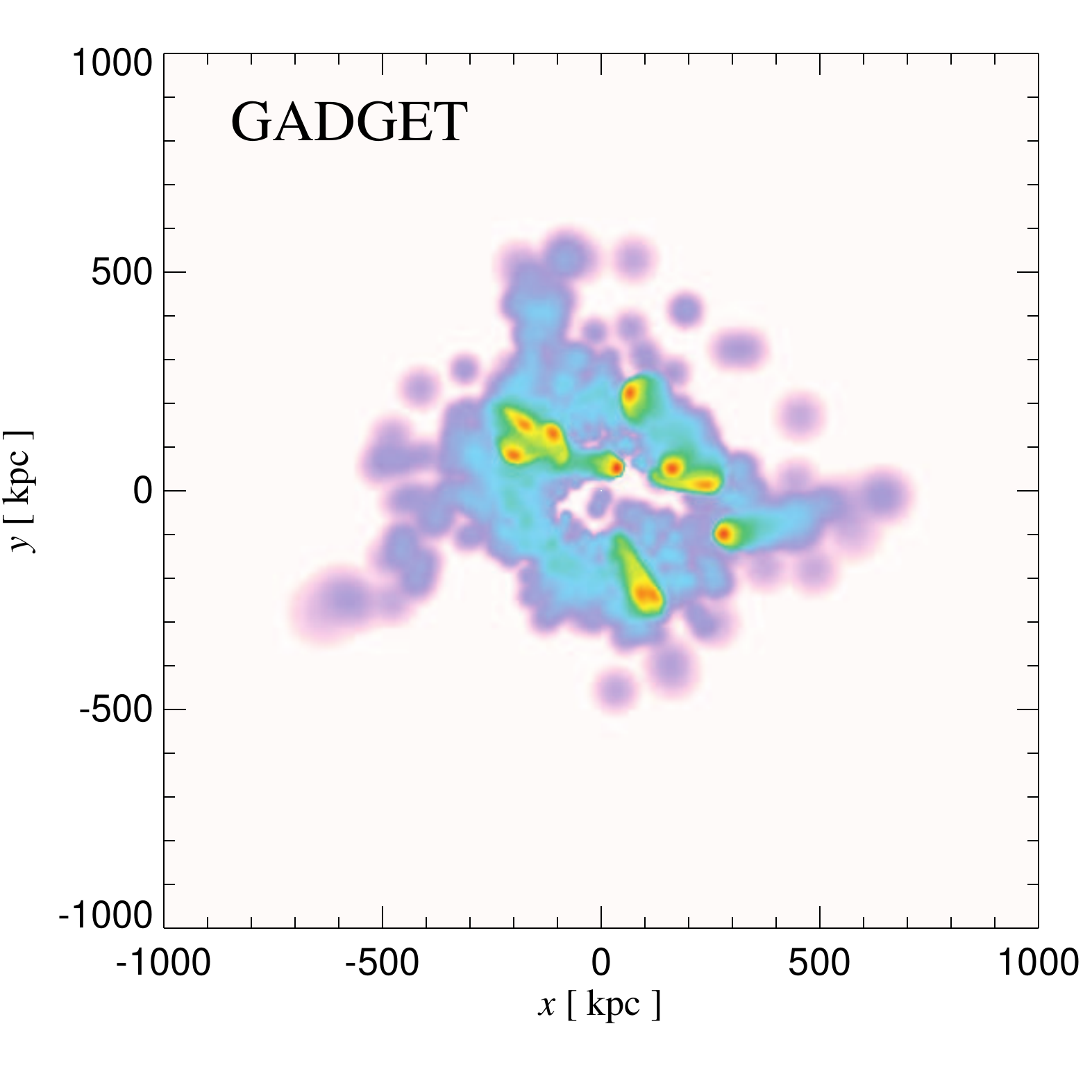}
\includegraphics[width=8.5truecm,height=8.5truecm]{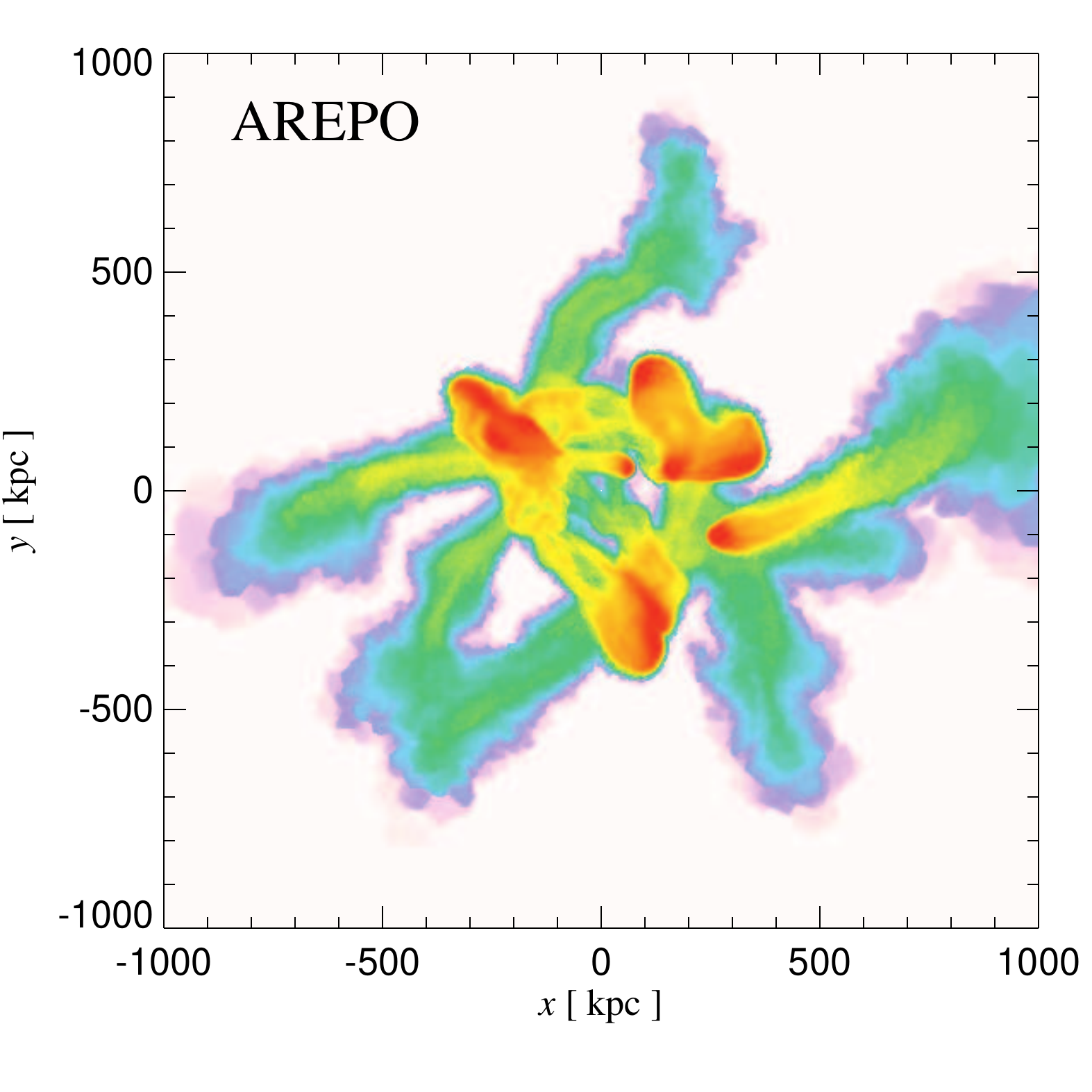}}}
\caption{Spatial distribution of gas at $t = 1.37\,{\rm Gyr}$ that was
  contained in the blobs at time $t = 0$. The plotted spatial domain
  is $2 \times 2 \times 2\,{\rm Mpc}$ (for details on map-making see
  the main text). Most of the material stays confined within blobs in
  the case of {\small GADGET}, with a relatively small fraction lost
  to their wake. The spatial distribution of blob material is markedly
  different in {\small AREPO}, showing much more stripped gas that
  creates prominent tails which extend up to several $100\,{\rm
    kpc}$.}
\label{FigGenBlobTracer}
\end{figure*}

\begin{figure*}\centerline{\hbox{
\includegraphics[width=9.5truecm,height=8.5truecm]{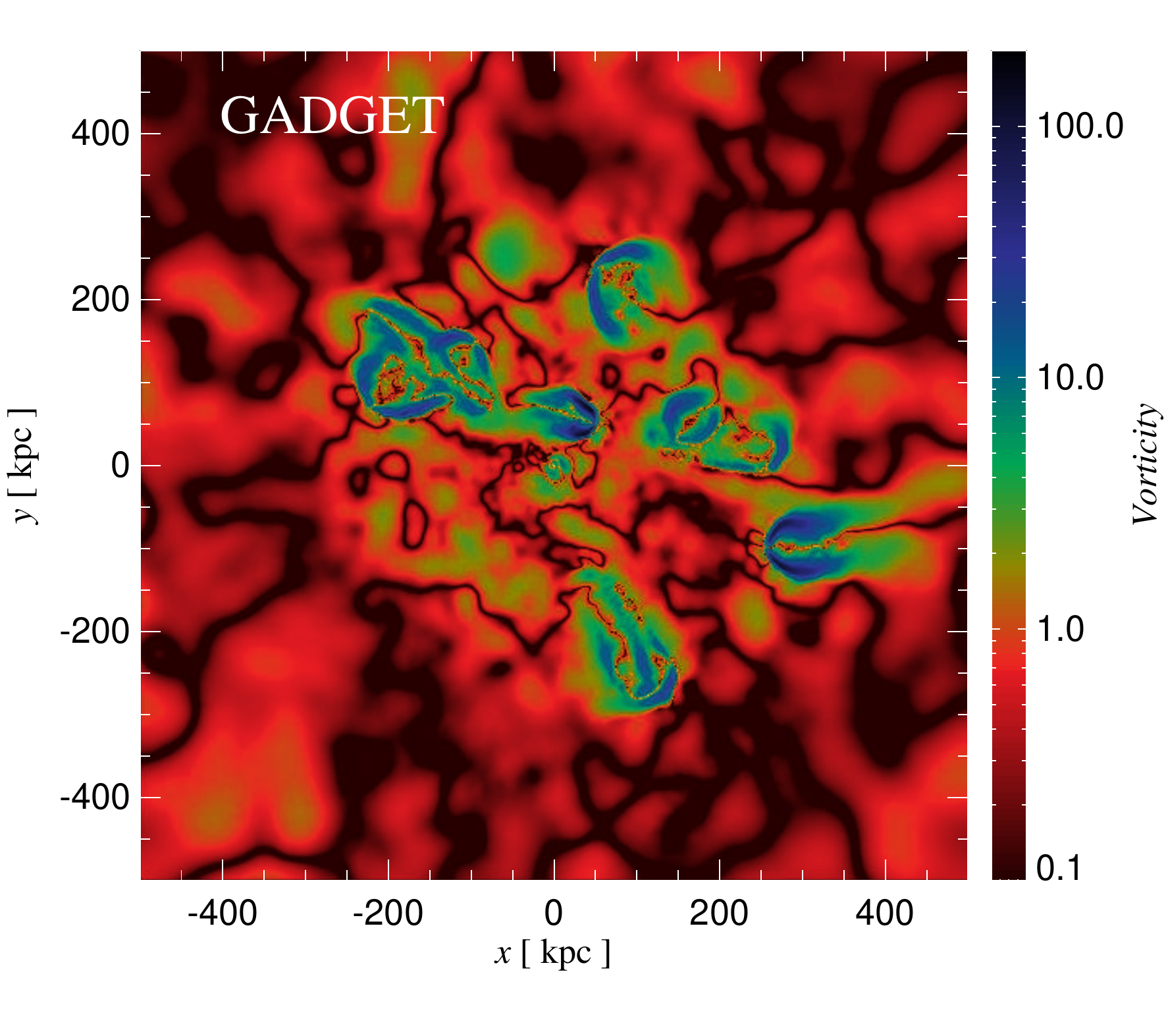}
\includegraphics[width=9.5truecm,height=8.5truecm]{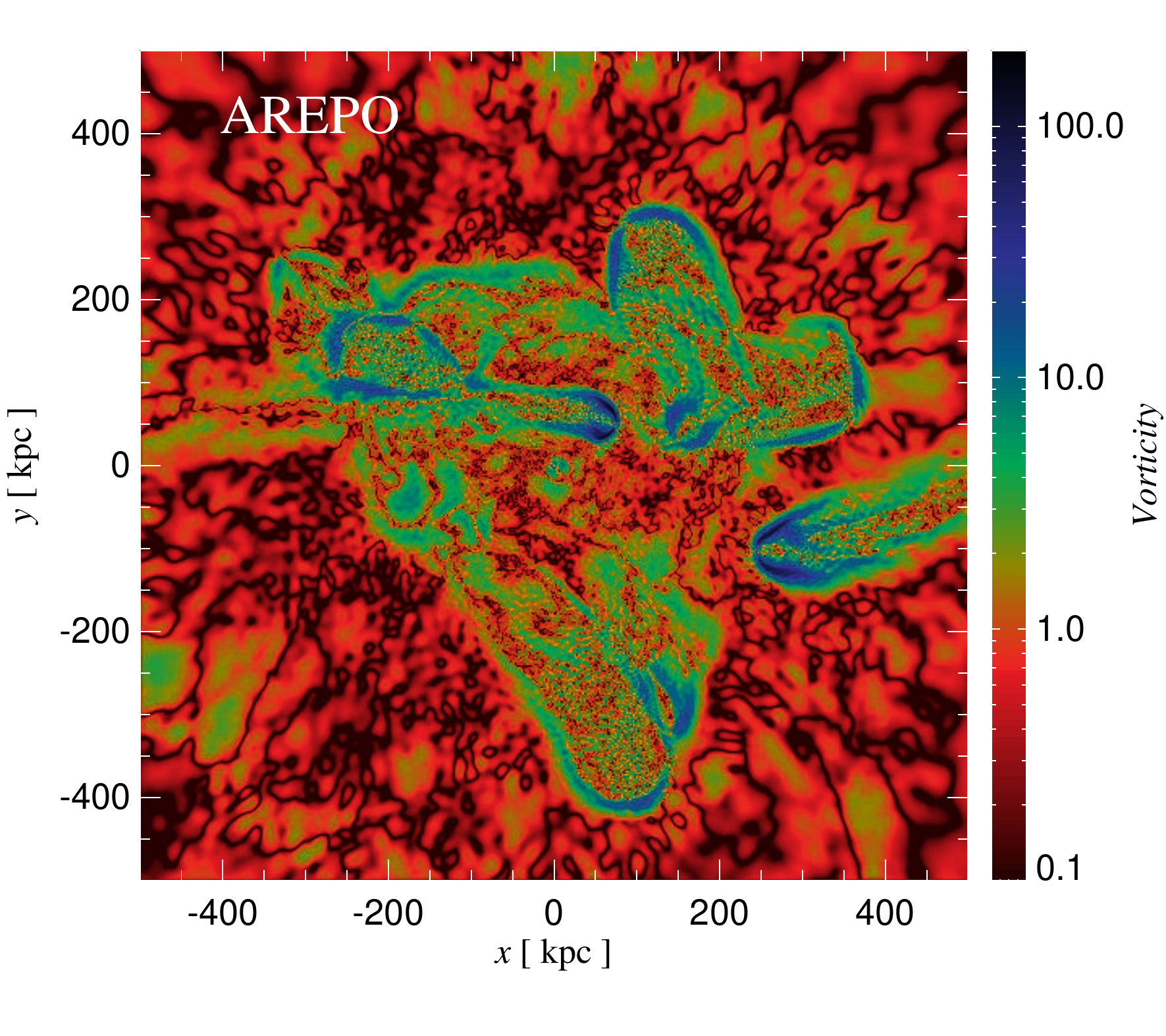}}}
\caption{Projected mass-weighted vorticity maps in units of $[{\rm
    km}\,{\rm s^{-1}}{\rm kpc}^{-1}]$ (absolute value of the
  $z$-component only) for a run with {\small GADGET} (left-hand panel)
  and with {\small AREPO} (right-hand panel). The maps are computed at
  time $t = 1.37\,{\rm Gyr}$ from the start of the simulation. The
  thickness of the projected region is $\Delta z = 1$Mpc. In the wake
  of the bow shocks produced by the moving blobs turbulence is
  generated. The spatial extent of the turbulent wakes is
  significantly larger in {\small AREPO}.}
\label{FigGenBlobVort}
\end{figure*}

We now devise a numerical experiment to capture the evolution of cold,
dense blobs in a more realistic setting, rather than a uniform density
windtunnel (see Section~\ref{Blob}). For this purpose we consider our
default isolated halo with a static, analytic Hernquist profile for
the dark matter component, gas in hydrostatic equilibrium, and we
include gas self-gravity, but neglect any radiative losses. We
additionally populate this halo with $10$ blobs, with the following
properties: the gas pressure within blobs is set to be $0.01$ of the
maximum intracluster pressure within $800\,{\rm kpc}$ radius from the
centre; the radius of each blob is $R_{\rm blob} = 20\,{\rm kpc}$; the
total mass of all blobs is $20\%$ of the total intracluster gas mass,
i.e.~$M_{\rm blobs} = 0.2 f_{\rm gas} M_{\rm halo}$; the adiabatic index
is the same for the blobs and for the surrounding gas, $\gamma =
5/3$. 

We place the blobs randomly within a spherical shell with
cluster-centric distance of $700\,{\rm kpc}$ and thickness of
$150\,{\rm kpc}$, and particles belonging to the blobs are drawn
randomly as well. Apart from positions, we also assign bulk velocities
to the blobs, while the reminder of the intracluster gas is initially
at rest. We set all three components of the velocity vector for each
blob, i.e. $v_{\rm r}$ (only inward radial velocity), $v_{\rm
  \theta}$, and $v_{\rm \phi}$, assuming a random distribution for
each velocity component starting from a characteristic velocity value
of $200\,{\rm km\,s^{-1}}$. Blob velocity values range then from $\sim
230\,{\rm km\,s^{-1}}$ to $\sim 510\,{\rm km\,s^{-1}}$, with the
average velocity of all blobs being $v_{\rm mean} \sim 400\,{\rm
  km\,s^{-1}}$. In this way, individual blobs will not reach the
cluster centre all at the same time, giving them more realistic orbits
than purely radial ones. Initially, the blobs are roughly in 
pressure equilibrium with the surrounding gaseous halo. We perform
this numerical experiment at two different resolutions: $N_{\rm gas} =
10^4$, $N_{\rm blob} = 10^3$ (low resolution simulation), and $N_{\rm gas}
= 10^5$, $N_{\rm blob} = 10^4$ (higher resolution simulation).

In Figure~\ref{FigGenBlobMap}, we show the time evolution of the dense blobs
moving through the isolated halo, for the higher resolution run. In the top
panels, the projected surface density map is plotted at the initial time,
where both the intracluster gas structure and the blob properties are exactly
the same in the two codes. Initially, for $t \le 1\,{\rm Gyr}$, the blobs are
moving on almost identical orbits in {\small GADGET} and {\small AREPO}, and
they also have very similar morphologies. As the blobs start approaching the
inner cluster region, well-defined bow shocks develop ahead of each blob
  and ram-pressure stripping ablates the blobs. Additionally, the
  Kelvin-Helmholtz and Rayleigh-Taylor instabilities arise which tend to
  disrupt the blobs on a characteristic timescale of several Gyrs, as
  discussed in Section~\ref{Blob} (but note that here gas is
  self-gravitating). The lower rate of ram pressure stripping and suppression
  of fluid instabilities in {\small GADGET} has a significant effect not only
  on blob morphologies, but also on their orbits. At $t = 1.37\,{\rm Gyr}$,
as illustrated in the middle panels, it is still possible to cross-identify
each blob in {\small GADGET} with the respective blob in the {\small AREPO}
run, but the blobs in {\small GADGET} have lost less material and thus appear
denser and some of them are closer to the centre. 

This different loss of blob material leads to systematically diverging orbits
due to higher buoyancy and dynamical friction forces acting on each blob in
{\small GADGET}. This can be clearly seen in the bottom panels where the blobs
in {\small GADGET} are markedly more concentrated and have essentially all
fallen to the innermost cluster region, while in {\small AREPO} they are at
much larger cluster-centric distances being gradually eroded. Moreover,
  given that the blobs in the moving mesh calculation are ablated more
  efficiently and that therefore the dynamical friction exerted on the blobs
  is lower, they have higher velocities when passing at the pericentre and
  thus can reach larger distances after the first passage, as visible in the
  bottom panels of Figure~\ref{FigGenBlobMap}. As the gas is self-gravitating,
  the blobs are also subject to tidal stripping when passing close to the
  innermost regions which contributes to the ablation of the blob material.

To illustrate more clearly the differences in mass loss of the blobs
in {\small GADGET} and {\small AREPO}, in
Figure~\ref{FigGenBlobTracer} we show projected maps of gas material
at $t = 1.37\,{\rm Gyr}$ (corresponding to the middle panels of
Figure~\ref{FigGenBlobMap}) which initially belonged to the blobs. To
compute these maps in the case of {\small GADGET}, we show integrated
mass along the line-of-sight ($\Delta z = 2 \,{\rm Mpc}$) for all
particles initially contained in the blobs. In the case of {\small
  AREPO}, we use a tracer field to follow spreading of fluid elements
initially within the blobs: each cell is characterized by an
additional scalar field {\it Tracer} which at $t=0$ is equal to $1$
for all blob cells and $0$ elsewhere. The tracer field essentially
evolves as a dye cast on the moving fluid, and at some time $t > 0$
the {\it Tracer} value indicates the mass fraction of the material
which initially was in the blobs for each cell. In the right-hand
panel of Figure~\ref{FigGenBlobTracer}, we plot the projected
density-weighted tracer field. The dynamical range is the same in both
panels, ranging from the maximum of the projected quantity to
$10^{-7}$ of this maximum value, using a logarithmic colour
mapping. As anticipated, cold dense blobs in {\small GADGET} lose much
less material while in the run with {\small AREPO} the lost blob
material is significantly more diffuse. In the wake of the infalling
blobs prominent tails develop, extending up to several $100\,{\rm
  kpc}$. The mass deposition of blob material is clearly much more
spatially extended in the moving-mesh code and occurs over larger
cluster-centric distances, allowing fluids with different entropies to
intermingle and to affect host halo properties on larger scales.

The formation of bow shocks in front of the moving blobs also implies that
vorticity will be generated due to the baroclinic term. To explore this issue
in more detail, in Figure~\ref{FigGenBlobVort} we show vorticity maps for the
run with {\small GADGET} (left-hand panel) and for the simulation with {\small
  AREPO} (right-hand panel) at $t = 1.37\,{\rm Gyr}$ (corresponding to the
middle panels of Figure~\ref{FigGenBlobMap}). The vorticity maps are
constructed by first evaluating projected mass-weighted velocity maps for the
$x$ and $y$-components, and then by taking the absolute value of the
$z$-component of the curl. Note that while the positions and the
  structure of the bow shocks are rather similar in the two codes, as expected,
 there is a striking difference between the spatial extent of the high
vorticity regions produced in the wake of the bow shocks (denoted with
  green colours), where turbulent motions should be generated. For example,
  focusing on the right-most blob centred on $x\sim 250\,{\rm kpc}$ and $y \sim
  -100\, {\rm kpc}$ in projection, the mean projected vorticity value in its wake is
 a factor of $\sim 2$ higher in the moving mesh calculation. This clearly
suggests that the suppression of vorticity generation in {\small GADGET} will
have an impact on the level of turbulence injected by curved shocks that are
associated with galaxy or subhalo motions through the intracluster medium
(ICM), producing a bias in the amount of non-thermal pressure support in
galaxy clusters \citep[see also e.g.][]{Dolag2005, Vazza2011b, Iapichino2011}.

\begin{figure}\centerline{
\includegraphics[width=9.truecm,height=8.truecm]{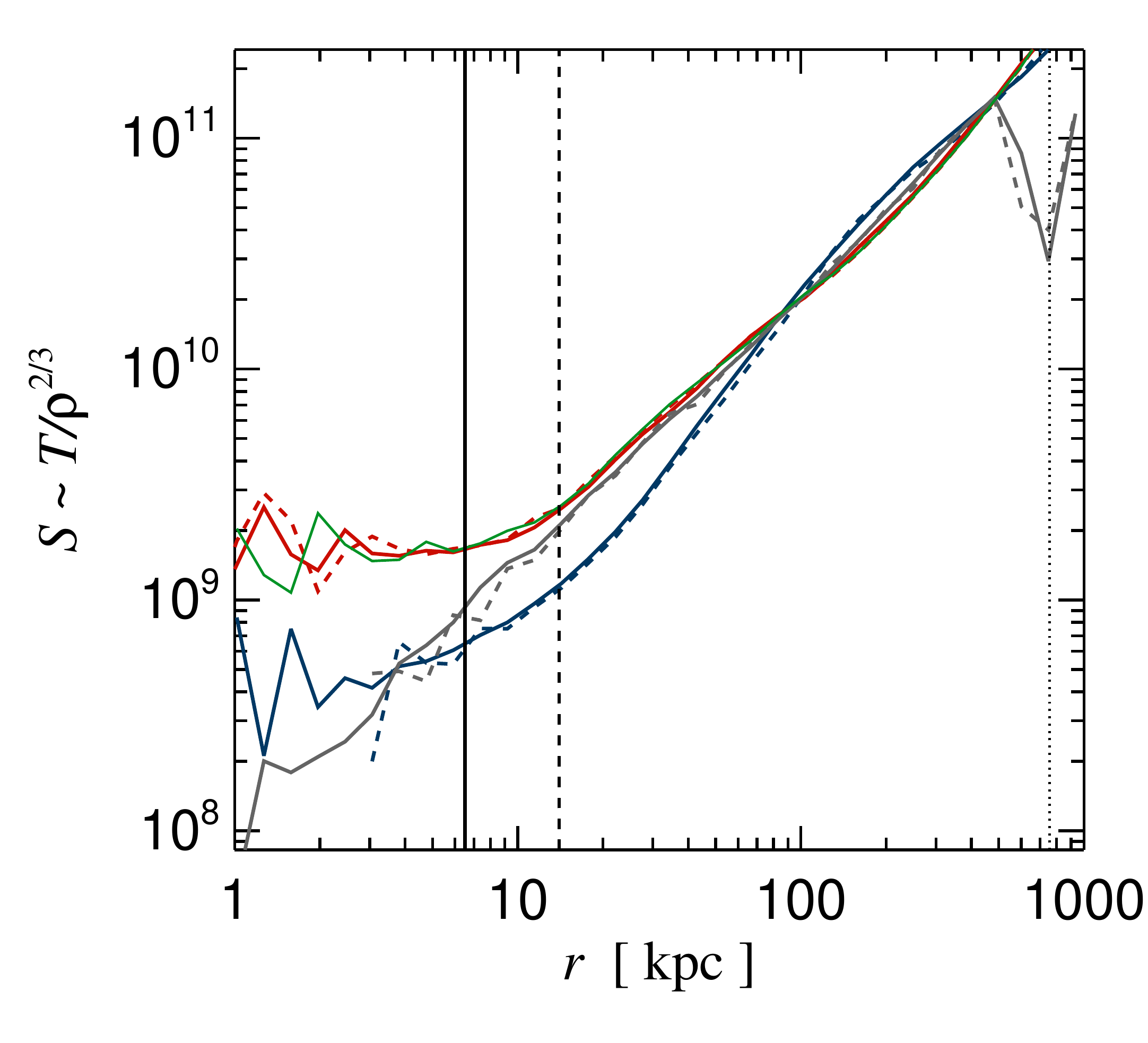}}
\caption{Radial entropy profiles at time $t \sim 10\,{\rm Gyr}$ after the
  start of the simulation, at which point all the blobs have been disrupted
  and the system has reached a new equilibrium. For each code ({\small
    GADGET}: blue lines; {\small AREPO}: red lines) two resolution simulations
  are shown: $N_{\rm gas} = 10^4$ and $N_{\rm blob} = 10^3$ (dashed lines),
  $N_{\rm gas} = 10^5$ and $N_{\rm blob} = 10^4$ (continuous lines). Grey
  curves with the same line-styles represent the initial entropy profiles of
  these simulations. Additional green line shows AREPO simulation with
    $N_{\rm gas} = 10^4$ and $N_{\rm blob} = 10^3$, but with the
    fixed gravitational softening as in {\small GADGET}. {\small
      AREPO} runs with adaptive and fixed softenings for the gas produce very
    similar entropy profiles. Vertical lines with the same line-styles denote
  the gravitational softenings and vertical dotted line the virial radius of
  the system. The interaction of the moving dense blobs with the intracluster
  medium leaves systematically different imprints on the gas entropy profiles
  when simulated with {\small AREPO} or with {\small GADGET}.}
\label{FigGenBlobProfile}
\end{figure}

As the blobs orbit at larger cluster-centric distances in {\small
  AREPO} for a longer time and mix with the surrounding medium more
efficiently than is the case for {\small GADGET}, they lead to higher
entropy generation over a wider range of radii. This is shown in
Figure~\ref{FigGenBlobProfile}, where we compute radial entropy
profiles for all intracluster gas including the blobs, at the final
time $t \sim 10\,{\rm Gyr}$ when the system has reached hydrostatic
equilibrium. Both in the low and high resolution runs with {\small
  AREPO} the gas entropy profile is significantly higher in the inner
regions up to $r \sim 100\,{\rm kpc}$. Instead, in {\small GADGET},
the entropy content of the dense blobs is increased less as they sink
towards the centre, such that they settle on a lower adiabat,
corresponding to smaller radii. We also note that these systematic
  differences in entropy profiles are entirely due to the different hydro
  solvers employed by {\small GADGET} and {\small AREPO}, and not due to the
  different choice of gravitational softenings for gas (fixed in {\small
    GADGET} and adaptive in {\small AREPO}, but with a floor set equal to the
  {\small GADGET} value). In fact, in Figure~\ref{FigGenBlobProfile}, the green
  line shows the entropy profile obtained from an identical low resolution
  {\small AREPO} simulation where instead of an adaptive a constant
  gravitational softenings is used for the gas, in the same way as in {\small GADGET}.

It is also instructive to analyze how the entropy profiles of the
intracluster medium evolve with time. In
Figure~\ref{FigGenBlobProfile} we show radial entropy profiles at the
initial time (grey lines) for both resolutions. The initial entropy
profile of the run with a higher particle number extends further
inwards due to the better spatial resolution, while some differences
in the outer regions are due to different positions of the blobs which
are drawn randomly. The kink in the initial entropy profile for $r \ge
500\,{\rm kpc}$ is caused by the blobs which populate this region and
have low entropy content with respect to the surrounding ICM. During
the first $0.5\,{\rm Gyr}$, the entropy profiles in {\small GADGET} and
{\small AREPO} do not change noticeably, remaining nearly identical to
each other and to the values prescribed by the initial conditions. At
$1\,{\rm Gyr}$, the gas entropy starts rising in the very centre in the run
with {\small AREPO} (it is higher by a factor of $\le 3$ with respect
to the initial value), while it takes almost twice as much time in the
run with {\small GADGET} before the central entropy rises by the same
amount. At that time, however, the central entropy profile in {\small
  AREPO} is already about a factor of $6$ higher than it was initially, and this
systematic difference in the central entropy values persists
until the final simulated time. 

At time $t \sim 2\,{\rm Gyr}$ another interesting process occurs: as
several blobs reach the cluster core region for the first time,
dissipation of their kinetic energy leads to a gradual expansion of
the whole gaseous halo (compare middle to bottom panels of
Figure~\ref{FigGenBlobMap}) which needs to readjust itself to find a new
equilibrium solution within the static dark matter potential.

Interestingly, in a recent paper by \citet{Vazza2011} non-radiative
cosmological simulations performed with the {\small GADGET} and {\small ENZO}
codes show that infalling satellites in {\small GADGET} sink to smaller radii
and are characterized by lower entropy content than is the case for {\small
  ENZO}, confirming the importance of the physical processes discussed here in
the full cosmological setting. This indicates that at least part of the
  systematic difference in central entropy values found between SPH and grid
  codes in the Santa Barbara cluster comparison project \citep{Frenk1999} is
  due to the different treatment of stripping and mixing of the cold material
  from the infalling satellites.

\subsection{Radiative gas cooling in isolated haloes} \label{RadHaloes}

\subsubsection{Radiative gas cooling in non-rotating haloes}\label{IsolatedHaloCool} 

We now investigate the time evolution of an isolated $10^{14}\, {\rm M_{\odot}}$ mass
halo subject to radiative cooling and star formation. We first simulate a halo
without any net rotation and compare gas and stellar properties of this system
between the two codes. The results illustrated here are for haloes with static
dark matter potentials, but we find very similar results if we consider live
haloes instead.

\begin{figure}\centerline{
\includegraphics[width=8.5truecm,height=8.2truecm]{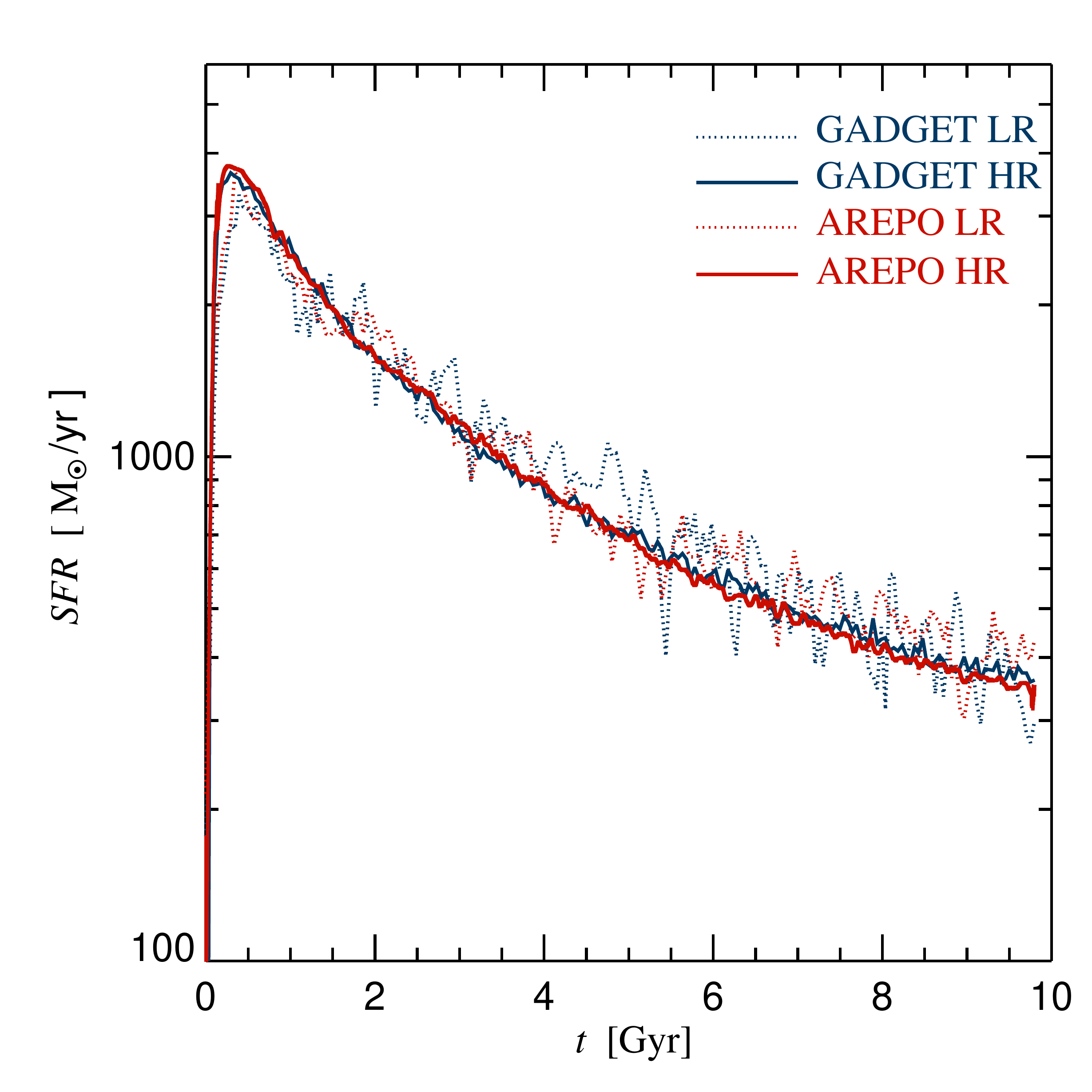}}
\caption{Time evolution of the star formation rate of a $M_{\rm vir} = 10^{14}\,
  {\rm M_{\odot}}$ isolated halo which radiatively cools and has negligible
  spin. Illustrated are low ($N_{\rm gas} = 10^4$; dotted lines) and high
  ($N_{\rm gas} = 10^6$; continuous lines) resolution simulations with {\small GADGET}
  (blue) and {\small AREPO} (red). The agreement in star formation rates is very good
  over the whole simulated time-span.}
\label{FigSFRCooling}
\end{figure}

In Figure~\ref{FigSFRCooling} we show star formation rates as a
function of time for our simulated object consisting of $N_{\rm gas} =
10^4$ or $N_{\rm gas} = 10^6$ particles/cells, calculated either with
{\small GADGET} or {\small AREPO}. Over the whole simulated time-span
of $10\,{\rm Gyr}$ the star formation rates are very similar in the
two codes. This indicates that not only do the implemented gas cooling
rates match very well, but also that the sub-grid model for star
formation and supernova feedback is consistent between the codes, even
at a relatively low numerical resolution \citep[a similar conclusion
is reached for mergers of isolated galaxies by][]{Torrey2011}.

\begin{figure}\centerline{
\includegraphics[width=8.5truecm,height=8.2truecm]{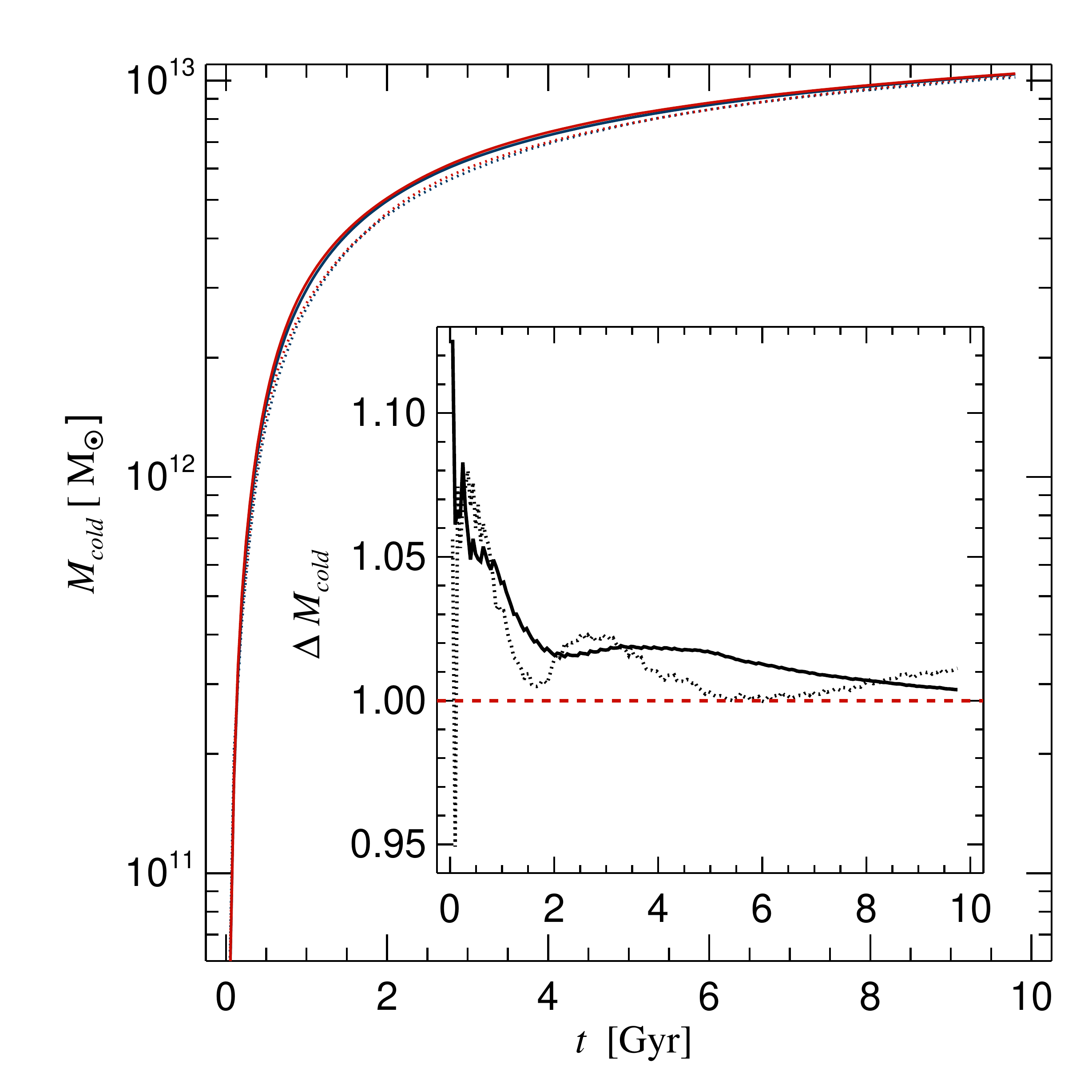}}
\caption{Time evolution of the cold baryonic mass (stars and gas with entropy $<
  10^5$ in internal units) for a $M_{\rm vir} = 10^{14}\, {\rm M_{\odot}}$
  isolated halo which radiatively cools and has negligible spin. Illustrated
  are low ($N_{\rm gas} = 10^4$; dotted lines) and high ($N_{\rm gas} = 10^6$;
  continuous lines) resolution simulations with {\small GADGET} (blue) and {\small AREPO}
  (red). In the inset, the ratio of $M_{\rm cold}$ values found in {\small AREPO}
  and {\small GADGET} is shown for both numerical resolutions. The agreement of
  the amount of cold baryons formed is extremely good.}
\label{FigMcoldNoRot}
\end{figure}

To demonstrate more clearly that gas radiative cooling and star
formation proceed in a very similar manner in our isolated haloes in
the absence of any net rotation, in Figure~\ref{FigMcoldNoRot} we plot
the total mass of cold baryons (stars and gas with entropy $< 10^5$ in
internal units) as a function of time for {\small GADGET} (blue lines)
and {\small AREPO} (red lines), at two different resolutions,
i.e.~$N_{\rm part} = 10^4$ (dotted lines) and $N_{\rm part} =
10^6$. For the same number of SPH particles as cells used in {\small
  AREPO} the amount of cold baryons matches to within a few percent
between the two codes, as is evident from the inset in the plot where
we show the ratio of $M_{\rm cold}$ values found with our moving-mesh
code and with {\small GADGET}. In the case of live haloes we find that
the total mass of cold baryons exhibits the same level of agreement.
 
\begin{figure}\centerline{
\includegraphics[width=8.5truecm,height=8.2truecm]{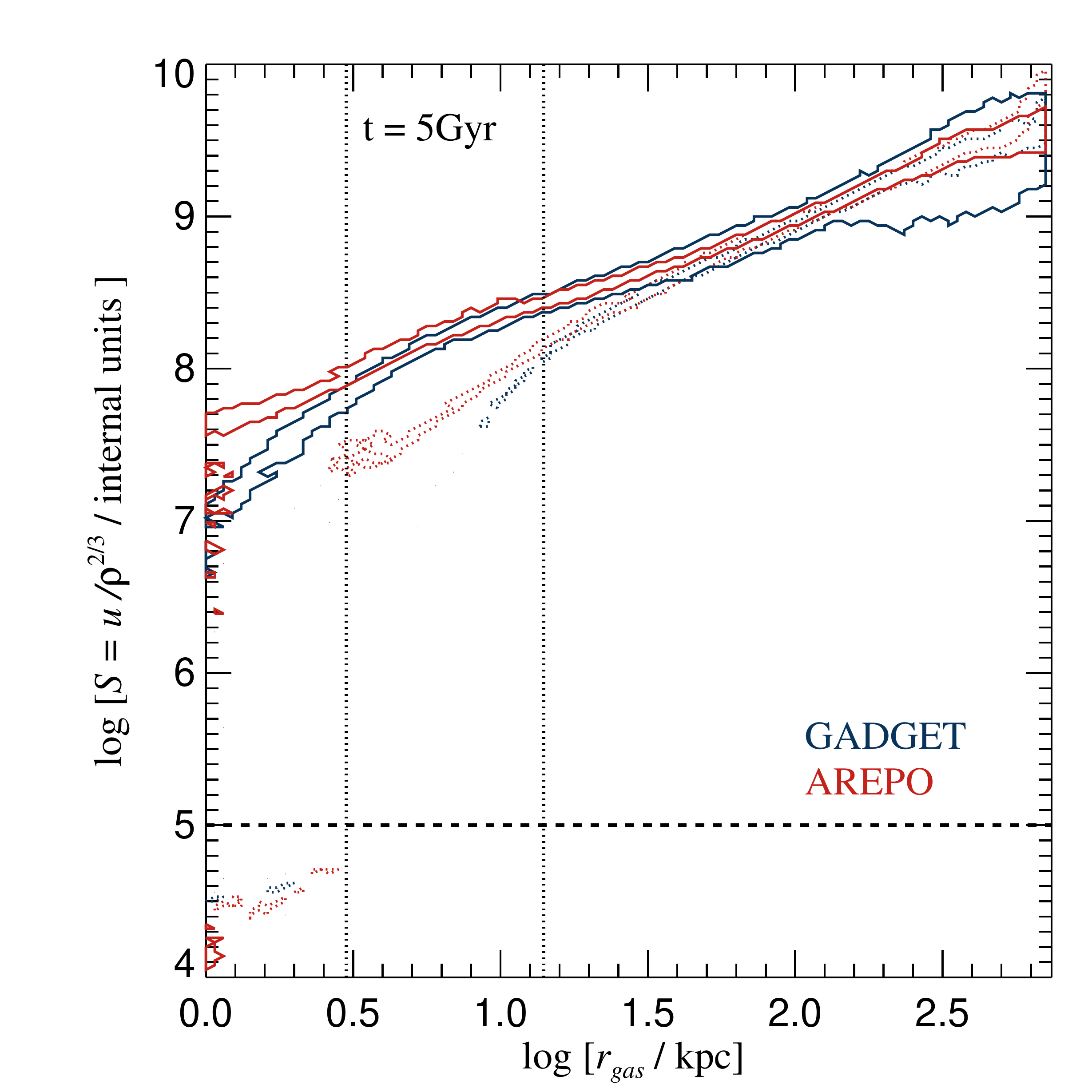}}
\caption{$2$D histogram of the gas entropy as a function of radial distance
  from the centre for {\small GADGET} (blue) and {\small AREPO} (red) at time
  $t=5\,{\rm Gyr}$ for $N_{\rm part} = 10^4$ (dotted lines) and $N_{\rm part}
  = 10^6$ (continuous lines). The gaseous halo is allowed to radiatively cool
  but there is no net rotation. The vertical dotted lines indicate the
  gravitational softening used in the {\small GADGET} run, which correspond
  to the floor values of the adaptive softenings in the moving mesh
  calculations.}
\label{FigEntCoolHist2d}
\end{figure}

Furthermore, in Figure~\ref{FigEntCoolHist2d} we show a $2$D histogram
of intracluster gas entropies as a function of cluster-centric
distance, after $5\,{\rm Gyr}$ from the start of the simulation for the
  low resolution and high resolution runs. The
distribution of gas entropies is almost identical in the two codes for
$r > r_{\rm soft}$ (note that below $r_{\rm soft}$ the simulation results are not
trustworthy due to the limited gravitational resolution on these scales). It can be seen that
there is some difference in the gas entropy close to the virial radius
of the system, which is caused by different boundary conditions
(vacuum for {\small GADGET} and a uniform low resolution grid for
{\small AREPO}).

These results are in line with the expectation that gas cooling and
condensation in this simulated system is determined entirely by the
gas properties at the cooling radius \citep{Bertschinger1989,
  White1991, Hernquist2003} which corresponds very closely between the
simulation codes.

\subsubsection{Radiative gas cooling in rotating
  haloes}\label{IsolatedHaloCoolRot} 

\begin{figure}\centerline{
\includegraphics[width=8.5truecm,height=8.2truecm]{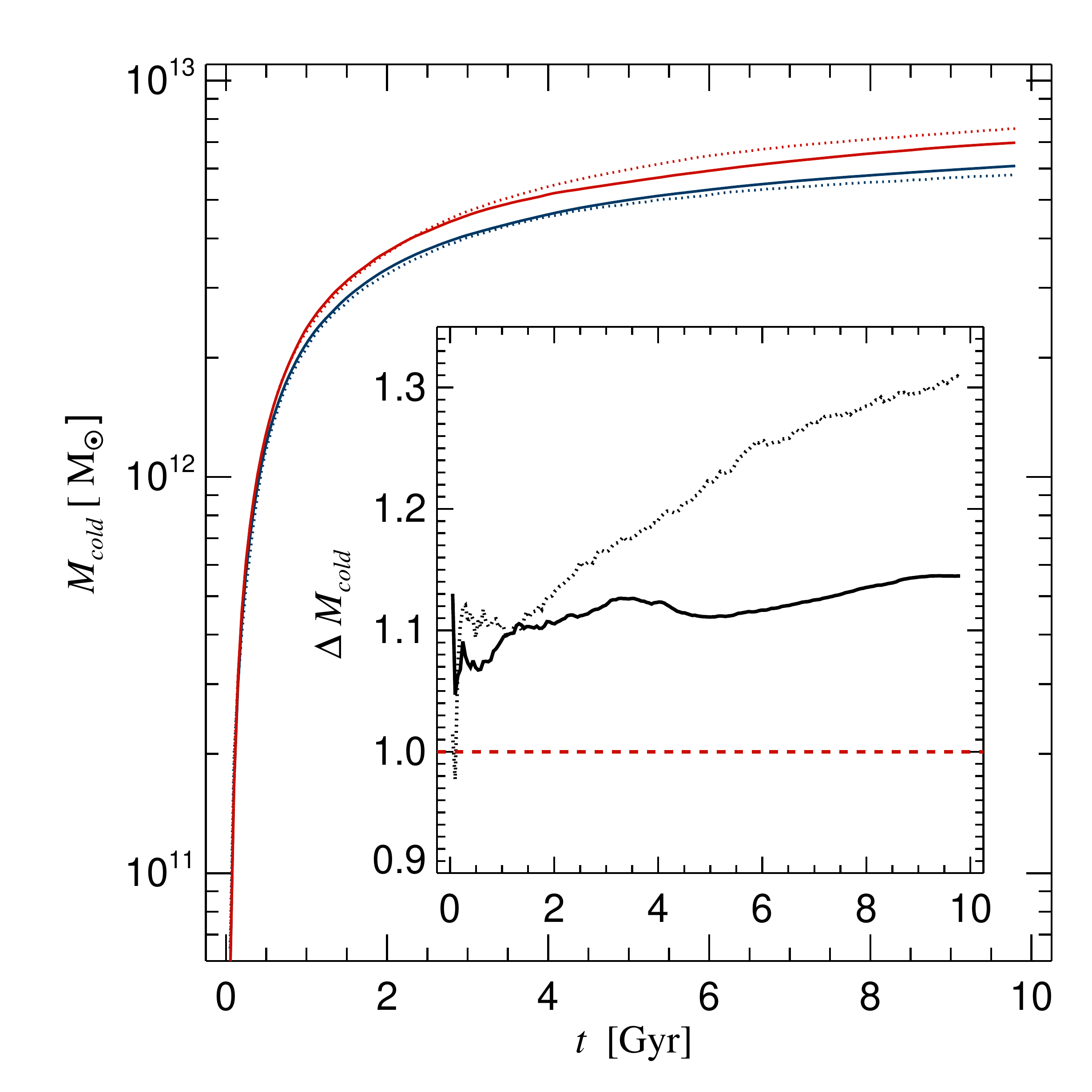}}
\caption{Time evolution of the cold baryonic mass (stars and gas with entropy $<
  10^5$ in internal units) for a $M_{\rm vir} = 10^{14}\, {\rm M_{\odot}}$
  isolated halo which radiatively cools and has a spin of $\lambda = 0.4$. Illustrated
  are low ($N_{\rm gas} = 10^4$; dotted lines) and high ($N_{\rm gas} = 10^6$;
  continuous lines) resolution simulations with {\small GADGET} (blue) and {\small AREPO}
  (red). In the inset, the ratio of $M_{\rm cold}$ values found in {\small AREPO}
  and {\small GADGET} is shown for both numerical resolutions. The agreement between
  the amount of cold baryons formed is poorer for $N_{\rm gas} = 10^4$, but
  improves for the high resolution run, with $M_{\rm cold}$ in {\small AREPO} being
  $\sim 10-15\%$ higher.}
\label{FigMcoldRot}
\end{figure}

Even though gas cooling and star formation proceed in a remarkably
similar way in {\small GADGET} and the moving-mesh code for haloes
with vanishing spins, this is not guaranteed to remain the case once
some degree of rotation is included. We therefore simulate exactly the
same isolated haloes as in the previous section, but imposing a
certain level of gas rotation within the static dark matter potential.
To highlight the effect, we use a large spin parameter equal to
$\lambda=0.4$.

Figure~\ref{FigMcoldRot} shows the time evolution of the total mass in
cold baryons (stars and gas with entropy $< 10^5$ in internal units)
for the rotating haloes simulated with {\small GADGET} (blue curves)
and {\small AREPO} (red curves) at different resolutions. Comparing
$M_{\rm cold}$ with our findings from Figure~\ref{FigMcoldNoRot} for
non-rotating haloes indicates that overall less gas cools from the hot
phase if the gas spins. This effect is not surprising given our
initial conditions. Even though the gas density and temperature
distribution are initially identical, in the simulations where there
is considerable spin the gas will be subject to centrifugal
accelerations, preventing it from collapsing radially. The (partial)
centrifugal support will tend to reduce the gas densities and hence
the cooling rates.
 
\begin{figure*}\centerline{\hbox{
\includegraphics[width=8.5truecm,height=8.2truecm]{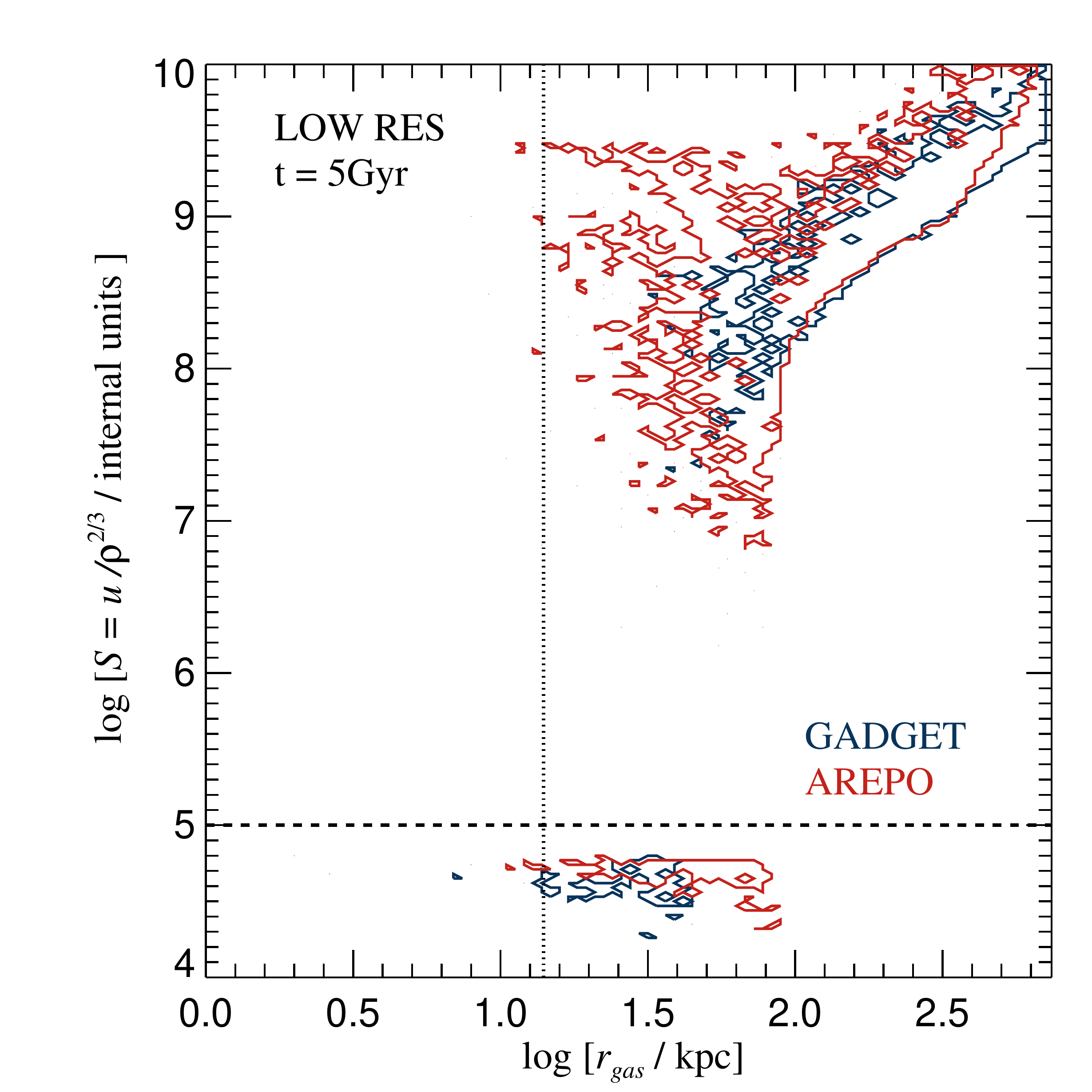}
\includegraphics[width=8.5truecm,height=8.2truecm]{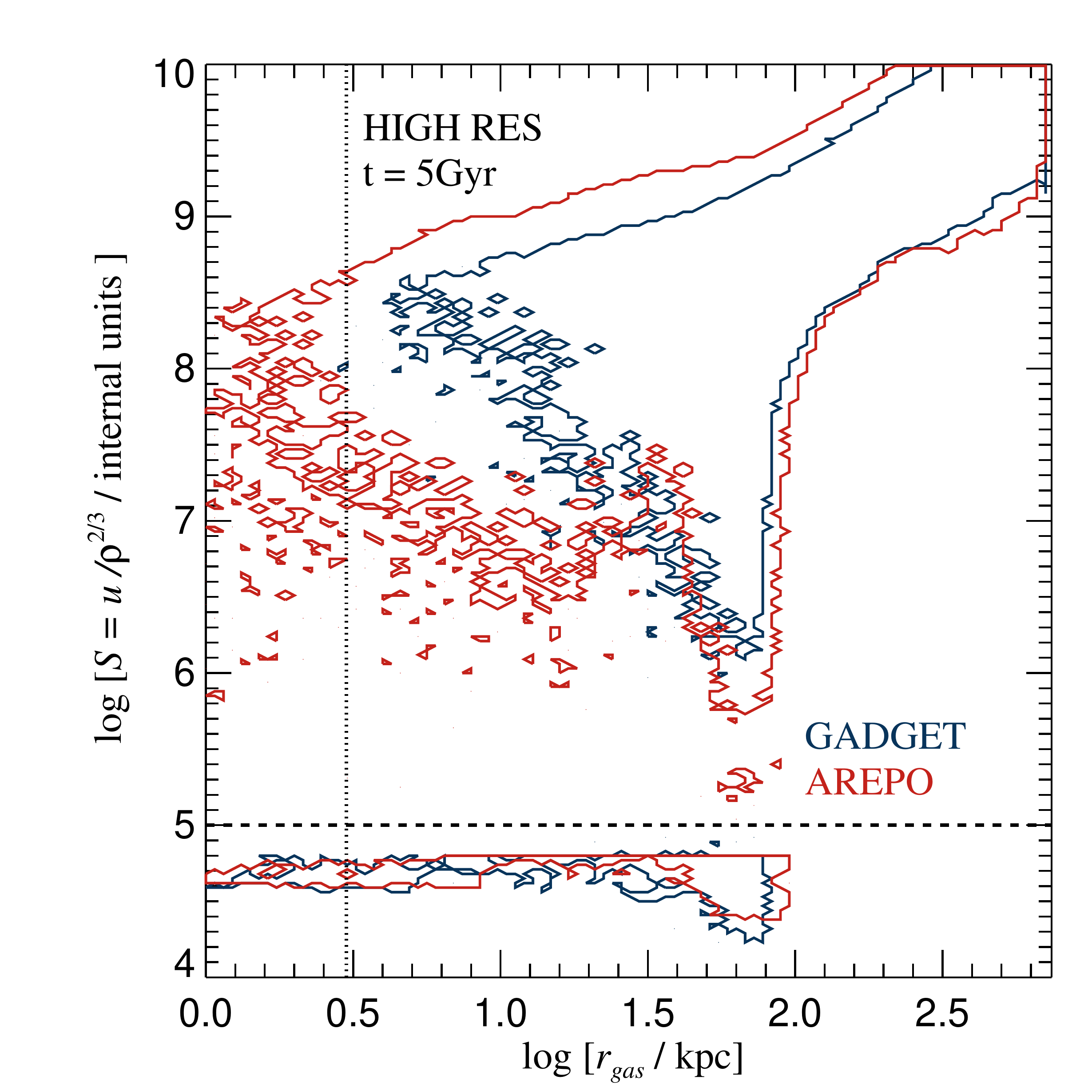}}}
\caption{$2$D histograms of gas entropy as a function of radial distance from
  the centre for {\small GADGET} and {\small AREPO} at $t=5\,{\rm Gyr}$ for
  the $10^{14}\, {\rm M_{\odot}}$ mass halo which radiatively cools and spins
  at low ($N_{\rm gas} = 10^4$; left-hand panel) and high ($N_{\rm gas} =
  10^6$; right-hand panel) resolution. The vertical dotted line is the
  gravitational softening used in the {\small GADGET} run, which
    correspond to the floor value of the adaptive softenings in the moving
    mesh calculations.}
\label{FigEntCoolRotHist2d}
\end{figure*}

More importantly, from Figure~\ref{FigMcoldRot} it can be seen that
there is poorer agreement in the amount of cold baryons between
{\small GADGET} and {\small AREPO} for a rotating gaseous halo. The
discrepancy is larger for the low resolution run with $N_{\rm gas} =
10^4$, where the final $M_{\rm cold}$ value after $10\,{\rm Gyr}$ is
about $30\%$ higher in the moving-mesh code. The reason for this
discrepancy is twofold: at low resolution, {\small GADGET}
underestimates the cooling rate somewhat, while {\small AREPO}
overestimates it. The SPH result turns out to be more stable at
  poor resolution than the mesh-based calculation. Here it is
  advantageous for SPH that even for few particles a clearly defined
  phase boundary between cold and hot gas is maintained (in fact, a
  `pressure blip' in SPH leads to a sampling gap at this boundary)
  whereas in {\small AREPO} this boundary is blurred at low
  resolution, causing slightly elevated cooling. This trend is also
confirmed by the gas radial velocities which are least negative in the
low resolution {\small GADGET} run and most negative in the low
resolution {\small AREPO} simulation. The radial velocities
systematically differ within a few $100\,{\rm kpc}$ and throughout
most of the simulated time-span. For $N_{\rm gas} = 10^6$, the radial
velocities obtained with the two codes are much closer, which also
translates into a better match of the amount of cold baryons, with
$M_{\rm cold}$ higher by $\sim 10-15\%$ in {\small AREPO}.

At least part of the remaining difference in $M_{\rm cold}$ between
{\small GADGET} and {\small AREPO} is driven by the interaction
between the gas which is cooling towards the centre and the cold gas
that is already in the disk. This is shown in
Figure~\ref{FigEntCoolRotHist2d}, where we plot $2$D histograms of the
gas entropy as a function of cluster-centric distance at $t=5\,{\rm
  Gyr}$ for the low (left-hand panel) and high resolution simulations
(right-hand panel). In the moving-mesh code the gas surrounding the
central disk has a range of entropy values, with some fluid elements
exhibiting very large entropies due to the shock in the immediate
vicinity of the disk. Conversely, in the runs with {\small GADGET} and
especially at low resolution, there is a clear gap between cold
material in the disk and hotter gas in the halo \citep[see
e.g.][]{Agertz2007,SpringelRev2010}. Similarly, as described in
Section~\ref{ColdInflow}, SPH particles are affected by artificial
viscosity in the converging part of the flow which can slightly offset
cooling losses. Note that for $N_{\rm gas} = 10^6$ the difference in
gas entropy structure around the disk is lower between the codes, as
expected, given that unwanted artificial viscosity effects are reduced
and that the gap between the cold and hot gas phase due to repulsive
pressure forces is smaller.
 
\begin{table*} 
\bc
\begin{tabular}{cccccccc} 
\hline
${\rm code}$ & $N_{\rm gas}$ & $R_{\rm gas,max}$ & $R_{\rm gas,HM}$ & $M_{\rm
  gas,HM}$ & $R_{\rm stars,HM}$ & $M_{\rm stars,HM}$ &  $M_{\rm cold}$ \\[3pt]
 & & $[{\rm kpc}]$ & $[{\rm kpc}]$ & $[10^{10} \, {\rm M_{\odot}}]$ & $[{\rm
    kpc}]$ & $[10^{10} {\rm M_{\odot}}] $ & $[10^{10} \, {\rm M_{\odot}}]$\\
\hline
${\rm {\small GADGET}}$ & $10^4$ & $40.0$ & $30.3$ & $27.0$ & $13.5$ & $215.3$ & $484.6$\\
${\rm {\small GADGET}}$ & $10^6$ & $79.0$ & $54.7$ & $9.8$ & $6.8$ & $238.3$ & $496.2$\\
${\rm {\small AREPO}}$ & $10^4$ & $79.0$ & $47.4$ & $27.7$ & $18.1$ & $267.9$ & $591.2$\\
${\rm {\small AREPO}}$ & $10^6$ & $85.0$ & $67.1$ & $12.8$ & $9.7$ & $263.1$ & $551.8$\\
\hline
\end{tabular}
\caption{Gaseous and stellar properties of the disk at $t=5\,{\rm Gyr}$ which
  forms in an isolated $10^{14} {\rm M_{\odot}}$ halo which radiatively cools
  and rotates. For simulations with two different resolutions with {\small
    GADGET} and {\small AREPO}, we list the maximum radial extent of the
  gaseous disk, its half-mass radius and the total mass within the half-mass
  radius, in the third, fourth and fifth columns, respectively. Stellar
  half-mass radius and the total stellar mass enclosed within it are shown in
  the sixth and seventh columns. In the eighth column the total mass of cold
  baryons is given, being equal to $2 \times (M_{\rm gas,HM} + M_{\rm
    stars,HM})$. All gas particles/cells with entropy less than $10^5$ (in
  internal units) are assumed to be part of the disk, while for the stellar
  disk we consider all star particles that form in the simulated volume.}
\label{TabDiskProp}
\ec
\end{table*}

From Figure~\ref{FigEntCoolRotHist2d} is it also evident that the extent of
the cold disk is different between the codes, especially at low
resolutions. To quantify this important effect, in Table~\ref{TabDiskProp} we
summarize the main properties of the forming disk. For $N_{\rm gas} =10^4$ the
half-mass radius of the gaseous disk in {\small GADGET} is almost a factor of
two smaller than in the higher resolution run, while in the moving mesh code
$R_{\rm gas,HM}$ is $70\%$ of the value we obtain with $N_{\rm gas} =
10^6$. This indicates that the convergence rate of the gas disk size is slower
in the case of {\small GADGET}, due to spurious transfer of angular momentum
from the cold to the hot phase \citep{Okamoto2003} and due to the artificial
viscosity in the case of poorly resolved disks (note, however, that the total
angular momentum is manifestly conserved in {\small GADGET}). In the case of
the stellar disks they are essentially not resolved in our low resolution runs
($R_{\rm stars,HM} \sim r_{\rm soft} = 14\,{\rm kpc}$), while the half-mass
radius is $\sim 40\%$ higher in {\small AREPO} at higher resolution. Contrary
to {\small GADGET}, total angular momentum conservation is not automatically
guaranteed in the moving mesh code, particularly for disks resolved with only
a small number of cells. Nonetheless, it is reassuring that $R_{\rm gas,HM}$
increases with higher resolution in {\small AREPO} (attesting that spurious
angular momentum transport inwards is probably small) and that the value of
$R_{\rm gas,HM}$ obtained with both codes for $N_{\rm gas} = 10^6$ is
relatively close. This indicates that the differences in the disk sizes
  in this numerical experiment are at least partly driven by resolution effects, and in
  particular by the slow convergence rate of {\small GADGET}
  simulations. Comparing the resolution requirements to obtain good agreement
  between the codes here with Section~\ref{RadHaloes} we note that in the case of
  cooling gaseous haloes with a significant spin more resolution elements are
  needed to follow accurately the thermo-dynamical interaction between the
  central cold disk and the inflowing hotter gas.
 
\subsubsection{Generalized blob test: radiative case}\label{GeneralBlobCooling}
\begin{table*} 
\bc
\begin{tabular}{cccccccccc} 
\hline
${\rm code}$ & $N_{\rm gas}$ & $N_{\rm blob}$ & $R_{\rm gas,max}$ & $R_{\rm gas,HM}$ & $M_{\rm
  gas,HM}$ & $R_{\rm stars,HM}$ & $M_{\rm stars,HM}$ &  $M_{\rm cold}$ \\[3pt]
 & & & $[{\rm kpc}]$ & $[{\rm kpc}]$ & $[10^{10} \, {\rm M_{\odot}}]$ & $[{\rm
    kpc}]$ & $[10^{10} {\rm M_{\odot}}] $ & $[10^{10} \, {\rm M_{\odot}}]$ \\
\hline
${\rm {\small GADGET}}$ & $10^4$ & $10^2$ & $22.0$ & $16.3$ & $12.9$ & $13.5$ & $184.4$ & $394.8$\\
${\rm {\small GADGET}}$ & $10^5$ & $10^3$ & $45.0$ & $17.9$ & $7.5$ & $15.8$ & $221.1$ & $457.1$\\
${\rm {\small GADGET}\_EOS}$ & $10^5$ & $10^3$ & $24.0$ & $7.6$ & $154.4$ & $/$ & $/$ & $308.9$\\
${\rm {\small GADGET}}$ & $10^6$ & $10^4$ & $56.0$ & $32.9$ & $5.4$ & $26.6$ & $235.1$ & $481.0$\\
${\rm {\small AREPO}}$ & $10^4$ & $10^2$ & $76.0$ & $37.8$ & $16.8$ & $17.8$ & $199.9$ & $433.5$\\
${\rm {\small AREPO}}$ & $10^5$ & $10^3$ & $71.0$ & $43.0$ & $17.3$ & $21.8$ & $261.3$ & $557.2$\\
${\rm {\small AREPO}\_EOS}$ & $10^5$ & $10^3$ & $76.0$ & $13.4$ & $244.7$ & $/$ & $/$ & $489.5$\\
${\rm {\small AREPO}}$ & $10^6$ & $10^4$ & $75.0$ & $41.5$ & $12.7$ & $32.7$ & $273.2$ & $571.9$\\
\hline
\end{tabular}
\caption{Gaseous and stellar properties of the disk at $t=5\,{\rm Gyr}$ that
  formed in an isolated $10^{14}\, {\rm M_{\odot}}$ halo which radiatively
  cools, rotates and contains $10$ orbiting substructures. For simulations
  with three different resolutions with {\small GADGET} and {\small AREPO} we
  list the maximum radial extent of the gaseous disk, its half-mass radius and
  the total mass within the half-mass radius, in the fourth, fifth and sixth
  columns, respectively. Stellar half-mass radius and the total stellar mass
  enclosed within it are shown in the seventh and eighth columns. In the ninth
  column the total mass of cold baryons is given, being equal to $2 \times
  (M_{\rm gas,HM} + M_{\rm stars,HM})$. All gas particles/cells with entropy
  less than $10^5$ (in internal units) and within $R_{\rm gas,max}$ are
  assumed to be part of the disk, while for the stellar disk we consider all
  star particles that form in the simulated volume (including the blobs). We
  also list the gaseous disk properties in two additional simulations, denoted
  by EOS, performed at intermediate resolution. In these simulations we adopt
  the standard sub-grid model for star formation, with the dense, cold gas
  lying on the effective equation-of-state, but we prevent any spawning of
  star particles.}
\label{TabDiskPropDMBlobs}
\ec
\end{table*}

Here we consider the evolution of cold, dense blobs embedded in a
galaxy cluster as in Section~\ref{GeneralBlob}, but now the whole
system is allowed to radiatively cool. To add an additional layer of
realism the blobs are constructed to be similar to cosmological
substructures: they are equipped with their own dark matter halo, and
stars may form out of their gas during the simulated time-span. More
specifically, each blob is represented by a live Hernquist dark matter
halo and gas in hydrostatic equilibrium. The virial mass of each blob
is $2 \times 10^{12} {\rm M_{\odot}}$, the scale length parameter is
$a = 41.4  \,{\rm kpc}$, and the gas fraction amounts to $f_{\rm gas} = 0.17$. As
before, we populate our default $10^{14} {\rm M_{\odot}}$ halo (that
has a static dark matter potential) with $10$ identical
substructures. The procedure for assigning blob positions and
velocities is the same as in Section~\ref{GeneralBlob}, but here
  we use a different random number seed which leads to the different
  initial positions and velocities with respect to
  Section~\ref{GeneralBlob}. The positions of blob centres are in the
  range of $\sim 650$ to $\sim 750\, {\rm kpc}$, while the
  characteristic blob velocities range from $200$ to $500\,{\rm
    km}\,{\rm s}^{-1}$. Also, in Section~\ref{GeneralBlob} the gas in the
  halo is initially at rest while here the gas has considerable angular
  momentum, which contributes to the relative velocity between the
  blobs and the gas. We simulate the evolution of this system at
three different resolutions: $N_{\rm gas} = 10^4$, $N_{\rm blob} =
10^2$ (low resolution), $N_{\rm gas} = 10^5$, $N_{\rm blob} = 10^3$
(intermediate resolution), and $N_{\rm gas} = 10^6$, $N_{\rm blob} =
10^4$ (high resolution), with gas self-gravity, and with cooling and
the sub-grid model for star formation.

Additionally, we perform two simulations at intermediate resolution where the
standard sub-grid model for star formation is included as well, but where we
simply prevent any star particles from being spawned out of the star-forming
phase, denoted by EOS. This simulation set-up is particularly useful for
following the thermodynamical evolution of cold and hot gas for many Gyrs
without the dense cold gas being subject to fragmentation that would likely
occur in pure cooling runs. For numerical experiments with {\small AREPO}, as
a default choice, we have not used mesh refinement and de-refinement. However,
we have performed extra runs at intermediate resolution where we adopt a
de-/refinement strategy to limit the mass range of gas cells (within a factor
of $2$ of the gas particle mass in the matching {\small GADGET} run) which
automatically imposes a narrow range of stellar masses as well. For these runs
we have increased the number of gas cells/particles in each blob to $2.5
\times 10^3$, such that the gas particle/cell mass in blobs is exactly
identical to the one in the parent halo (which is optimal for our
de-/refinement method). These test runs recover very closely all of our
results with the default set-up, confirming that N-body heating effects
  (e.g. due to the spectrum of star particle masses) are not very important
here.

\begin{figure*}\centerline{\vbox{\hbox{
\includegraphics[width=8.truecm,height=7.truecm]{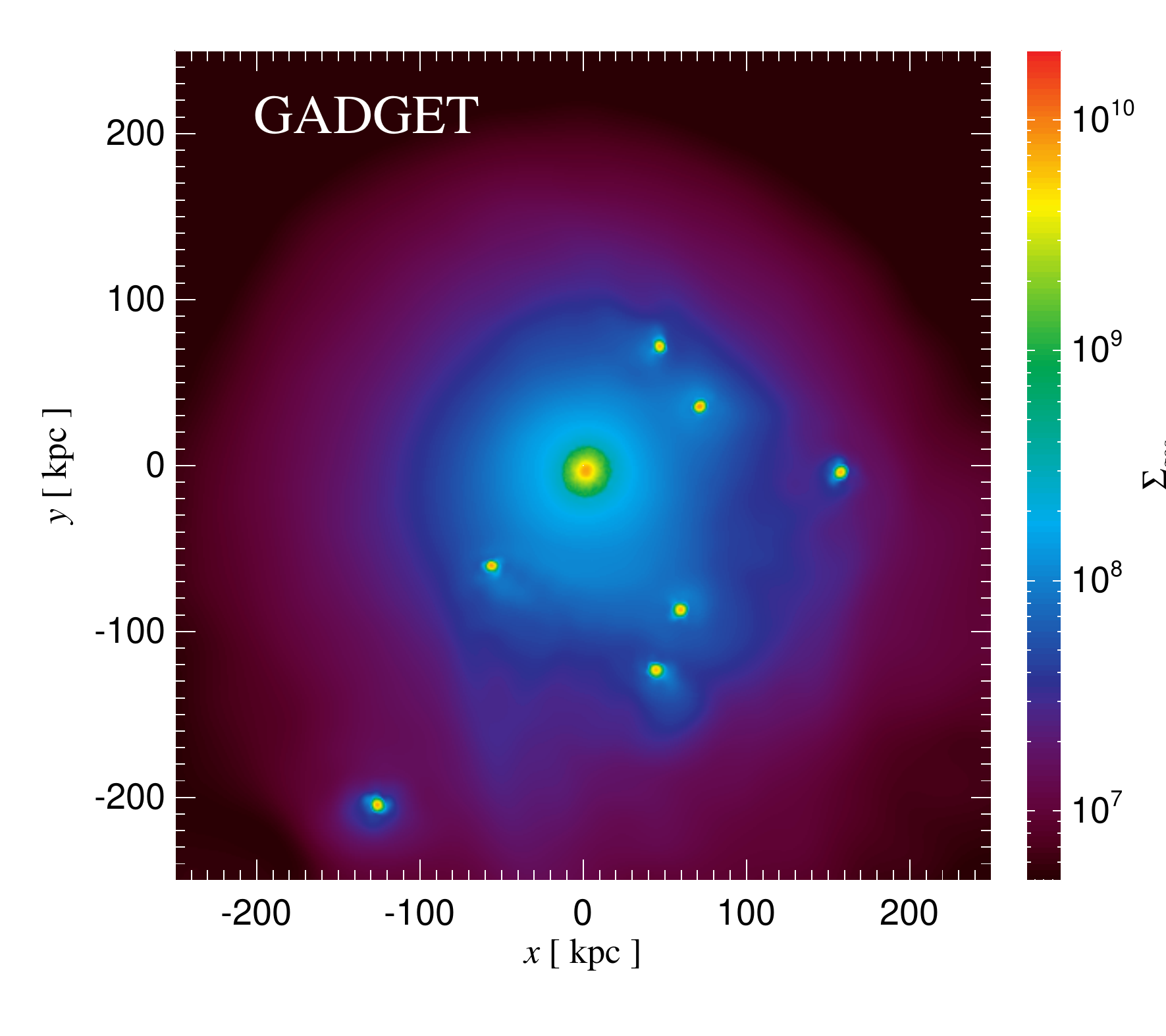}
\includegraphics[width=8.truecm,height=7.truecm]{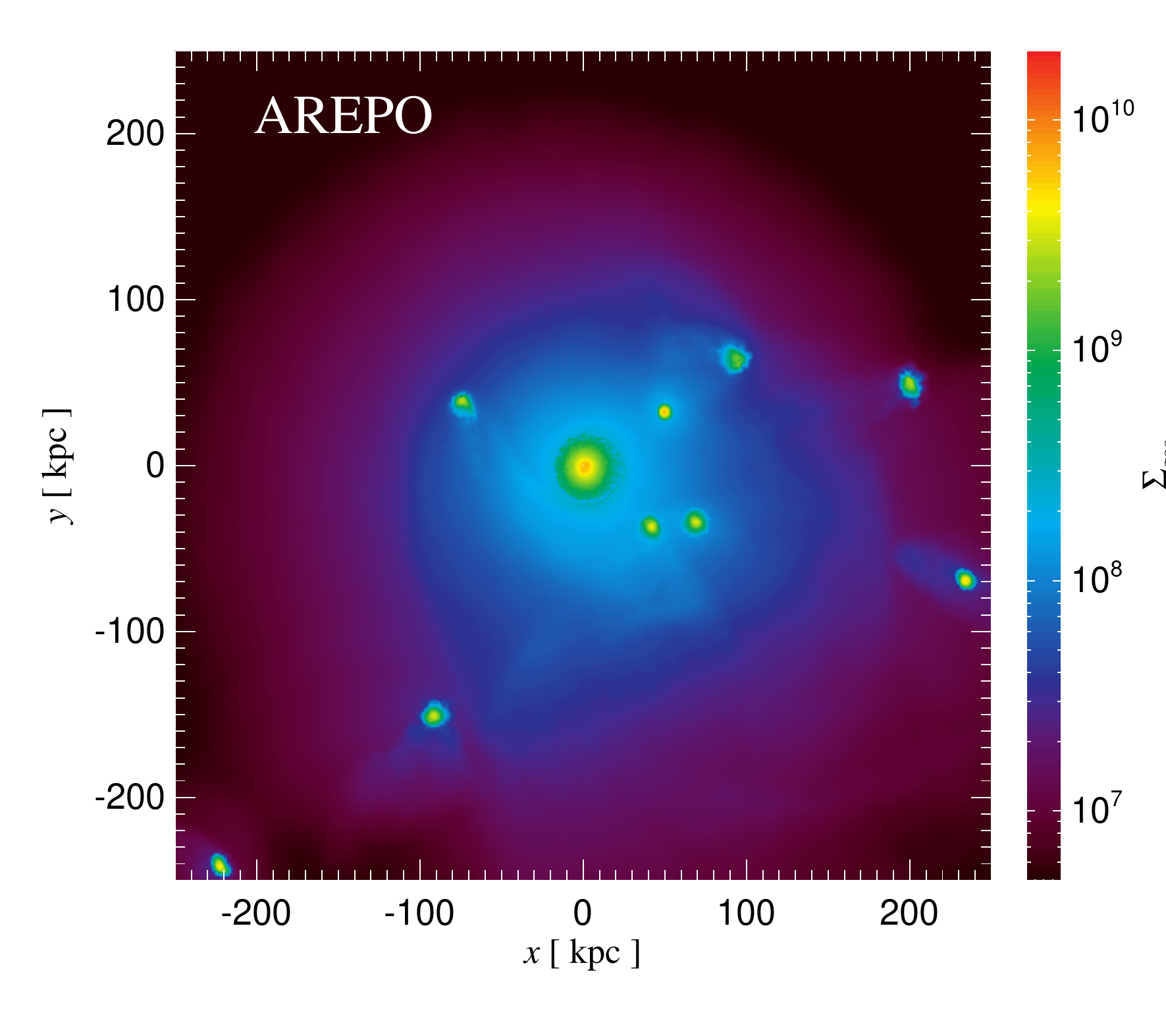}}
\hbox{
\includegraphics[width=8.truecm,height=7.truecm]{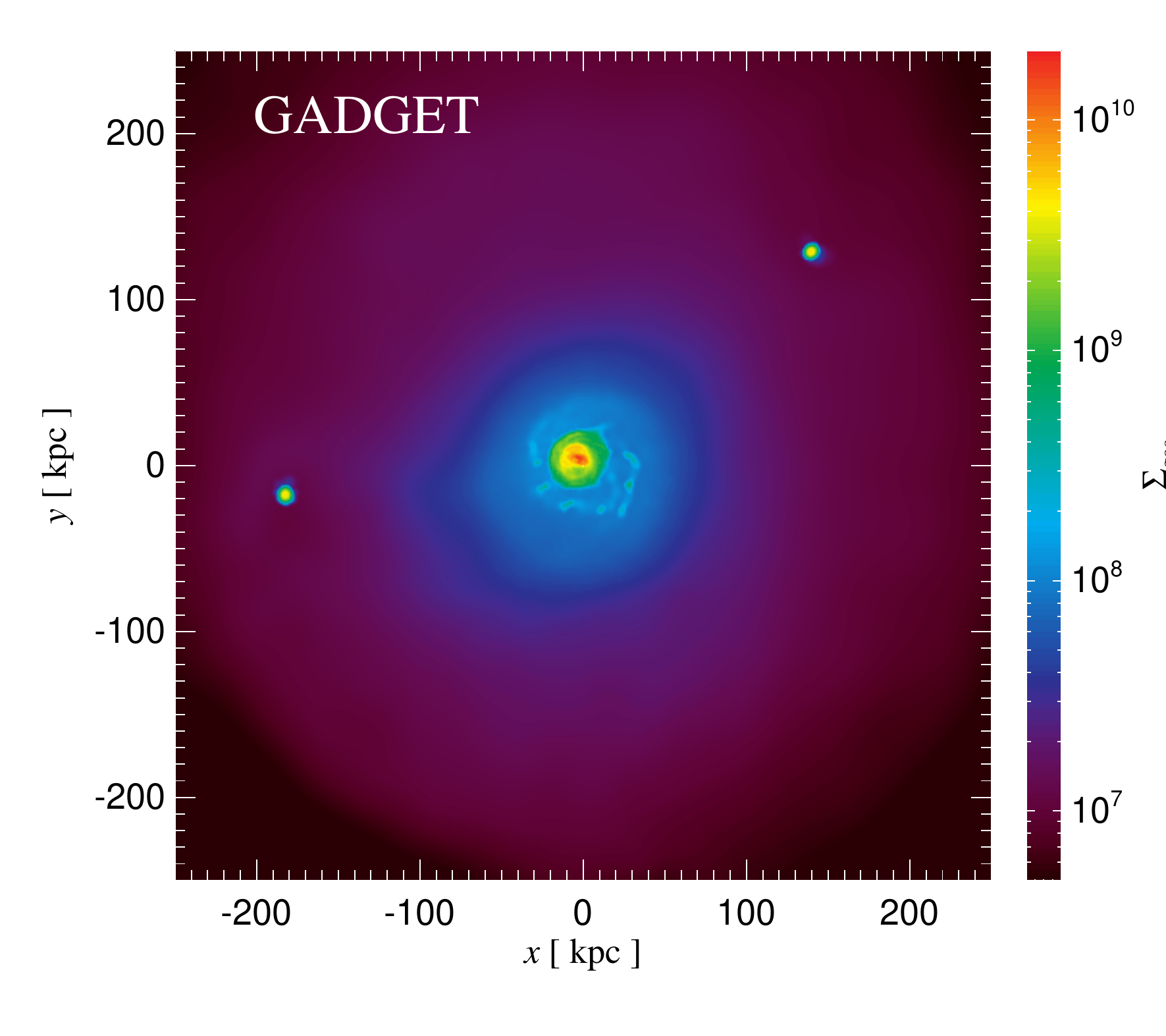}
\includegraphics[width=8.truecm,height=7.truecm]{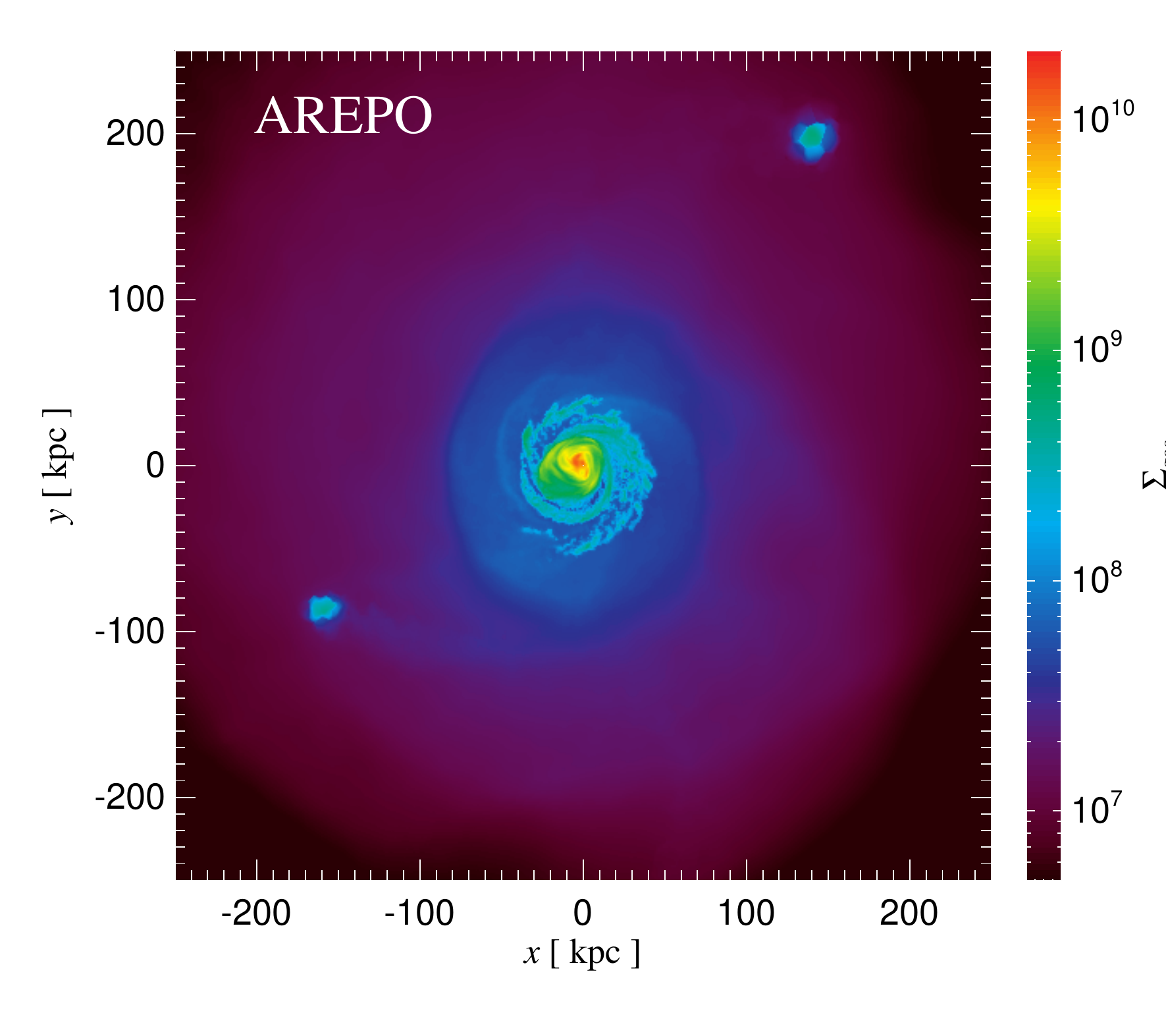}}
\hbox{
\includegraphics[width=8.truecm,height=7.truecm]{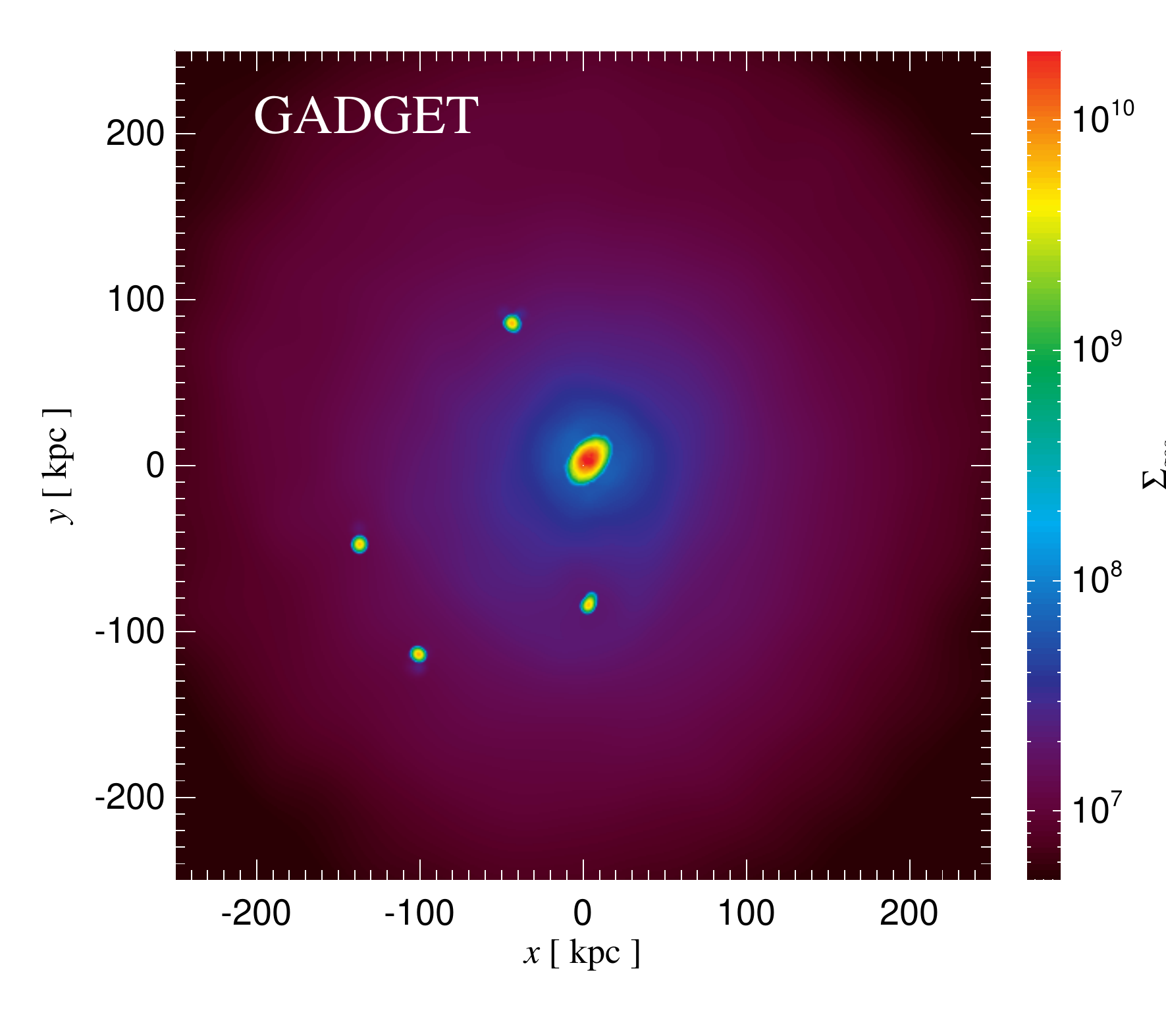}
\includegraphics[width=8.truecm,height=7.truecm]{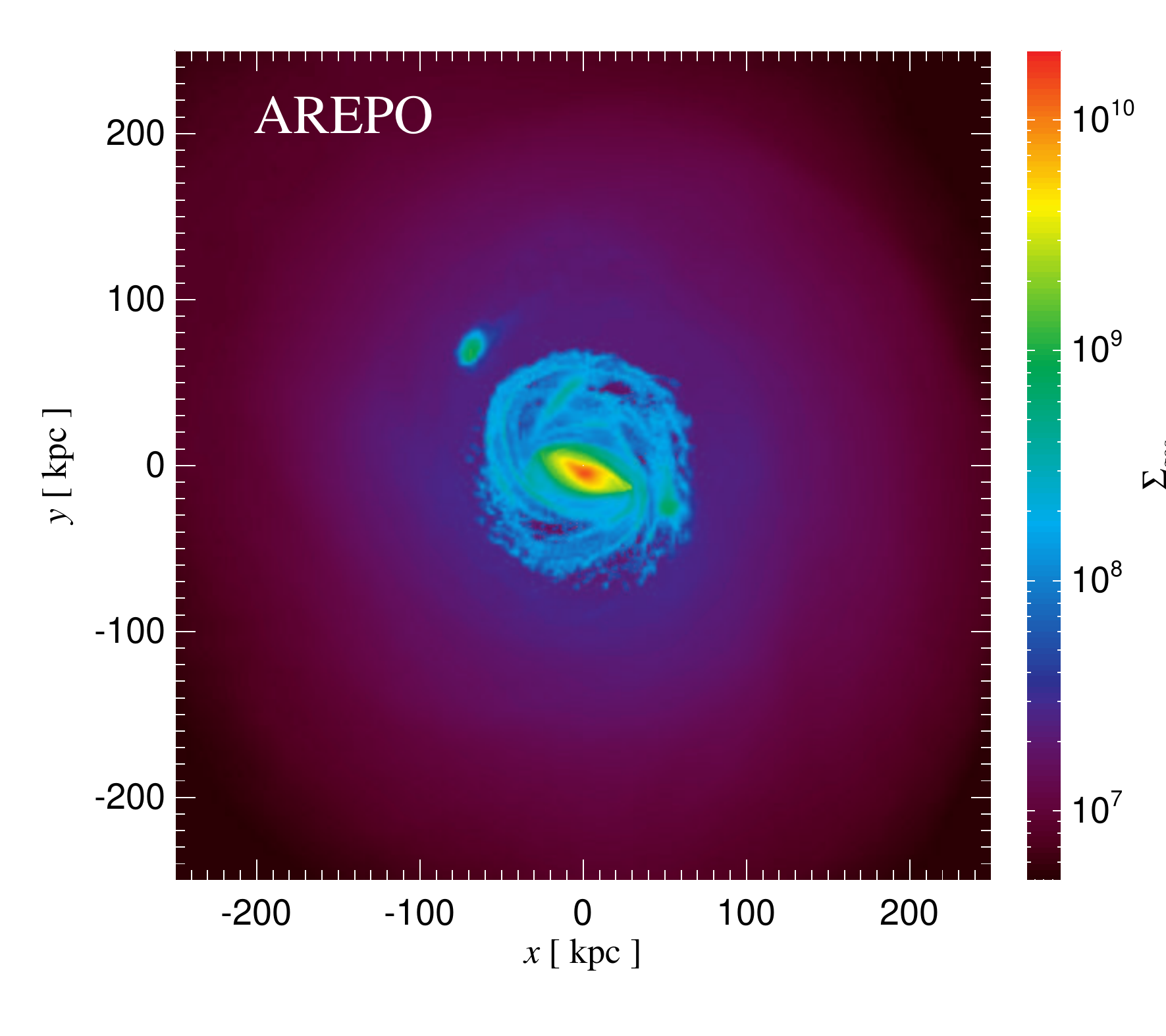}}}}
\caption{Projected surface density maps in units of $[{\rm
    M_{\odot}}{\rm kpc}^{-2}]$ at times $t = 0.9\,{\rm Gyr}$ (top
  panels), $t = 2.3\,{\rm Gyr}$ (central panels), and $t = 6\,{\rm
    Gyr}$ (bottom panels) for a $M_{\rm vir} = 10^{14}\, {\rm
    M_{\odot}}$ isolated halo which rotates, radiatively cools, and
  has $10$ orbiting substructures. The thickness of the projection is
  $\Delta z = 2\,{\rm Mpc}$. Although we have used our sub-grid model
  for star formation in these simulations, spawning of star particles
  has been intentionally prevented here. Gas stripping from the
  orbiting blobs is found to be very different in the two numerical
  techniques. In {\small GADGET}, blobs largely survive, and when they
  interact with the central disk they tend to disrupt it, whereas in
  {\small AREPO}, the blobs are more easily shredded and have a less
  damaging effect on the forming disk. In fact, some of the stripped
  blob material ends up contributing to the extended gaseous disk.}
\label{FigRhoGasDMBlobs}
\end{figure*}

In Figure~\ref{FigRhoGasDMBlobs}, we show projected surface density
maps at $t = 0.9\,{\rm Gyr}$, $t = 2.3\,{\rm Gyr}$, and $t = 6\,{\rm
  Gyr}$ for simulations where we use the sub-grid model for star
formation, but prevent the spawning of star particles. Initially the
evolution of the blobs proceeds in a very similar fashion in
simulations with {\small GADGET} and {\small AREPO}. However, already
after less than a Gyr (see top panels of
Figure~\ref{FigRhoGasDMBlobs}) blobs in the moving mesh code are much
more affected by ram pressure and dynamical fluid instabilities, which cause efficient
gas stripping. As the blobs reach the very inner regions, they
interact with the forming disk. In {\small GADGET} simulations, the
blobs have a significantly more damaging effect on the disk, simply
because they are less gas depleted. This is clearly visible in the
central panels of Figure~\ref{FigRhoGasDMBlobs}, where in {\small
  GADGET} feeble spiral features are present, while in {\small AREPO}
the disk is more extended and spiral arms are well
developed. Additionally, in the moving-mesh simulation a large
fraction of the stripped material is deposited in the main halo, and
is gradually accreted onto the central disk, promoting its
growth. Blobs in {\small GADGET} are more coherent and eventually lose
their angular momentum due to hydro-/dynamical friction.  As they merge with
the central disk, an ellipsoidal structure is formed (see bottom
panels of Figure~\ref{FigRhoGasDMBlobs}). Note that due to the static
dark matter halo the dynamical friction force arises only due to the
gas in the main halo. Thus, we expect the blobs will lose their
angular momentum on an even faster time-scale in simulations with live
dark matter haloes, as is the case for cosmological runs. After
$6\,{\rm Gyrs}$, the difference between {\small GADGET} and {\small
  AREPO} simulations is significant: while in the latter case an extended
disk forms, with gaseous spiral arms reaching up to $60\,{\rm kpc}$
away from the centre, in {\small GADGET} we are left with a flattened,
amorphous blob.
      
\begin{figure*}\centerline{\hbox{
\includegraphics[width=8.truecm,height=7.truecm]{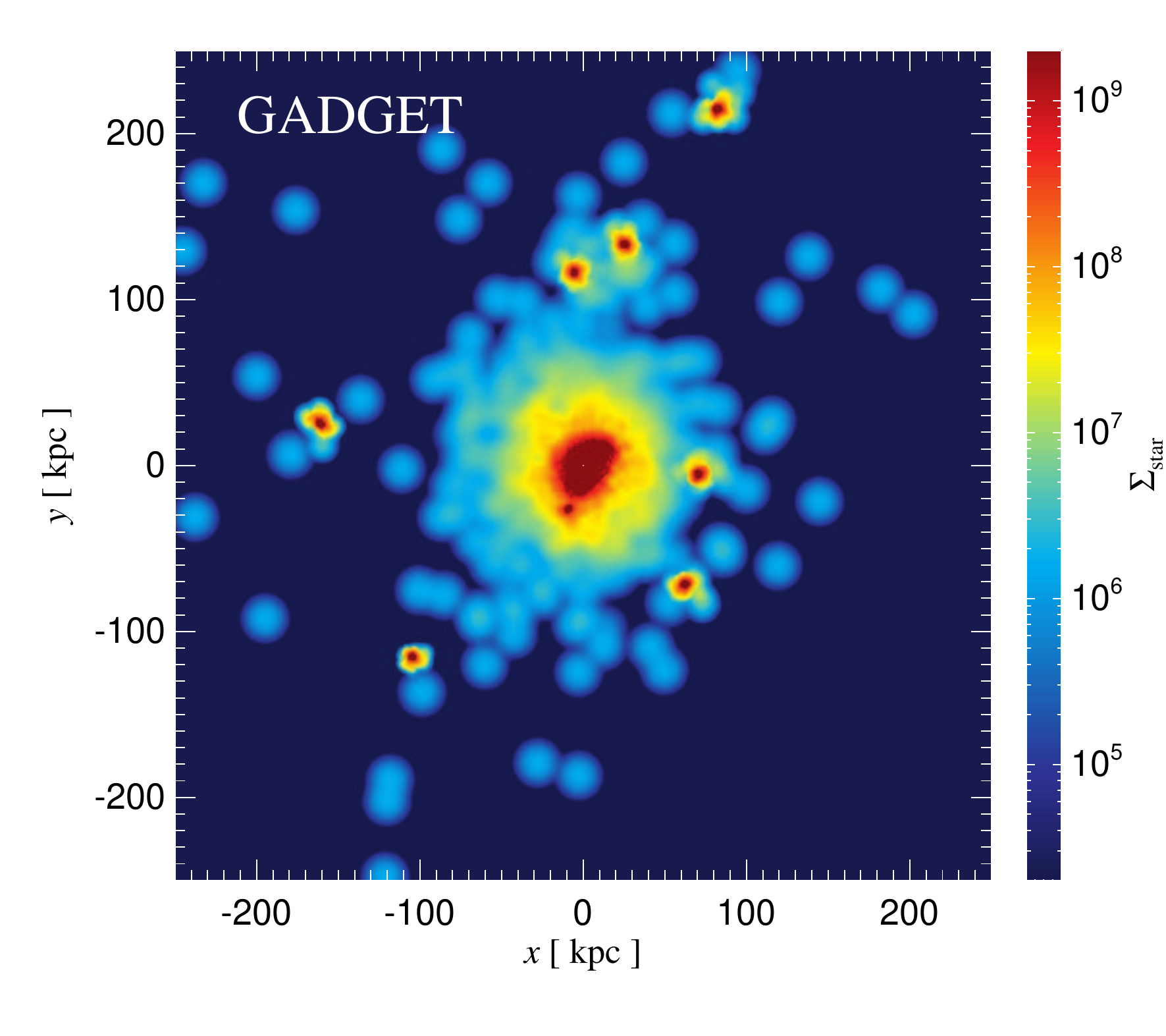}
\includegraphics[width=8.truecm,height=7.truecm]{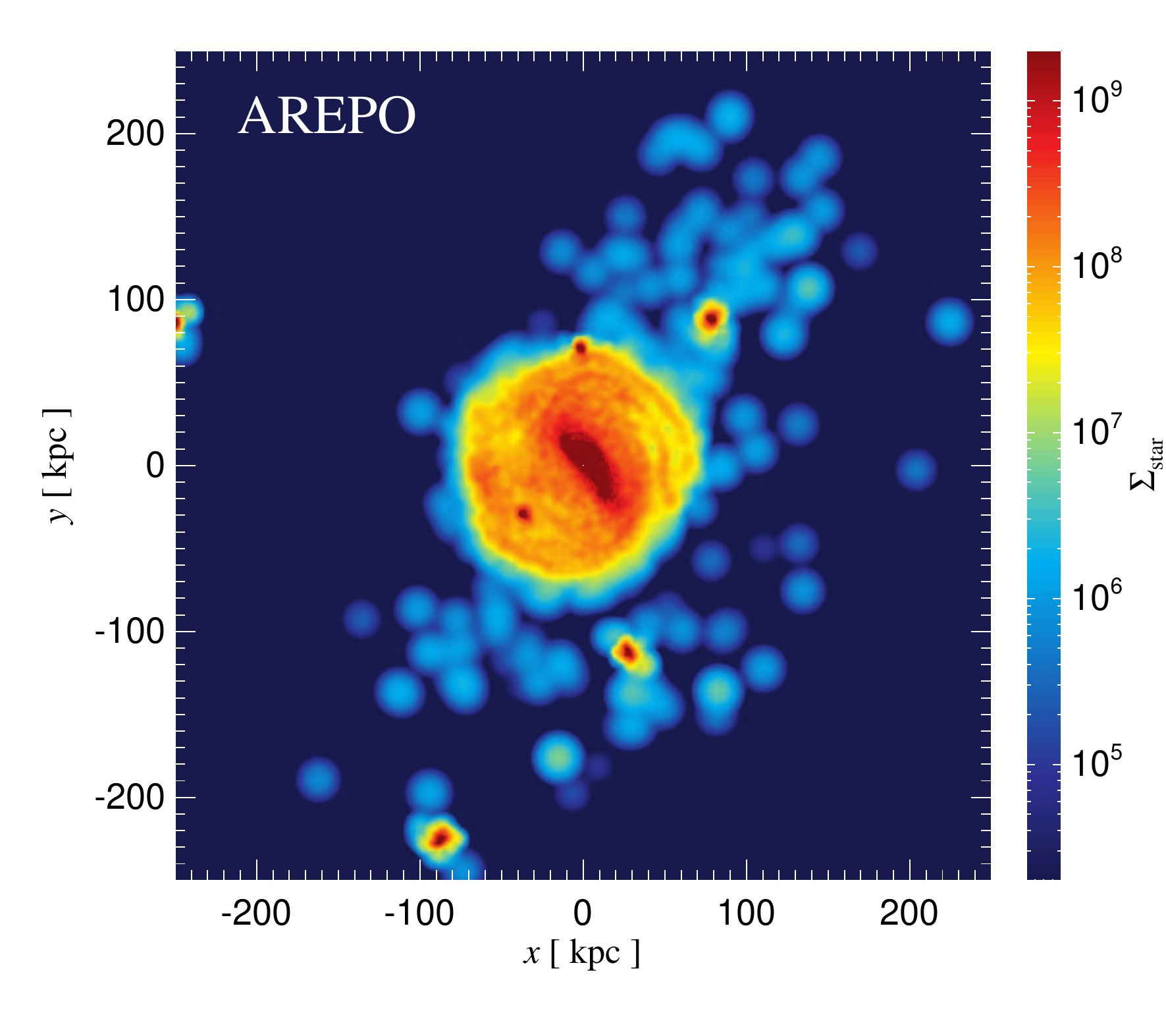}}}
\caption{Projected stellar density maps in units of $[{\rm
    M_{\odot}}{\rm kpc}^{-2}]$ at $t = 6\,{\rm Gyr}$ for a $M_{\rm
    vir} = 10^{14}\, {\rm M_{\odot}}$ isolated halo which rotates,
  radiatively cools, forms stars, and has $10$ orbiting
  substructures. The maps have been constructed from our intermediate
  resolution runs and the thickness of the projection is $\Delta z =
  2\,{\rm Mpc}$. While in {\small GADGET} a centrally concentrated
  disk forms that is surrounded by a feeble thick structure, in the simulation
  with {\small AREPO}, a well-defined extended stellar disk assembles
  with a central bar.}
\label{FigRhoStarsDMBlobs}
\end{figure*}

\begin{figure*}\centerline{\hbox{
\includegraphics[width=8.5truecm,height=8.2truecm]{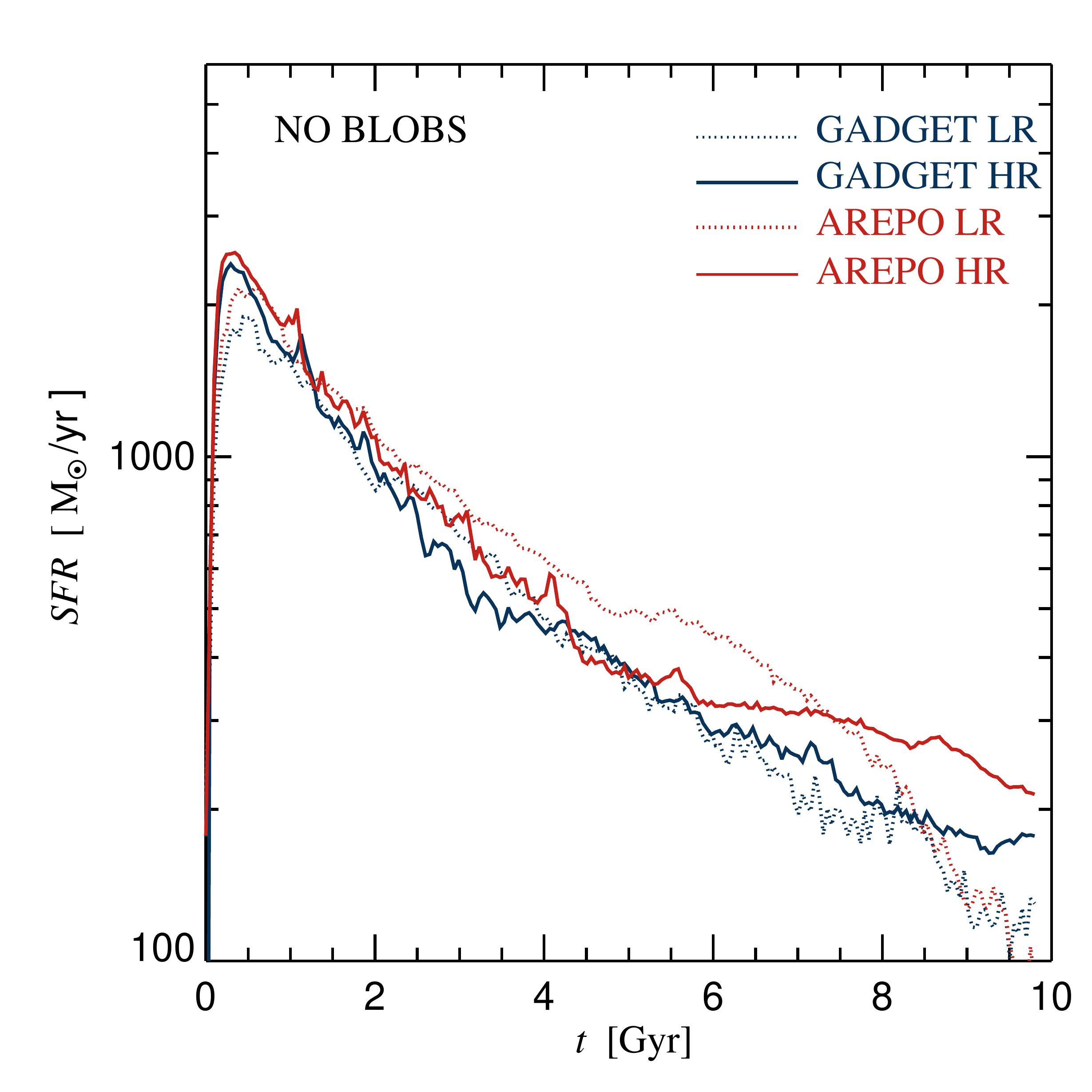}
\includegraphics[width=8.5truecm,height=8.2truecm]{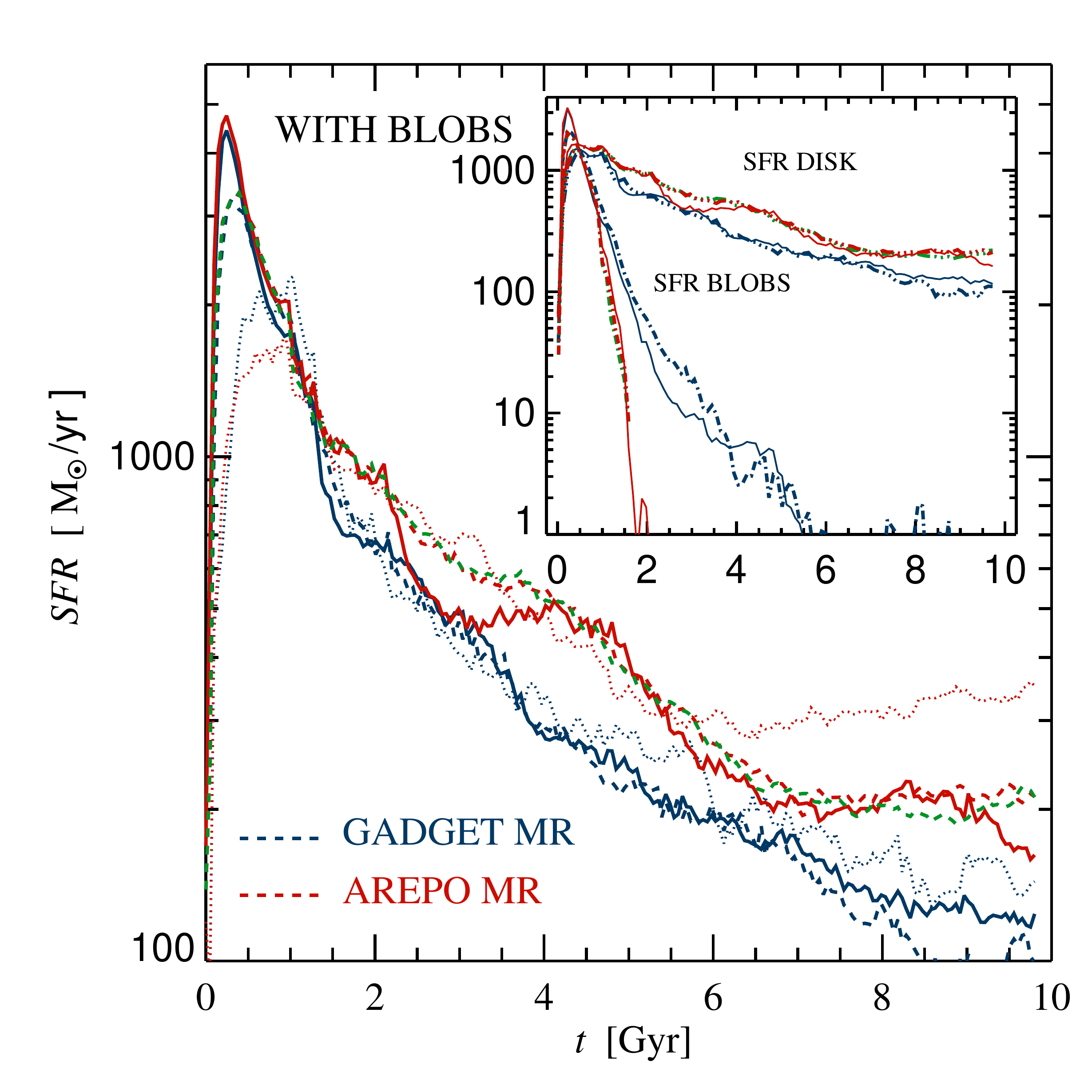}}}
\caption{Time evolution of the star formation rate of a $M_{\rm vir} =
  10^{14}\, {\rm M_{\odot}}$ isolated halo which radiatively cools and
  rotates. In the left-hand panel we show simulations without blobs, while in
  the right-hand panel the haloes contain $10$ orbiting substructures. In the
  right-hand panel, the curves give results from simulations with three
  different resolutions, with {\small GADGET} and {\small AREPO}, using
  $N_{\rm gas} = 10^4$ (LR), $N_{\rm gas} = 10^5$ (MR), and $N_{\rm gas} =
  10^6$ (HR) particles/cells. In the left-hand panel, only low and high
  resolution simulations are shown. In the simulations without blobs, the star
  formation rates are somewhat higher in {\small AREPO}, but the relative
  difference increases when we include the blobs. Higher star formation
    rates in {\small AREPO} emanate from the central disk which contains a
    greater amount of cold, star-forming gas which fills a larger area. In
  the inset plot, we compute the total star formation rate occurring within
  the blobs (dot-dashed and continuous lines) and the one coming from the disk
  (triple dot-dashed and continuous lines) for our intermediate and high
  resolution simulations with {\small GADGET} (blue curves) and {\small AREPO}
  (red curves). In the right-hand panel we also show intermediate
    resolution AREPO run with the fixed gravitational softening for the gas
    (green lines) which produces very similar SFRs (both in the disk and in
    the blobs) as the simulation with the adaptive softening.}
 \label{FigSFRDMBlobs}
\end{figure*}

In numerical experiments where we do not suppress star particles from forming,
we find very similar results. In Figure~\ref{FigRhoStarsDMBlobs}, we show
projected stellar density maps at $t = 6\,{\rm Gyr}$ using our intermediate
resolution simulations (which both in resolution and in time match the bottom
panels of Figure~\ref{FigRhoGasDMBlobs}). Clearly, the stellar distribution in
{\small GADGET} is more centrally concentrated, forming a small disk. On the
contrary, a well-defined extended stellar disk assembles in the moving-mesh
calculation, endowed with a central bar. We quantify the properties of gaseous
and stellar disk in Table~\ref{TabDiskPropDMBlobs}. The convergence rate of
the gas disk size in {\small GADGET} is even slower than in the simulations
without blobs, because additionally to the spurious transfer of the angular
momentum the damaging effect of the blobs is the highest in the low resolution
runs, where gas stripping is largely suppressed. This leads to larger
systematic difference in disk sizes between the two codes with
  resolution, but for the highest resolution runs {\small AREPO} disks are
  only about $\sim 30\%$ larger.

The interaction of cold blobs with the surrounding medium influences the
global star formation rates as well. In Figure~\ref{FigSFRDMBlobs}, we compare
the star formation histories in simulations without blobs (left-hand panel;
see Section~\ref{IsolatedHaloCoolRot}) with the numerical tests where haloes
are populated with $10$ substructures (right-hand panel). In the case without
blobs, the star formation rates are somewhat higher in {\small AREPO}
  especially for the low resolution run due to the more efficient gas cooling
  and thus larger amounts of cold gas available for star formation. For the
  high resolution run the differences in star formation rates between the
  codes are smaller, given that difference in the amount of cold gas between
  the codes amount to $\sim 30\%$. The difference in the star formation rates
between the codes becomes more pronounced once we include the
blobs. Initially, as can be seen from the inset plot where we show the star
formation rate from the blobs and from the disk separately, the star formation
rate within blobs contributes more than $50\%$ to the total star formation
rate, and it is very similar in both codes. This induces a much higher peak in
the total star formation rate in the right-hand panel with respect to the
simulations without blobs. For $t>1\,{\rm Gyr}$ the star formation rate in the
{\small AREPO} blobs dramatically drops, and at $\sim 2\,{\rm Gyr}$ it is
truncated altogether. This is due to ram-pressure stripping and
dynamical fluid instabilities that efficiently remove the gas from the
blobs. In fact, we can crudely estimate a characteristic Kelvin-Helmholtz
  timescale at the initial time using the equation from Section~\ref{Blob}
  (but note that here gas is self-gravitating), where for the blob density we
  take the average density within the blob radius $R_{\rm blob}$ (which we
  varied from the scale length parameter to the virial radius) and for the
  surrounding medium density we take the typical halo density at the position
  of the blobs. With these assumptions we obtain typical values for $t_{\rm
    KH}$ in the range of $1-3\,{\rm Gyr}$, depending on the blob positions,
  relative velocities and the choice of $R_{\rm blob}$. These $t_{\rm KH}$
  values are comparable to the timescale on which star formation within the
  blobs is suppressed in {\small AREPO}. In {\small GADGET} simulations, star
formation proceeds in the blobs even until $9\,{\rm Gyr}$s, albeit at a
progressively reduced rate. For $t > 1\,{\rm Gyr}$ the total star formation
rate comes mainly from the central regions. The central disks that form in the
{\small AREPO} simulations have higher star formation rates over many Gyrs,
which are typically larger by a factor of $2$ than the {\small GADGET}
results. Even in our highest resolution moving mesh simulation the star
  forming disk has roughly twice as large amount of cold gas, a difference
  which originates from the interaction of hotter infalling gas with the gas
  in the disk, as described in Section~\ref{IsolatedHaloCoolRot}. However,
  also the star-forming gas is distributed over a larger area. In fact, while
  within the half-mass radius of the {\small GADGET} disk the star formation
  rate is $\sim 40\%$ higher in the moving mesh run, outside of it is a factor
  $\sim 2.3$ higher, contributing about half to the total {\small AREPO} star
  formation rate. The reason why cold, star forming gas in {\small GADGET} is
  filling a smaller area in the disk and is more confined to the dense arm
  segments and blobs (the so called ``string-of-pears'', which can be also
  seen in the middle panel of Figure~\ref{FigRhoGasDMBlobs}), is due to the
  SPH surface tension originating at the interface between cold and hot media
  in relative motion. In the right-hand panel of
  Figure~\ref{FigSFRDMBlobs} we also show an identical intermediate resolution
  {\small AREPO} run, but with fixed gas gravitational softening, the same as
  in the matching {\small GADGET} run. It can be seen that the choice of gas
  gravitational softening in {\small AREPO}  does not affect our results in
  any significant way: both the evolution of the SFR in the central disk and
  in the orbiting blobs remains very similar, indicating that the differences
  that we see between the two codes are indeed entirely driven by
  hydrodynamical effects.

These findings have immediate consequences for more realistic
astrophysical situations. For example, \cite{Puchwein2010} have
simulated a high-resolution sample of galaxy clusters with {\small
  GADGET} finding that up to $30\%$ of intracluster stars form in
dense cold blobs -- remnants of infalling satellites. Suppressed
dynamical instabilities in {\small GADGET} enhance the probability of
survival for these dense blobs, which can then serve as sites of star
formation, thus possibly over-predicting the number of intracluster
stars. In fact, if we compute the total mass of cold baryons $M_{\rm
  cold}$ (stars plus gas above the density threshold for star
formation) in two cosmological simulations (for further details see
Paper~I) where in one we prevent the spawning of stars and in the
other we allow it (the simulations are otherwise identical), $M_{\rm
  cold}$ is found to be very similar in {\small GADGET} regardless of
whether stars are formed or not. Instead, in the simulations with
{\small AREPO}, we find that $M_{\rm cold}$ is systematically reduced
at lower redshifts if the spawning of stars is prevented. This
demonstrates explicitly that very low entropy material formed in
{\small GADGET} cannot easily be shredded and mixed with higher
entropy gas (at least in the absence of additional feedback processes
such as galactic winds or black hole heating), so that $M_{\rm cold}$
is preserved \citep[a similar conclusion has been reached
independently by][]{Hess2011}. In contrast, in the simulations with
{\small AREPO} if star formation is switched-off some of the cold
material can be stripped out of galaxies due to dynamical
instabilities, returning it to diffuse form and lowering $M_{\rm
  cold}$. Importantly, this also implies that the number of stars
formed in {\small AREPO} will be much more sensitive to the
characteristic timescale for star formation and to the numerical
resolution (needed to resolve fluid instabilities) than is the case
for {\small GADGET}. In a recent paper by \citet{Agertz2011} it
  has been shown that lower star formation efficiency leads to the
  production of larger disks in cosmological simulations.  While
  this result is in line with our findings, it is important to
  recognize that the actual cause is quite different: in the study by
  \citet{Agertz2011}, differences in the physical modelling of star
  formation in cosmological simulations affect the disk sizes, while
  here the cause lies in different accuracies of the hydro
  solvers involved.

\subsection{Gravitational N-body heating}\label{N-bodyHeating}

\begin{figure}\centerline{
\includegraphics[width=8.3truecm,height=21.truecm]{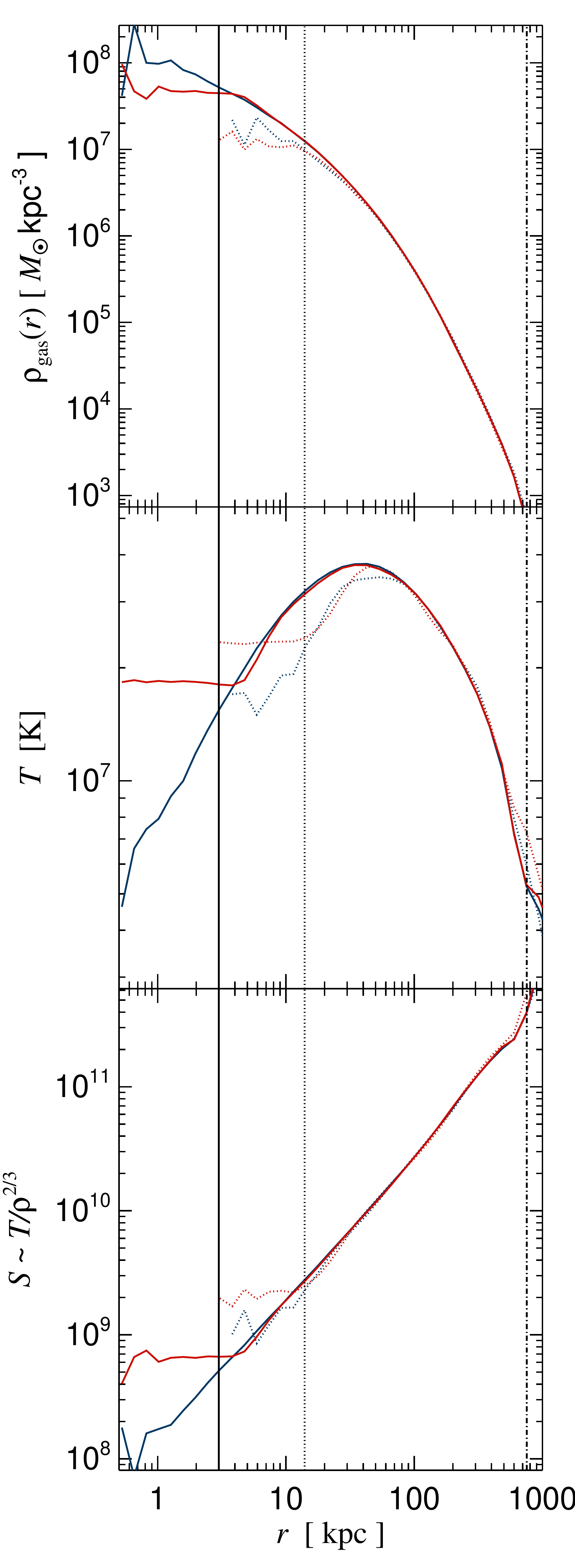}}
\caption{Radial profiles of gas density, temperature and entropy at
  time $t \sim 10\,{\rm Gyr}$ for {\small GADGET} (blue lines) and for
  {\small AREPO} (red lines). For each code we show two different
  resolution runs: $N_{\rm gas} = 10^4$ and $r_{\rm soft} = 14.0
  \,{\rm kpc}$ (dotted lines), and $N_{\rm gas} = 10^6$ and $r_{\rm
    soft} = 3.0 \,{\rm kpc}$ (continuous lines). The vertical black
  lines with the same style indicate the softening scale of the dark
  matter potential, whereas vertical dot-dashed lines show the
  virial radius of the system.}
\label{EquilHaloDMNoRot}
\end{figure}

\begin{figure}\centerline{
\includegraphics[width=8.3truecm,height=8.truecm]{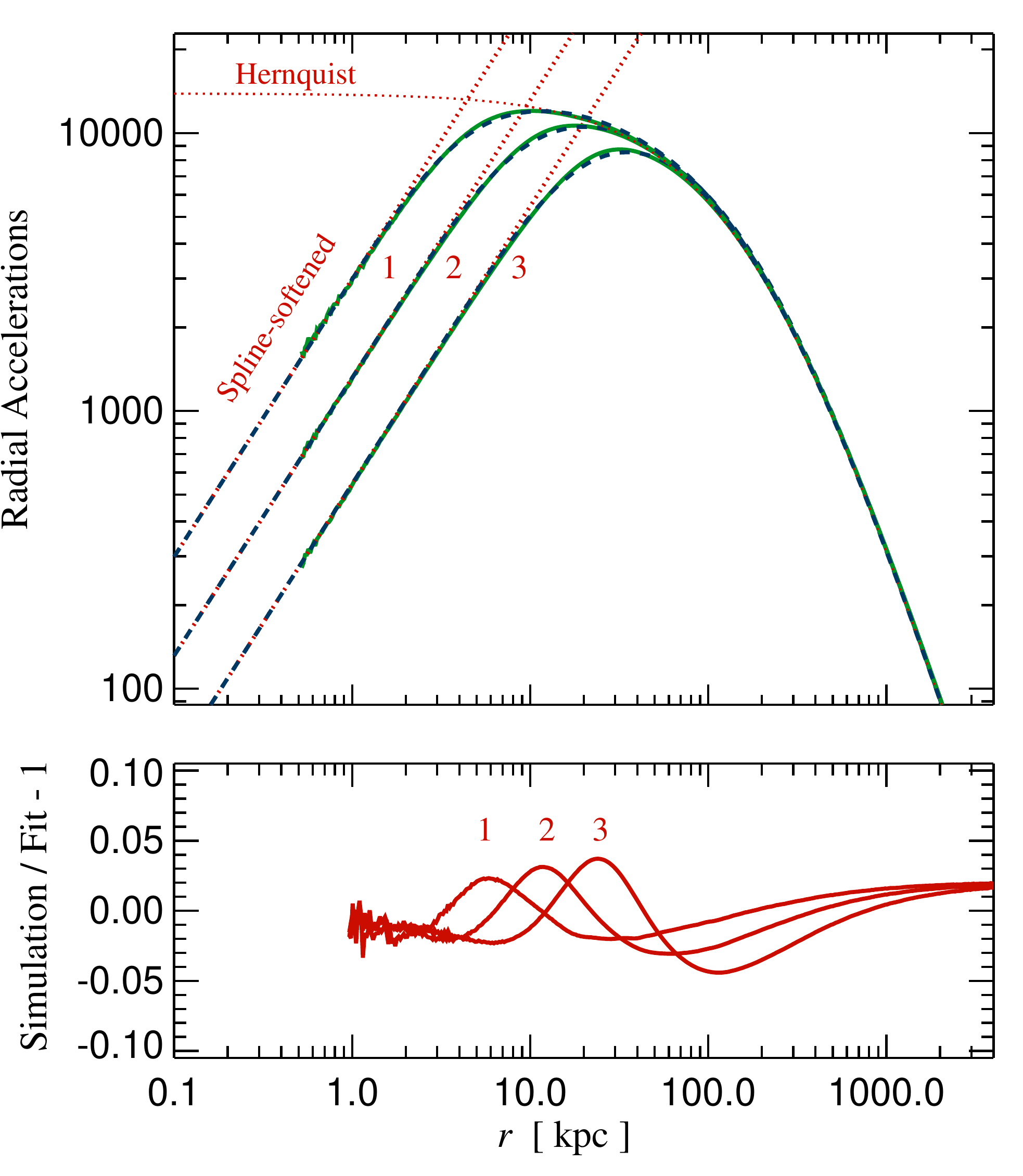}}
\caption{Comparison of fitting functions for spline-softened
  gravitational accelerations (equation~\ref{EqConvolved}; blue dashed
  lines) with actual gravitational accelerations obtained from very
  high resolution runs with matching $r_{\rm soft}$ values (green
  continuous lines). The limiting cases of spline-softened
  accelerations for small and large radii are illustrated with dotted
  red lines. The bottom panel shows the residuals of our fitting
  functions over the relevant range of radii.}
\label{FigHernquistFit}
\end{figure}

We now investigate potential artificial gas heating due to the Poisson
noise induced by the finite number of particles in dark matter
haloes. \citet{Steinmetz1997} outlined the analytic theory of this
effect, and confirmed it with non-radiative and radiative numerical
experiments that quantified gravitational N-body noise present in
structure formation simulations. For an equilibrium system they
defined a characteristic N-body heating timescale which is
proportional to the cube power of the dark matter velocity dispersion
and inversely proportional to the dark matter particle mass and dark
matter density. Due to this inverse proportionality to the dark matter
density, N-body heating is expected to be strongest in the innermost
regions of haloes. From their analysis it follows that the dark matter
particle mass adopted in numerical simulations should be lower than a
critical mass which is of order of few times $10^9 {\rm M_{\odot}}$
for galaxy clusters (see their equation [10]), otherwise the radiative
cooling losses may be overwhelmed by spurious heating.

Since nowadays the typical mass resolution of hydrodynamical
cosmological simulations has improved dramatically, with simulated
galaxy clusters containing several times $10^6$ up to $10^7$ particles in
zoom-in runs \citep{Dolag2009, Sijacki2009, Vazza2010}, possible
numerical artifacts due to a grainy dark matter distribution are
rarely addressed in the literature \citep[but see e.g.][]{Kay2000,
  Borgani2006, VazzaEntropy2011}. Recently, \citet{Springel2010} has pointed out that
mesh based codes are potentially more severely affected by
gravitational N-body heating than SPH codes, due to their better
ability to detect weak shocks, which in this case is an unwanted
feature.

Here we perform a number of idealized test problems aimed at
explicitly addressing the N-body heating problem, and in particular to
understand whether there are systematic differences between {\small
  GADGET} and {\small AREPO} in this respect. Note that intentionally
all previous tests (except for Section~\ref{GeneralBlobCooling}) have
been performed either without any dark matter component or with a
static dark matter potential to avoid such possible Poisson-noise
imprints.

We first consider our standard isolated halo, where we replace the
static dark matter potential with a live dark matter halo in which
dark matter particles have a characteristic velocity dispersion for
the given mass of the system (but there is no net rotation). As in the
case of the rigid potential, we populate the halo with gas particles
in hydrostatic equilibrium which are initially at rest. We include
self-gravity of the gas component and evolve the system
non-radiatively for $10\,{\rm Gyr}$.

\begin{figure*}\centerline{\hbox{
\includegraphics[width=8.3truecm,height=20.truecm]{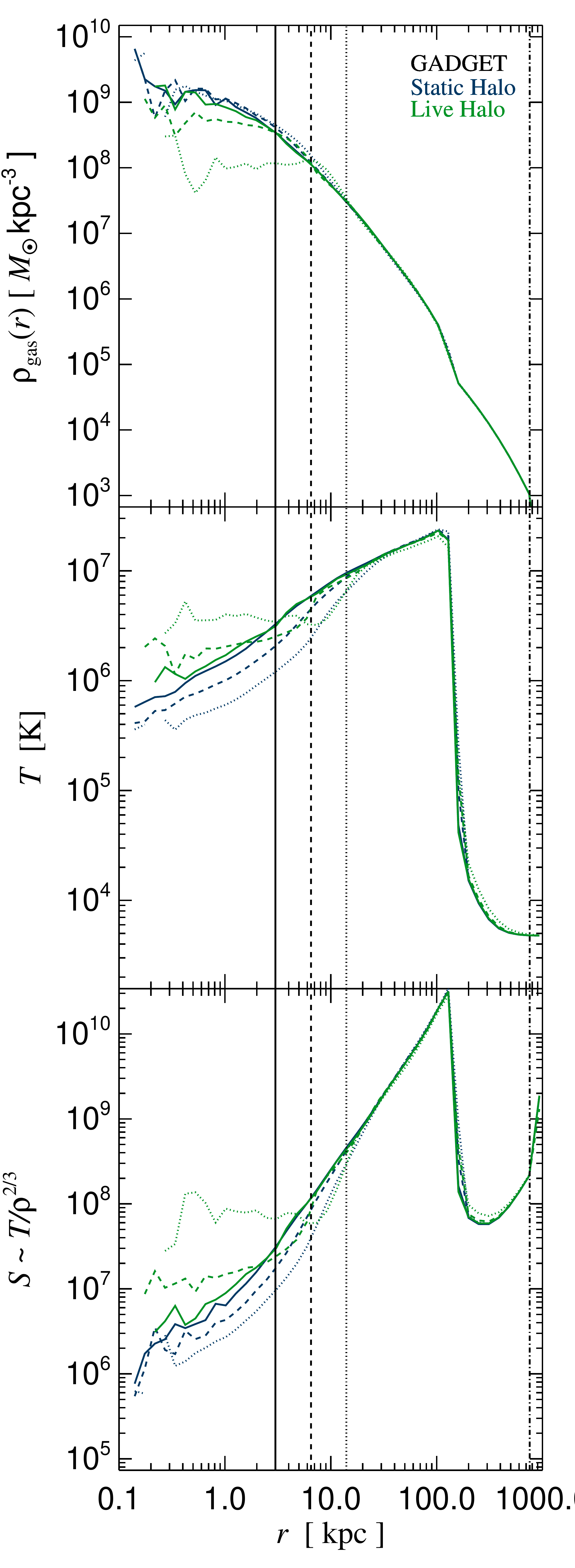}
\includegraphics[width=8.3truecm,height=20.truecm]{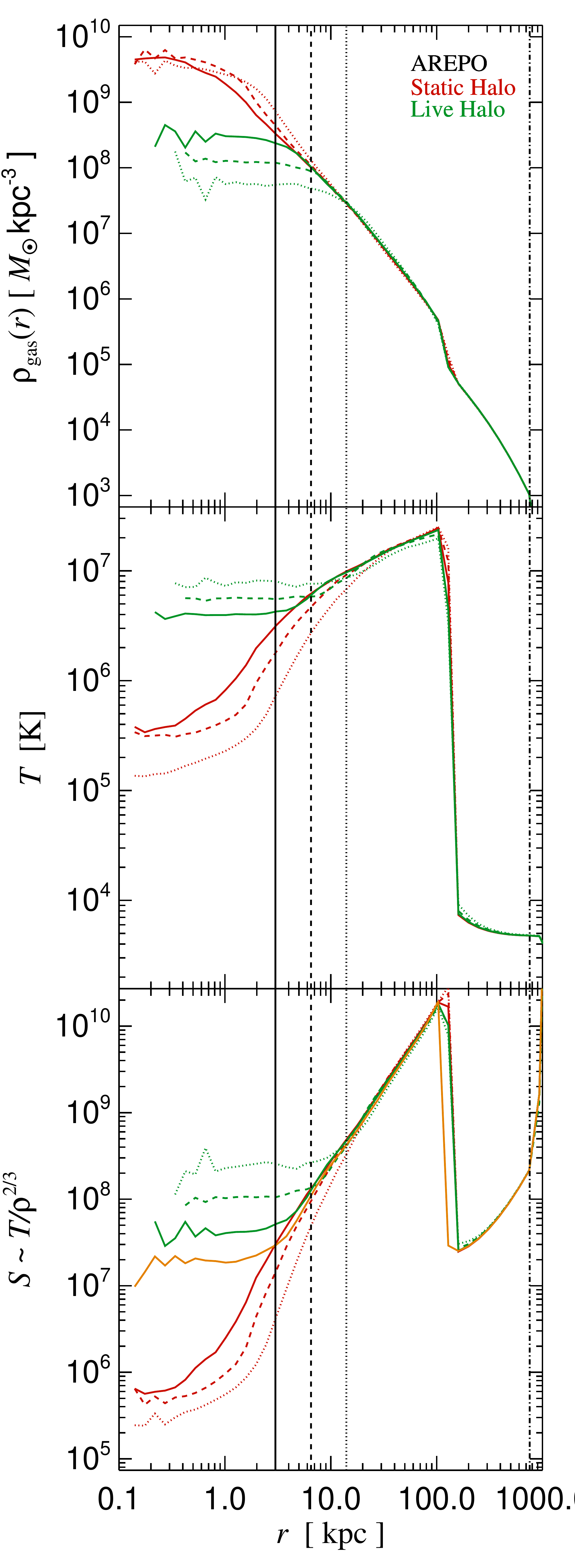}}}
\caption{Radial profiles of gas density, temperature and entropy at
  time $t = 0.3\,{\rm Gyr}$ for {\small GADGET} (left-hand panels;
  blue lines: static dark matter haloes; green lines: live dark matter
  haloes) and for {\small AREPO} (right-hand panels; red lines: static
  dark matter haloes; green lines: live dark matter haloes).  For each
  code we show three different resolution runs: $N_{\rm DM} = 10^4$
  and $r_{\rm soft} = 14.0 \,{\rm kpc}$ (dotted lines), $N_{\rm DM} =
  10^5$ and $r_{\rm soft} = 6.5 \,{\rm kpc}$ (dashed lines) and
  $N_{\rm DM} = 10^6$ and $r_{\rm soft} = 3.0 \,{\rm kpc}$ (continuous
  lines), while $N_{\rm gas} = 10^6$ is kept fixed. In the right-hand
  panel we also show the entropy profile obtained with the moving-mesh
  code in the entropy mode for our highest resolution simulation
  (continuous orange line). The vertical black lines with the same
  style indicate the softening scales of the dark matter
  potential. The black vertical dot-dashed lines denote the virial
  radius of the system.}
\label{FigColdInflowLive}
\end{figure*}

In Figure~\ref{EquilHaloDMNoRot}, we show radial profiles of gas
density, temperature and entropy at the final time for both codes at
two resolutions: $N_{\rm gas} = N_{\rm DM} = 10^4$ particles (dotted
lines) and $N_{\rm gas} = N_{\rm DM} = 10^6$ particles (continuous
lines). The centre of the system is defined by the position of the
most bound particle. For cluster-centric distances greater than the
gravitational softening value, $r_{\rm soft}$, the entropy profiles of
{\small GADGET} and {\small AREPO} agree to within $\sim 20\%$ for the
low resolution runs, and to better than $10\%$ for the high resolution
simulation. For $r < r_{\rm soft}$, the differences between the codes
in the entropy profile amount up to a factor of $2$, with the gas
entropy in {\small AREPO} being systematically higher. An extra test
simulation performed with {\small AREPO} in its dual entropy mode
(where entropy instead of energy is explicitly conserved for cells
whose Riemann problems with adjacent cells all have Mach number less
than $1.1$) resulted in identical radial profiles as for the {\small
  AREPO} simulation in the standard energy mode. This clearly
indicates that the small entropy core which develops in {\small AREPO}
is not caused by weak shocks generated by the grainy dark matter
potential. Rather, central gas particles/cells acquire small scale
velocities when simulated with live haloes. In the case of the
moving-mesh code, these velocities lead to fluid mixing and
temperature equilibration in the innermost regions, while in {\small
  GADGET} such mixing does not occur. Note, however, that in the
gravitating systems, $r_{\rm soft}$ represents the minimum spatial
scale below which the simulation results are not trustworthy due to
discreteness effects. Therefore, provided that gravitational softening
lengths are chosen conservatively, the gas properties of galaxy
clusters in an equilibrium configuration simulated with {\small
  GADGET} and {\small AREPO} match very well.

We now consider a more challenging problem, where we simulate the infall of cold
gas into dark matter haloes. This is the same problem discussed in
Section~\ref{ColdInflow}, but here we compare outcomes from live versus static
dark matter haloes. To match numerical experiments with rigid and live haloes as
closely as possible we adopt the following procedure. 

For live haloes, the motion of dark matter particles is prevented in the  code
to avoid local deformations of the gravitational potential which will not be
present in the static case. Effectively, in this way the distribution of dark
matter is ``frozen'', and live haloes are coarse grained representations of
static haloes.   

For static haloes, we convolve the analytic Hernquist potential with a spline
softened potential in the centre, such that gas particles feel exactly the
same gravitational acceleration as in the case of live haloes, where the
central potential needs to be softened to minimize effects from two-body
encounters. For the Hernquist potential, gravitational accelerations are given
by  \be g(r) = - \frac{G M_{\rm vir}}{(r + a)^2}\,.   \ee   We modify the
gravitational acceleration felt by gas particles in the code, viz.   \be g(r)
= - \frac{G M_{\rm vir}}{ \big(\frac{32 \pi r}{3 h_0^3 M_{\rm vir}}\big)^{-1}
  + (r + a)^2 \exp(-h_1/r)}\,,  \label{EqConvolved} \ee  where $h_0$ and $h_1$
are free coefficients (to first approximation $h_0 \sim h_1 \sim r_{\rm
  soft}$). We determine the value of $h_0$ and $h_1$ for three different
resolution runs by fitting equation~(\ref{EqConvolved}) to the gravitational
acceleration of a live ``frozen'' halo with $10^7$ dark matter particles
simulated three times assuming $r_{\rm soft}$ values appropriate for the
numerical resolutions we want to investigate. 

In Figure~\ref{FigHernquistFit} we illustrate the outcome of this
method. In the upper panel, the green lines denote the radial
gravitational acceleration of the live ``frozen'' halo with $10^7$
dark matter particles, adopting $r_{\rm soft} = 3$, $6.5$ and
$14\,{\rm kpc}$ (indicated by numbers $1$, $2$ and $3$),
respectively. Blue dashed lines are our best fit functions to these
simulations using equation~(\ref{EqConvolved}). Dotted red lines
indicate the radial acceleration in the case of the analytic Hernquist
profile and of the spline-softened potential, which are the limiting
cases of equation~(\ref{EqConvolved}) for large and small radii,
respectively. The bottom panel of Figure~\ref{FigHernquistFit} shows
how accurate our modified gravitational acceleration is as a function
of cluster-centric distance.
  
With the procedure described above we perform simulations of radial infall of
cold gas into live (``frozen'') haloes. We keep the number of gas resolution
elements the same and equal to $N_{\rm gas} = 10^6$ (corresponding to $r_{\rm
  soft, gas} = 3\,{\rm kpc}$ for {\small GADGET}, while gravitational
softenings are adaptive in {\small AREPO} but with a floor of $3\,{\rm
    kpc}$), while we increase the number of dark matter particles from
$N_{\rm DM} = 10^4$ ($r_{\rm soft, DM} = 14\,{\rm kpc}$), to $10^5$ ($r_{\rm
  soft, DM} = 6.5\,{\rm kpc}$) and $10^6$ ($r_{\rm soft, DM} = 3\,{\rm
  kpc}$). For each live halo we run a matching static halo simulation where
the number of gas particles/cells is kept the same, i.e.~$N_{\rm gas} =
10^6$. In all runs, gas self-gravity is neglected and there are no radiative
losses. In close encounters of gas and dark matter, the effective
gravitational softening is the maximum between $r_{\rm soft, gas}$ and $r_{\rm
  soft, DM}$, such that $r_{\rm soft, DM}$ is the relevant length scale of the
problem.

In Figure~\ref{FigColdInflowLive} we show radial profiles of gas
density, temperature and entropy at time $t = 0.3\,{\rm Gyr}$
(analogous to Figure~\ref{FigColdInflow}). The left-hand panels are
for simulations of live and static haloes with {\small GADGET}, while
the right-hand panels show the results with {\small AREPO}. Due to
gravitational N-body heating there are systematic differences in the
central gas properties. The central gas density is lower, while the
central gas temperature and entropy are higher for the live haloes,
due to spurious transfer of energy from dark matter to gas, which
heats the gas and makes it expand. In general, N-body heating effects
are largest for $N_{\rm DM} = 10^4$ and smallest for $N_{\rm DM} =
10^6$, and they are confined to spatial regions within $r_{\rm soft,
  DM}$, which is reassuring. We can see by comparing the left-hand to
the right-hand panels that {\small AREPO} is indeed much more
sensitive to spurious gravitational heating, as discussed in
\citet{Springel2010}, with the central entropy boosted by two orders
of magnitude. In the right-hand panel of
Figure~\ref{FigColdInflowLive}, we also show radial entropy profile
obtained with {\small AREPO} in the dual entropy mode (where entropy
instead of energy is explicitly conserved for cells with all Mach
numbers less than $1.1$) with $N_{\rm DM} = 10^6$ particles (orange
continuous line). This indicates that at least part of the central
entropy core is generated by the weak shocks which gas cells
experience when moving through the grainy dark matter potential.

At the final time $t = 2.45\,{\rm Gyr}$, when the system has reached
an equilibrium state, the difference between the matching live and
static halo runs for $r > r_{\rm soft, DM}$ is $\sim 10\%$, $\sim
5\%$, and $\sim 5\%$, respectively, for low, intermediate and high
resolution in {\small GADGET}, while it is somewhat higher in {\small
  AREPO}, i.e.~$\sim 15-20\%$, $10\%$, and $5\%$, respectively. For $r
< r_{\rm soft}$ the ratios of central entropy values are $\le 10$,
$\le 4$ and $\le 2$ in {\small GADGET} and $\le 300$, $\le 150$ and
$\le 25$ in {\small AREPO}, for low, intermediate and high resolution
runs. Even though the amount of artificial heating is significantly
larger in {\small AREPO} than it is for {\small GADGET}, the
departures between static and live halo radial profiles always occur
within $r_{\rm soft, DM}$ for both codes. Provided that gravitational
softening values for the dark matter component are chosen cautiously,
our numerical experiments hence indicate that the systematic
discrepancy in central entropy values found between SPH and mesh-based
codes for the Santa Barbara cluster comparison project
\citep{Frenk1999, Springel2010} is unlikely to be due to effects from
a grainy dark matter potential during cosmological structure
formation.

\section{Discussion and Conclusions} \label{Conclusions}

In this study we have carried out a detailed comparison between the
SPH code {\small GADGET} and the new moving-mesh code {\small AREPO}
on a number of hydrodynamical test problems, which are crucial for
understanding cosmological simulations of galaxy formation. In a
purely hydrodynamical regime without gas self-gravity or an external
gravitational potential we have first carried out a set of numerical
experiments previously considered in the literature, some of which
have rarely been shown for SPH codes, such as the $2$D implosion
test. We have then focused on idealized non-radiative galaxy cluster
simulations, specifically aimed towards benchmarking differences in
hydro solvers for problems with shocks and fluid instabilities. In
simulations where radiative losses were included we have analyzed the
amount of baryons which cool from the hot halo atmospheres in {\small
  GADGET} and {\small AREPO}, both for rotating and non-rotating
haloes. In the former case, we have also studied how the central
baryonic disks form, and how orbiting substructures affect the disk
morphology and the star formation rate. Finally, we have constructed
special test problems designed to gauge the effect of gravitational
N-body heating on the gaseous properties of the haloes. Our main
conclusions are as follows:

\begin{itemize}

\item While post-shock fluid properties are captured well both in {\small
  GADGET} and {\small AREPO} in the case of $1$D shock tube tests with high
  Mach numbers, the shocks are significantly broader in {\small GADGET}, and
  substantial post-shock oscillations develop, largely because of an
  inadequate treatment of the initial contact discontinuity. {\small AREPO} in
  the moving-mesh mode preserves the contact discontinuity much more
  accurately, but it is broadened if we employ a static mesh, due to larger
  numerical diffusivity in this case. Note that the more accurate
    treatment of the contact discontinuity in the moving mesh code can lead to
    a larger `wall heating' effect \citep{Rider2000} than in static grid
    codes, which tend to wash-out at some level the initial start-up errors at
    the contact discontinuity \citep[see also description of the Noh problem
      in][]{Springel2010}.

\item For shocks with complicated geometries in multi-dimensions, differences
  between {\small GADGET} and {\small AREPO} are more striking. Even though
  the global fluid properties remain similar, the sampling of the fluid
  properties is much noisier in {\small GADGET}, and the development of
  dynamical fluid instabilities is inhibited. These problems are not specific
  to {\small GADGET}, but are inherent to the standard SPH method as a whole
  \citep[see also][and discussion in Section~\ref{OtherSPH}]{Agertz2007,
    SpringelRev2010}. The suppression of fluid instabilities reduces the
  amount of entropy generation by mixing and artificially prolongs the
  lifetime of gaseous structures which are moving through a medium with a
  different density. Additionally, vorticity generation in the wake of curved
  shocks due to the baroclinic source term is largely suppressed in {\small
    GADGET}, which directly impacts angular momentum transfer by vortices and
  the level of generated turbulence.

\item These fundamental differences between {\small GADGET} and {\small AREPO}
  identified in simple hydrodynamical test problems affect the properties of
  gas in more realistic, cosmologically motivated simulations as
  well. Specifically, in non-radiative idealized simulations of merging galaxy
  clusters and in simulations of isolated haloes with orbiting gaseous
  substructures we find that in {\small AREPO}: {\it i)} an entropy core is
  produced in the centre due to more efficient fluid mixing, {\it ii)} the gas
  stripping rate from the orbiting substructures is larger, and {\it iii)}
  more vorticity is produced in the wake of curved shocks. These findings
    are in line with results of previous works which simulated similar
    problems with AMR codes \citep[e.g.][]{Mitchell2009, Vazza2011,
      VazzaEntropy2011}. Moreover, unphysical dissipation of shocks and
  subsonic turbulence in {\small GADGET}, as shown by \citet{Bauer2011} (see
  also Section 4.2 of Paper I), leads to the heating of the halo outskirts
  rather than of the central regions.

\item In radiative simulations of isolated haloes without any net rotation,
  gas cooling and condensation into stars proceeds in a very similar fashion
  in {\small GADGET} and {\small AREPO}, given that the gas properties at the
  cooling radius match closely. However, for spinning haloes there is a net
  difference in the total amount of cold baryons, which is higher in {\small
    AREPO}, and this difference persists at a level of about $10-15\%$ in our
  highest resolution runs. Baryonic disks which form due to dissipative
  collapse of rotating gas are systematically larger in {\small AREPO} at low
  resolution, and the convergence rate of the gas disk sizes is higher.
 
\item In numerical experiments where we follow the interaction between
  a forming central disk and $10$ orbiting substructures in an
  isolated halo which radiatively cools, the final disk morphology is
  significantly different. While in {\small AREPO} an extended disk is
  produced with well developed spiral arms and a central bar, in
  {\small GADGET} the disk is more centrally concentrated and
  amorphous. Orbiting substructures are much more efficiently stripped
  of their gas content in the moving-mesh calculation and the material
  is incorporated into the host halo atmosphere. Instead, in {\small
    GADGET} gaseous substructures are more coherent, thus they lose
  more angular momentum from hydro-/dynamical friction, and when passing
  through the central disk they induce morphological transformations.

\item While star formation is more readily truncated in infalling
  substructures due to gas stripping, extended gaseous disks in {\small AREPO}
  have significantly larger star formation rates for many Gyrs than is
  the case for {\small GADGET}.  

\item Due to its better ability to detect weak shocks, {\small AREPO}
  is more sensitive to gravitational N-body heating
  \citep{Springel2010}. This is confirmed by our specifically designed
  numerical experiments, which allow us to quantify the magnitude of
  spurious gas heating for simulations of haloes with a finite number
  of dark matter particles with respect to the analytic dark matter
  potentials. Nonetheless, the spatial extent of this artificial
  heating is reassuringly constrained to lie within the gravitational
  softening length for typical set-ups, which is a scale below which
  simulation results are not trustworthy due to resolution effects anyway.

\end{itemize}

The numerical experiments presented in this study clearly demonstrate
that several important shortcomings of the SPH solver not only affect
idealized test problems but are equally detrimental in more realistic
setups relevant for structure formation, and ultimately in full
cosmological simulations. This is especially the case because of: {\it
  i)} the complicated flows involving multiphase media which are the
norm in cosmological simulations, and {\it ii)} the hierarchical
nature of structure formation where low mass systems are always poorly
resolved. In both of these regimes the hydrodynamical solver of the
standard SPH method exhibits the largest inaccuracies. On the other
hand, our numerical tests confirm and significantly extend the
findings of \citet{Springel2010} that the moving-mesh code {\small
  AREPO} delivers a physically more accurate representation of the
evolution of inviscid gases. Note that in the current work we
  deliberately kept the modelling of the baryon physics at a very
  simple level, so as to isolate the differences between the hydro
  solvers in as clean a manner as possible. While inclusion of more
  realistic feedback mechanisms, such as supernovae winds and AGN
  heating, will likely modify the properties of the simulated galaxies
  significantly, it is of prime importance to disentangle numerical
  inaccuracies of the hydro solver from the uncertainties of the
  feedback physics modelling. It is hence clear that cosmological
simulations with {\small AREPO} have the potential to provide a much
more realistic description of structure formation in the Universe (see
also Papers~I and II), something we will explore in more depth in
forthcoming studies.

\section*{Acknowledgements} 
We would like to thank Andrew MacFadyen and Daniel Eisenstein for very useful
discussions and suggestions on the topic, and Manfred Kitzbichler and Diego
Mu{\~n}oz for carefully reading the manuscript. We would like to thank
  the anonymous referee for many constructive suggestions, which helped to
  improve the presentation of the results. DS acknowledges NASA Hubble
Fellowship through grant HST-HF-51282.01-A. DK acknowledges NASA Hubble
Fellowship through grant HST-HF-51276.01-A. The computations in this paper
were performed on the Odyssey cluster supported by the FAS Science Division
Research Computing Group at Harvard University.

\bibliographystyle{mn2e}

\bibliography{paper}

\appendix

\section{Isolated haloes in hydrostatic
  equilibrium} \label{AppIsolatedHaloEquil}

\begin{figure}\centerline{
\includegraphics[width=7.3truecm]{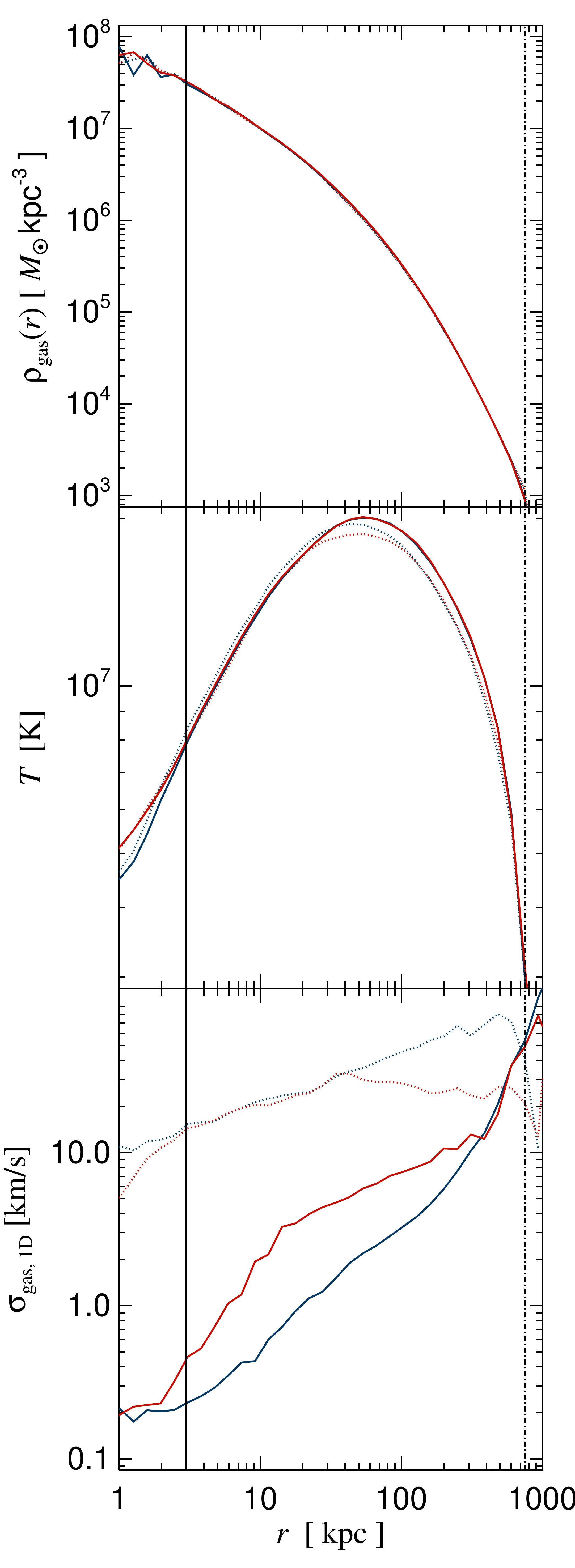}}
\caption{Radial profiles of gas density, temperature and $1$D velocity
  dispersion at time $t = 0.05\,{\rm Gyr}$ (dotted lines) and $t = 2.45\,{\rm
    Gyr}$ (continuous lines) for {\small GADGET} (blue lines) and {\small
    AREPO} (red lines) with $N_{\rm gas} = 10^6$ in a static dark matter halo
  with a Hernquist profile. Vertical black lines indicate the softening scales
  of the gas equal to $r_{\rm soft} = 3.0 \,{\rm kpc}$. The black vertical
  dot-dashed line denotes the virial radius of the system.}
\label{A1}
\end{figure}

Here we show how gaseous spheres placed in hydrostatic equilibrium
within a static Hernquist dark matter halo evolve with time to determine
the level of accuracy of our initial conditions. The general set-up is
described in detail in Section~\ref{IsolatedHaloEquil}. Specifically
for the figures presented here we assume that the gas is
self-gravitating and that it has initially zero velocity.

\begin{figure}\centerline{
\includegraphics[width=7.3truecm]{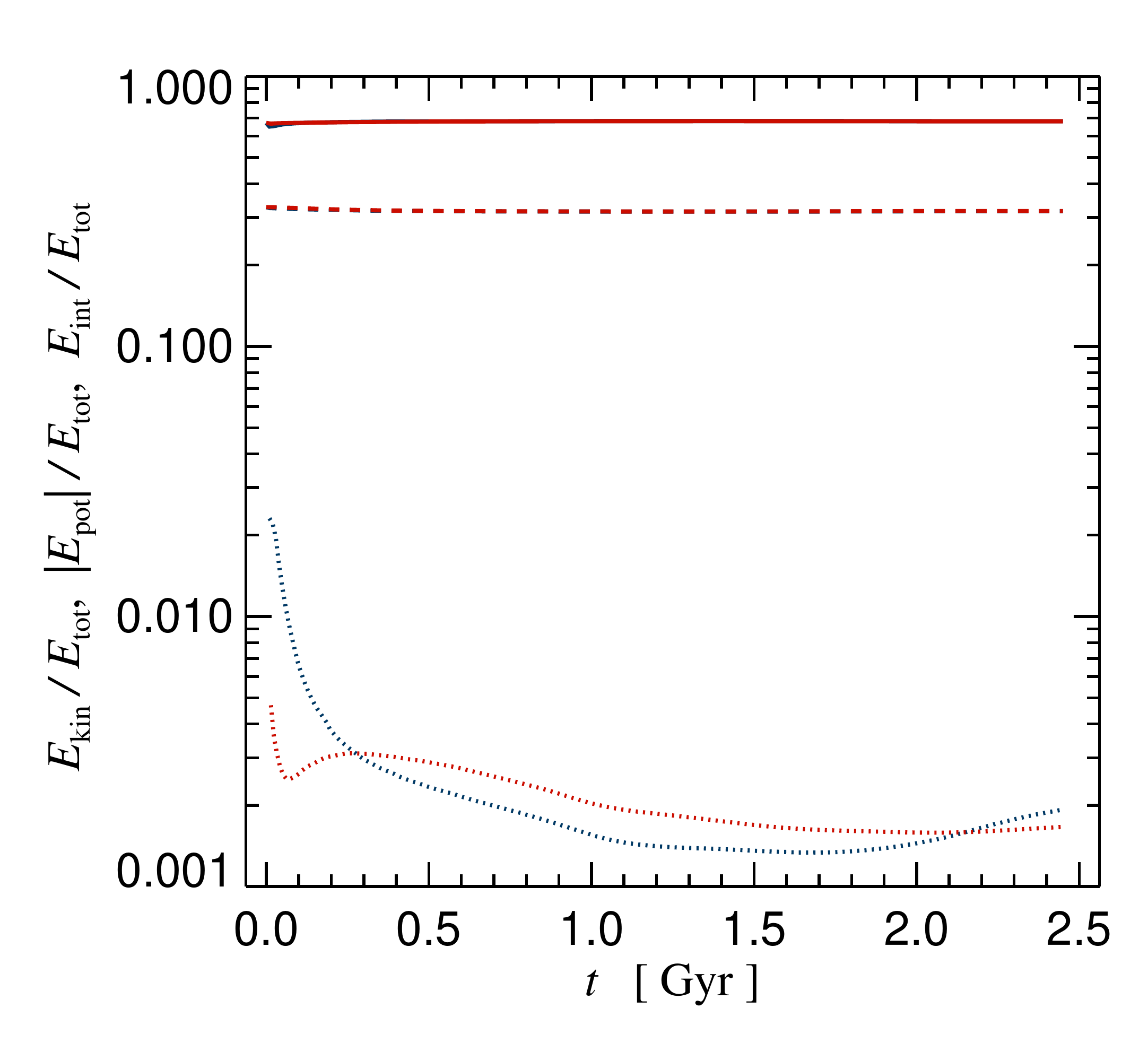}}
\caption{Time evolution of the ratio of gas kinetic (dotted lines), potential
  (dashed lines), and internal (continuous lines) to the total gas energy in
  {\small GADGET} (blue) and {\small AREPO} (red) simulations of isolated halos in
  hydrostatic equilibrium. Gas is self-gravitating and represented by $N_{\rm
    gas} = 10^6$ resolution elements and is evolved within a static Hernquist
  dark matter halo.}
\label{A2}
\end{figure}

In Figure~\ref{A1}, we show gas density, temperature and $1$D velocity
dispersion radial profiles at $t = 0.05\,{\rm Gyr}$ (dotted lines) and
$t = 2.45\,{\rm Gyr}$ (continuous lines) for {\small GADGET} (blue
lines) and {\small AREPO} (red lines) with $N_{\rm gas} = 10^6$
resolution elements. As discussed in Section~\ref{IsolatedHaloEquil},
due to the Poisson sampling of the gas positions in the initial
conditions they are not perfectly relaxed, which leads to the
development of small-scale random motions, as evidenced by a non-zero
gas velocity dispersion. Over time these residual gas velocities are
dissipated, as can be seen from the bottom panel of
Figure~\ref{A1}. This leads to a slight readjustment of the
temperature distribution, which is very similar in both codes. Note,
however, that the kinetic energy of the random gas motions is at all
times a very small fraction of both the potential and the internal energy, as
illustrated in Figure~\ref{A2}, amounting to only about $0.2\%$ of the
total gas energy after $2.45\,{\rm Gyr}$.

\end{document}